\documentclass[12pt]{article}
\usepackage{epsfig}
\usepackage{amssymb}
\usepackage{axodraw}

\def\lsim{\mathrel{\rlap{\lower4pt\hbox{\hskip1pt$\sim$}}
    \raise1pt\hbox{$<$}}}                
\def\gsim{\mathrel{\rlap{\lower4pt\hbox{\hskip1pt$\sim$}}
    \raise1pt\hbox{$>$}}}                

\newcommand{\alphas}{\alpha_{\mathrm{s}}}

\newcommand{\ee}{\mathrm{e}^+\mathrm{e}^-}
\newcommand{\xF}{x_{\mathrm{F}}}
\newcommand{\pt}{p_{\perp}}
\newcommand{\pthat}{\hat{p}_{\perp}}

\newcommand{\ptmin}{\hat{p}_{\perp\mathrm{min}}}
\newcommand{\kt}{k_{\perp}}
\newcommand{\Dy}{{\Delta y}}
\newcommand{\cl}{\mathrm{cl}}
\renewcommand{\b}{\mathrm{b}}
\renewcommand{\c}{\mathrm{c}}
\renewcommand{\d}{\mathrm{d}}
\newcommand{\e}{\mathrm{e}}
\newcommand{\g}{\mathrm{g}}
\newcommand{\p}{\mathrm{p}}
\newcommand{\q}{\mathrm{q}}
\newcommand{\s}{\mathrm{s}}
\renewcommand{\t}{\mathrm{t}}
\renewcommand{\u}{\mathrm{u}}
\renewcommand{\B}{\mathrm{B}}
\newcommand{\D}{\mathrm{D}}
\renewcommand{\H}{\mathrm{H}}
\newcommand{\Q}{\mathrm{Q}}
\newcommand{\W}{\mathrm{W}}
\newcommand{\Z}{\mathrm{Z}}
\newcommand{\bbar}{\overline{\mathrm{b}}}
\newcommand{\cbar}{\overline{\mathrm{c}}}
\newcommand{\dbar}{\overline{\mathrm{d}}}
\newcommand{\pbar}{\overline{\mathrm{p}}}
\newcommand{\qbar}{\overline{\mathrm{q}}}
\newcommand{\sbar}{\overline{\mathrm{s}}}
\newcommand{\ubar}{\overline{\mathrm{u}}}
\newcommand{\Bbar}{\overline{\mathrm{B}}}
\newcommand{\Dbar}{\overline{\mathrm{D}}}
\newcommand{\Qbar}{\overline{\mathrm{Q}}}

\newcommand{\Py}{{\sc{Pythia}}}

\newenvironment{Itemize}{\begin{list}{$\bullet$}%
{\setlength{\topsep}{0.2mm}\setlength{\partopsep}{0.2mm}%
\setlength{\itemsep}{0.2mm}\setlength{\parsep}{0.2mm}}}%
{\end{list}}
\newcounter{enumct}
\newenvironment{Enumerate}{\begin{list}{\arabic{enumct}.}%
{\usecounter{enumct}\setlength{\topsep}{0.2mm}%
\setlength{\partopsep}{0.2mm}\setlength{\itemsep}{0.2mm}%
\setlength{\parsep}{0.2mm}}}{\end{list}}
 
\newlength{\abstwidth}
\newlength{\captivewidth}
\newcommand{\captive}[1]{\rule{5mm}{0mm}%
\begin{minipage}{\captivewidth}%
\caption[small]{#1}\end{minipage}}
 
\begin{document}
 
\sloppy

\pagestyle{empty}

\begin{flushright}
LU TP 00--16 \\
hep-ph/0005110\\
May 2000
\end{flushright}

\vspace{\fill}

\begin{center}
{\LARGE\bf Production and Hadronization}\\[4mm]
{\LARGE\bf of Heavy Quarks}\\[10mm]
{\Large E. Norrbin\footnote{emanuel@thep.lu.se} and %
T. Sj\"ostrand\footnote{torbjorn@thep.lu.se}} \\[3mm]
{\it Department of Theoretical Physics,}\\[1mm]
{\it Lund University, Lund, Sweden}
\end{center}
 
\vspace{\fill}
 
\begin{center}
{\bf Abstract}\\[2ex]
\begin{minipage}{\abstwidth}
Heavy long-lived quarks, i.e. charm and bottom, are frequently 
studied both as tests of QCD and as probes for other physics 
aspects within and beyond the standard model. The long life-time 
implies that charm and bottom hadrons are formed and observed.
This hadronization process cannot be studied in isolation, but 
depends on the production environment. Within the framework of 
the string model, a major effect is the drag from the other end 
of the string that the $\c/\b$ quark belongs to. In extreme cases, 
a small-mass string can collapse to a single hadron, thereby 
giving a non-universal flavour composition to the produced hadrons. 
We here develop and present a detailed model for the charm/bottom 
hadronization process, involving the various aspects of string 
fragmentation and collapse, and put it in the context of several 
heavy-flavour production sources. Applications are presented from 
fixed-target to LHC energies.
\end{minipage}
\end{center}
 
\vspace{\fill}

\clearpage
\pagestyle{plain}
\setcounter{page}{1} 

\section{Introduction} 

The light $\u$, $\d$ and $\s$ quarks can be obtained from a number 
of sources: valence flavours in hadronic beam particles, perturbative
subprocesses and nonperturbative hadronization. Therefore the
information carried by identified light hadrons is highly ambiguous.
The charm and bottom quarks have masses significantly above the 
$\Lambda_{\mathrm{QCD}}$ scale, and therefore their production should 
be perturbatively calculable. That is, they are not expected to be 
produced at any significant rate in nonperturbative processes 
\cite{AGIS}, and they do not occur as valence flavours of the 
commonly used beam particles. A priori, they are therefore excellent 
probes of the underlying hard dynamics, whether that involves standard 
QCD processes or various kinds of new physics. They can also be 
identified in the data by techniques such as secondary vertices, 
prompt leptons or kinematical constraints (e.g. the small 
$m_{\D^*} - m_{\D} - m_{\pi}$).

In order to understand the character of the perturbative process that
occurred, it is often necessary to know not only that a $\c/\b$ quark
was produced, but also its original energy and direction of motion.
The assumption is then often made that the observed charm/bottom 
hadron accurately reflects the original quark.
For instance, the quark momentum distribution may be scaled down by a 
convolution with the Peterson fragmentation function \cite{Peterson},
while the direction is assumed unchanged. In other cases, like for the
study of CP violation in the $\B^0$--$\Bbar^0$ system, the produced 
hadrons are at the focus of attention, and a simple ansatz is that the 
composition of $\B$ hadrons is independent of the production
environment and symmetric between particles and antiparticles. 

The assumption, that heavy quarks remain rather unaffected by the
environment in which they are produced, has some support in the
Heavy Quark Effective Theory (HQET) \cite{HQET}, although that 
framework is more concerned with questions of decay than of production. 
However, even if many of the effects we will consider here indeed die out 
roughly inversely to the mass of the heavy quark --- the
$\Lambda/m_{\Q}$ behaviour of HQET --- evidence is that the numerator 
$\Lambda$ of such a scaling relation often is large. Thus the large 
lifetime differences between the $\D$ mesons has an analogue in large
production asymmetries between $\D$ mesons in hadronic collisions
\cite{oldobs,WA82,WA92,E769,E791}. Even with a reduction by a factor
2--3 when moving from charm to bottom, large effects are therefore 
expected also in the latter case. Only top hadrons would have been 
reasonably immune, had the top quark been long-lived enough for top 
hadrons to form.

In order to study and understand the sizeable deviations from the 
simple picture, of perturbation theory results only minimally modified,
it is necessary to have a realistic framework for nonperturbative
effects. The Lund string fragmentation model \cite{AGIS} will be the
starting point here. In this picture, the colour confinement field
between a quark and an antiquark (or a diquark) is squeezed into a 
tube-like region, corresponding to a linear confinement potential.
This couples the hadronization of a  $\c/\b$ quark to the flavour
content and momentum of the other string end, i.e. provides an explicit 
dependence on the production environment. An extreme case is the string
collapse, where the mass of a string only allows the production of one 
single hadron, which by necessity therefore combines the  $\c/\b$ quark 
with the flavour at the other string end, often one of 
the valence flavours of the incoming beams. Also less extreme situations
can give noticeable effects, and only in the limit of large string masses 
can one expect to recover a simple description. However, this description 
can still be counterintuitive, in that the hadronization process can
`speed up' as well as `slow down' the heavy hadron relative to the
heavy quark. In terms of the easily observable consequences, like
hadron flavour asymmetries, we therefore go from huge effects at 
fixed-target experiments to tiny ones at the LHC, but in less obvious 
ways the effects remain large also at high energies. Furthermore, 
events at large energies tend to be composed of several strings,
so even at the LHC a fraction of small-mass strings is produced.

The presence of such effects is implicit in the Lund model, and was 
studied qualitatively early on \cite{ABG}. It is only the more recent 
higher-precision data sets \cite{WA82,WA92,E769,E791} that allow quantitative 
comparisons to be carried out, and thereby give the possibility to 
`fine-tune' the model with respect to nonperturbative parameters.
A study of this kind was presented in a recent letter \cite{our}, where we 
concentrated on the production of charm at fixed-target energies. In this 
longer description, we will expand the study in several directions: by 
considering more heavy-flavour production mechanisms, by including bottom 
as well as charm production, by covering a larger range of energies and by
addressing a larger set of observables. In the process, further 
improvements have also been made in the model.

This paper is organized as follows. Section 2 contains the model description,
from the perturbative production mechanisms to the various domains of the
hadronization process. Sections 3 and 4 present some results. In section 3
the emphasis is on distributions that help explain how the basic aspects of 
the model work, with little regard whether distributions are observable or not.
In section 4 the emphasis is shifted to observable results for charm or bottom
production at some current or planned detectors, although explanations will 
sometimes rely on non-observable distributions. Finally, section 5 contains 
a summary and outlook.

\section{Model description} 
\label{sec:model}
Based on the concept of factorization, we here subdivide the process in two 
distinct phases, the perturbative one, where the heavy quarks are produced, 
and the nonperturbative one, where these quarks hadronize. Heavy-flavour
cross sections thus are completely determined by the former phase, while
observable event properties reflect a combination of the two. 

\subsection{Perturbative aspects}

\subsubsection{Production mechanisms}

Several different production mechanisms can be envisaged for heavy flavours.
Here we will concentrate on QCD processes in hadron--hadron collisions.
The $\mathcal{O}(\alphas^2)$ leading-order (LO) graphs, 
$\q\qbar \to \Q\Qbar$ and $\g\g \to \Q\Qbar$ \cite{pertcharm0}, 
Fig.~\ref{prodgraphs}a-b, then form the starting point for the continued 
discussion.

\begin{figure}[t] 
\begin{picture}(210,100)(-20,0)
\SetWidth{1.2}
\Gluon(30,10)(90,30){7}{3}\Text(20,10)[r]{$\g$}
\Line(90,30)(150,10)\Text(160,10)[l]{$\Q$}
\Gluon(30,90)(90,70){7}{3}\Text(20,90)[r]{$\g$}
\Line(90,70)(150,90)\Text(160,90)[l]{$\Qbar$}
\Line(90,30)(90,70)
\Text(90,0)[]{(a)}
\end{picture}
\begin{picture}(210,100)(-20,0)
\SetWidth{1.2}
\Line(30,20)(70,50)\Text(20,20)[r]{$\q$}
\Line(30,80)(70,50)\Text(20,80)[r]{$\qbar$}
\Gluon(70,50)(110,50){7}{2}
\Line(110,50)(150,80)\Text(160,80)[l]{$\Qbar$}
\Line(110,50)(150,20)\Text(160,20)[l]{$\Q$}
\Text(90,0)[]{(b)}
\end{picture}\\
\begin{picture}(210,100)(0,0)
\SetWidth{1.2}
\Gluon(30,10)(90,30){7}{3}\Text(20,10)[r]{$\g$}
\Line(90,30)(170,10)\Text(180,10)[l]{$\Q$}
\Gluon(30,90)(90,70){7}{3}\Text(20,90)[r]{$\g$}
\Line(90,70)(170,90)\Text(180,90)[l]{$\Qbar$}
\Line(90,30)(90,70)
\Gluon(114,24)(170,40){7}{3}\Text(180,40)[l]{$\g$}
\Text(90,0)[]{(c)}
\end{picture}
\begin{picture}(210,150)(-20,0)
\SetWidth{1.2}
\Gluon(10,10)(90,30){7}{4}\Text(00,10)[r]{$\g$}
\Gluon(90,30)(150,10){7}{3}\Text(160,10)[l]{$\g$}
\Gluon(10,90)(35,80){7}{1}\Text(00,90)[r]{$\g$}
\Line(35,80)(150,120)\Text(160,120)[l]{$\Qbar$}
\Line(35,80)(90,70)
\Line(90,70)(150,90)\Text(160,90)[l]{$\Q$}
\Gluon(90,30)(90,70){7}{2}
\Text(90,0)[]{(d)}
\end{picture}\\
\begin{picture}(210,100)(0,0)
\SetWidth{1.2}
\Gluon(30,10)(90,30){7}{3}\Text(20,10)[r]{$\g$}
\Gluon(90,30)(170,10){7}{4}\Text(180,10)[l]{$\g$}
\Gluon(30,90)(90,70){7}{3}\Text(20,90)[r]{$\g$}
\Gluon(90,70)(115,75){7}{1}
\Line(115,75)(170,55)\Text(180,55)[l]{$\Q$}
\Line(115,75)(170,95)\Text(180,95)[l]{$\Qbar$}
\Gluon(90,30)(90,70){7}{2}
\Text(90,0)[]{(e)}
\end{picture}
\begin{picture}(230,170)(-60,0)
\SetWidth{1.2}
\Gluon(-30,10)(90,30){7}{6}\Text(-40,10)[r]{$\g$}
\Gluon(90,30)(150,10){7}{3}\Text(160,10)[l]{$\g$}
\Gluon(-30,115)(10,100){7}{2}\Text(-40,115)[r]{$\g$}
\Line(10,100)(50,85)
\Line(10,100)(150,140)\Text(160,140)[l]{$\Qbar$}
\Line(50,85)(150,110)\Text(160,110)[l]{$\Q$}
\Gluon(50,85)(90,70){7}{2}
\Gluon(90,70)(150,80){7}{3}\Text(160,80)[l]{$\g$}
\Gluon(90,30)(90,70){7}{2}
\Text(90,0)[]{(f)}
\end{picture}\\[2mm]
\captive{Examples of heavy-flavour production diagrams. (a,b) Leading order.
(c) Pair creation (with gluon emission). (d) Flavour excitation. (e) Gluon 
splitting. (f) Events classified as gluon splitting but of 
flavour-excitation character.
\label{prodgraphs}}
\end{figure}

One way to proceed is to add next-to-leading order (NLO) perturbative 
processes, i.e the $\mathcal{O}(\alphas^3)$ corrections to the above 
\cite{pertcharm1}. 
New graphs are $\q\qbar \to \Q\Qbar\g$, $\q\g \to \Q\Qbar\q$ and
$\g\g \to \Q\Qbar\g$. Additionally the leading-order processes are modified 
by virtual corrections. Depending on the choice of cut-off parameters, the 
latter may give negative differential cross sections in some regions of phase
space. The divergences disappear in sufficiently inclusive distributions,
so much phenomenological insight can be gained \cite{pertcharm2}. 
However, with our currently available set of calculational tools, the NLO 
approach is not so well suited for the exclusive Monte Carlo studies we have 
in mind here, where hadronization is to be added on to the partonic picture. 
Furthermore, also the NLO results, although exact to $\mathcal{O}(\alphas^3)$, 
would be modified in yet higher orders, e.g. by the resummed effects of 
multiple gluon emission \cite{resummation}.

As an alternative, the parton-shower (PS) approach offers a different set of 
approximations. It is not exact even to $\mathcal{O}(\alphas^3)$, but it 
catches the leading-log aspects of the multiple-parton-emission phenomenon. 
Especially when one goes to higher energy this can offer many advantages. 
The PS approach is based on a probabilistic picture, wherein the overall 
$2 \to n$ partonic process is subdivided into three stages: initial-state 
cascades, hard scattering and final-state cascades. The hard scattering 
is here defined as the $2 \to 2$ sub-diagram that contains the largest 
virtuality, i.e. corresponds to the shortest-distance process. It is 
important to respect this in order to avoid double-counting, as will 
become apparent in the following. Heavy-flavour events 
can then be subdivided into three classes, which we will call pair creation,
flavour excitation and gluon splitting. The names may be somewhat misleading,
since all three classes create pairs at $\g \to \Q\Qbar$ vertices, but it is 
in line with the colloquial nomenclature. 

The three classes are characterized as follows.
\begin{Enumerate}
\item Pair creation is when the hard subprocess is one of the two LO processes
above. Showers do not modify production cross sections, but shift kinematics,
Fig.~\ref{prodgraphs}c. 
For instance, in the LO description, the $\Q$ and $\Qbar$ have to emerge 
back-to-back in azimuth in order to conserve momentum, while the parton shower 
allows a net recoil to be taken by one or several further partons. 
\item Flavour excitation is when a heavy flavour from the parton distribution
of one beam particle is put on mass shell by scattering against a parton of 
the other beam, i.e. $\Q\q \to \Q\q$ or $\Q\g \to \Q\g$, 
Fig.~\ref{prodgraphs}d. When the $\Q$ is not a
valence flavour, it must come from a branching $\g \to \Q\Qbar$ of the
parton-distribution evolution. In most current-day parton-distribution
parameterizations, heavy-flavour distributions are assumed to vanish for
virtuality scales $Q^2 < m_{\Q}^2$. The hard scattering must therefore 
have a virtuality above $m_{\Q}^2$. When the initial-state shower is 
reconstructed backwards \cite{backwards}, the $\g \to \Q\Qbar$ branching
will be encountered, provided that $Q_0$, the lower cutoff of the shower,
obeys $Q_0^2 < m_{\Q}^2$. Effectively the processes therefore become
at least $\g\q \to \Q\Qbar\q$ or $\g\g \to \Q\Qbar\g$, with the possibility 
of further emissions. In principle, such final states could also be obtained 
in the above pair-creation case, but the earlier advertised requirement on 
the hard scattering to be more virtual than the showers avoids double-counting.
\item Gluon splitting is when a $\g \to \Q\Qbar$ branching occurs in the 
initial- or final-state shower, and no heavy flavours enter the hard 
scattering, Fig.~\ref{prodgraphs}e. Here the dominant source is gluons in 
the final-state showers, since timelike gluons emitted in the initial state 
are restricted to a smaller maximum virtuality. Except at high energy, 
most gluon splittings in the 
initial state instead result in flavour excitation, already covered above.
An ambiguity of terminology exists with initial-state evolution chains where
a gluon first branches to $\Q\Qbar$ and the $\Q$ later emits another gluon 
that is the one to enter the hard scattering, Fig.~\ref{prodgraphs}f. From 
an ideological point of view, this is flavour excitation, since it is related 
to the evolution of the heavy-flavour parton distribution. From a practical 
point of view, however, we will choose to classify it as gluon splitting, 
since the hard scattering does not contain any heavy flavours. 
\end{Enumerate}  
In summary, the three classes above are then characterized by having 2, 1 or 
0, respectively, heavy flavours in the final state of the hard subprocess. 
Of course, all this assumes that only one heavy-flavour pair is produced in an 
event --- one could have e.g. double flavour excitation $\Q\Q' \to \Q\Q'$ ---
which normally is a good first approximation. Only in high-$\pt$ processes at
high energies do profuse shower evolution make the multiple gluon-splitting 
process relevant. 
     
To the above heavy-flavour sources, one could add the creation in decays of
heavier resonances, such as $\Z^0 \to \b\bbar$, $\W^+ \to \c\sbar$, 
$\H^0  \to \b\bbar$, $\t \to \b \W^+$ and, of course, $\b \to \c$. In the 
current paper we will have little to say about these. However, $\c$ and $\b$ 
production at LEP1 clearly provides the basis that we can build on here, 
by testing both the showering and the hadronization of heavy flavours,
although in a different environment.
For primary-produced heavy flavours, everything appears to be well 
understood in the framework of our
models. Some discrepancies have been noted in the rate of hard gluon emission
off $\b$'s \cite{batLEP}, i.e. in the region where the shower is not expected 
to be perfect anyway, and even so discrepancies are tiny compared with typical 
uncertainties in hadronic collisions. Rather more worrisome is the observed 
rate of secondary heavy-flavour production, i.e. what we have called gluon 
splitting above. There the LEP observations exceed the rate predicted by
shower programs \cite{Pythia,Herwig}, and also by analytical calculations 
\cite{Mikebc}, by maybe as much as 50\% \cite{splitatLEP, Mikebc}. 
The error bars are large, however, so the true excess 
could be lower. The possibility of higher rates already exists in some models
\cite{Ariadne}, and one could imagine modifications to others. Currently the
data is too poor to tell much about  whether the shape agrees or not with 
models. We will therefore assume that only the rate could be a problem, 
and then any effect in hadronic collisions could be absorbed under the 
general heading of $K$ factors, i.e. a rescaling in rate by higher-order 
corrections.

\subsubsection{Parton-shower particulars}

The perturbative shower approach is implemented in the {\Py} program 
\cite{Pythia} that we will use for the studies in this paper.

Pair creation is easy to generate by itself, by allowing only the two
hard processes $\q\qbar \to \Q\Qbar$ and $\g\g \to \Q\Qbar$, using the LO
matrix elements with quark masses included. The full phase space can be
populated, i.e. down to $\pthat = 0$, since the quark mass provides the soft 
cutoff. The $Q^2$ scale of the process, used to set the range of allowed 
shower evolution as described below, is here taken to be
$Q^2 = m_{\Q}^2 + \pthat^2$.
 
Flavour excitation can be obtained by only sampling the heavy flavour
$\Q/\Qbar$ from one of the incoming hadrons (a standard option of the program)
while allowing all flavours from the other hadron. The two sides of the event
are covered by two separate runs, added for the final results. We have not
implemented any special matrix elements for the scattering of one heavy quark
against another massless parton; instead massless matrix elements are used.
Since the heavy-flavour parton distributions vanish for scales 
$Q^2 < m_{\Q}^2$, where we associate 
$Q^2 = \pthat^2 = \hat{t}\hat{u}/\hat{s}$
for massless kinematics, it follows that $\hat{s} > 4  m_{\Q}^2$. The mass
corrections to the matrix elements are therefore not expected to be very large.
(In practice, massive four-vectors are constructed from the massless ones by a
scaling-down of the three-momenta, in the rest frame of the subprocess, while
preserving the energy: $p_i^2 = 0$, ${p'}_i^2 = m_i^2$, 
$\mathbf{p}'_i = \alpha \mathbf{p}_i$, $E'_1 + E'_2 = E_1 + E_2$.
The actual $\pthat$ may thus end up somewhat below the nominal cut at 
$m_{\Q}$.) The normal backwards shower evolution from the hard subprocess is 
then supposed to find a preceding $\g\to \Q\Qbar$ branching.
 
In the earlier {\Py} versions, this often failed, and a heavy quark
was allowed to form part of the beam-remnant flavour content that entered the
nonperturbative description. We have now studied this phenomenon and recognized
it as coming from the constrained kinematics that exists inside the shower.
To see this point, consider a branching $\g\to \Q\Qbar$ where $\Q$ takes a 
fraction $z$ of the lightcone momentum $p_+ = E + p_z$ of the gluon, and is 
virtual with $p_{\Q}^2 = -Q^2$. The recoiling $\Qbar$ is part of the final 
state and must thus have mass $m_{\Q}$, or above that if it radiates 
final-state gluons. Then conservation of $p_- = m_{\perp}^2/p_+$ gives
\begin{equation}
\frac{- Q^2 + \pt^2}{z} + \frac{m_{\Q}^2 + \pt^2}{1 - z} = 0 ~,
\end{equation}
and the requirement of a physical transverse momentum in the branching, 
$\pt^2 \geq 0$, translates into
\begin{equation}
z \leq \frac{Q^2}{m_{\Q}^2 + Q^2} ~.
\end{equation}
It may then become kinematically impossible to find a gluon with 
$x_{\g} = x_{\Q}/z < 1$. Since many common parton distribution 
parameterizations do not respect the above kinematics constraints, 
we have introduced a further explicit check, where kinematically 
impossible configurations are rejected, and the cross section is reduced 
accordingly. Even when a $\Q$ is formally in the allowed region, one may 
feel threshold effects that make event generation less efficient. As a 
first approximation for this region, the shower is forced to
`try harder' to find a  $\g\to \Q\Qbar$ branching, without any loss of cross 
section. The end result is that, in the new program version, no $\c$ or $\b$ 
quarks remain in the beam remnant, but are always constructed as coming from a 
shower branching.

Gluon splitting cannot easily be generated by itself, since it could appear
e.g. several steps down in a gluonic cascade, which cannot easily be 
predetermined. Instead it is necessary to attempt to generate the full QCD 
jet cross section, down to some lower $\ptmin$ cut, and then pick up all 
events that `happen' to contain the heavy flavour. Some events fall under 
the heading of pair creation or flavour excitation and are thus removed in 
a second step.

Note that the kinematics machinery here is based on massless quarks in the 
hard scattering, with some post-facto modifications for heavy quarks, so 
the alternative pair-creation description obtained here is less precise 
than the one in point 1 above. Masses are included in the description of 
the shower branching $\g\to \Q\Qbar$, however.

Showers should not populate kinematical regions already covered by the 
hard scattering. This requirement is not easy to implement exactly.
One reason is that several different sets of constraints can appear,
such as from virtuality ordering and angular ordering. Here we therefore
satisfy ourselves with an approximate matching of $Q^2_{\mathrm{max}}$
and $M^2_{\mathrm{max}}$, the maximally allowed virtualities of spacelike 
and timelike showers, respectively, to $Q^2 = \pthat^2 = 
\hat{t}\hat{u}/\hat{s}$, the conventional hard-scattering scale.
This matching is generic to all branchings in showers, but obviously
we give special attention to implications for the heavy-quark production 
vertices.

With massless kinematics, one may sensibly assume 
$Q^2 \leq Q^2_{\mathrm{max}},M^2_{\mathrm{max}} \leq 4 Q^2$.
The lower limit would be appropriate for a $t$-channel graph,
where $-\hat{t} \approx Q^2$ sets the maximal virtuality.
The upper limit is more relevant for an $s$-channel graph, where
$\hat{s} \geq 4 \pthat^2$ sets the scale. So the above range 
translates into an uncertainty in the amount of shower evolution.  
However, we can try to be more specific. Timelike parton showers 
are evolved in terms of $M^2$, the squared mass of the
propagating parton. It is thus akin to the $\hat{s}$ scale of the hard
scattering, and $M^2_{\mathrm{max}} = 4 Q^2$ is the preferred choice.
With both $\Q$ and $\Qbar$ having a mass at or above $m_{\Q}$, the 
heavy-flavour production threshold at $M^2_{\mathrm{max}} = 4 m_{\Q}^2$  
then corresponds to $Q^2 = m_{\Q}^2$, which agrees with the threshold
for pair creation at the hard scattering. Spacelike parton showers instead 
are evolved in a spacelike virtuality, analogous to $\hat{t}$, and the 
reasonable choice then is $Q^2_{\mathrm{max}} = Q^2$. Again, this gives 
a matching threshold for `flavour excitation' both in the shower and at the 
hard scattering, at $m_{\Q}^2$ for most heavy-flavour parton distribution
parameterizations. Specifically, note that an initial-state shower branching
does not have to produce two heavy quarks on mass shell, but only one at a 
time.

\subsubsection{Parameters}

The main formal parameters in the perturbative description are the heavy-quark
masses. They enter in the description of hard scatterings and parton showers
alike, both directly as mass terms in matrix elements or splitting kernels and
indirectly in the description of the phase space. Therefore cross sections are 
especially sensitive to the value selected. Also the nonperturbative 
phenomenology is significantly affected. In \cite{our}, we chose to 
standardize on 
$m_{\c} = 1.5$~GeV. Based on conventional mass formulae \cite{deruj}, 
\begin{equation}
\label{bmass}
\frac{3 m_{\D^*} + m_{\D}}{4} - m_{\c} = \frac{3 m_{\B^*} + m_{\B}}{4} - 
m_{\b} ~,
\end{equation}
we then obtain $m_{\b} = 4.8$~GeV. 

Also the choice of parton distributions gives some leeway, especially since 
the gluon distribution is not yet so well constrained at small $x$ and 
moderately small $Q^2$, where a non-negligible amount of the total 
charm/bottom cross section at high energies comes from. Unless otherwise 
specified, we have used the CTEQ 5L
parameterized distributions for the proton \cite{CTEQ5}, with 
$\Lambda^{(4)} = 0.192$~GeV. For the pion we rely on the GRV LO (updated) sets
\cite{GRVpi}. As already noted, the default factorization scale is 
$Q^2 = \pthat^2$.  

\subsection{Nonperturbative aspects}

The way string fragmentation affects charm production was described in 
\cite{our}. We will here develop the main points and outline the current 
status of our modeling, which is slightly modified since the previous 
publication.

The partonic state that is to be hadronized consists, at the very least, of 
the outgoing partons from the hard scattering and of the beam-remnant partons. 
Furthermore, a realistic study has to include the additional partons produced 
by initial- and final-state showers, and by the possibility of having
several hard parton--parton interactions in the same event. These
aspects increase in importance with increasing energy, and have to be included
in the event description, but will not be at the focus of attention. 

\subsubsection{Colour flow}

In the string model, confinement is implemented by spanning strings between 
the outgoing partons. These strings correspond to a Lorentz-invariant 
description of a linear confinement potential, with string tension 
$\kappa \approx 1$~GeV/fm. Each string piece has a colour charge at one end 
and its anticolour at the other. The double colour charge of the gluon 
corresponds to it being attached to two string pieces, while a quark is 
only attached to one. A diquark is considered as being in a colour 
antitriplet representation, and thus behaves (in this respect) like an 
antiquark. Then each string contains a colour triplet endpoint, a number 
(possibly zero) of intermediate gluons and a colour antitriplet end. An event
will normally contain several separate strings.

\begin{figure}[t]
\begin{picture}(210,130)(-10,-10)
\Text(100,-5)[]{(a)}
\Text(10,30)[r]{$\p$}
\Text(10,100)[r]{$\pbar$}
\Line(15,30)(25,30)
\Line(15,100)(25,100)
\GOval(30,30)(20,5)(0){0.5}
\GOval(30,100)(20,5)(0){0.5}
\Gluon(33,45)(75,65){4}{4}
\Gluon(33,85)(75,65){4}{4}
\Gluon(75,65)(120,65){4}{4}
\Line(120,65)(160,45)
\Line(120,65)(160,85)
\Text(165,45)[l]{$\bbar$}
\Text(165,85)[l]{$\b$}
\Line(33,15)(160,15)
\Line(33,115)(160,115)
\Text(165,15)[l]{$\u\d$}
\Text(165,115)[l]{$\ubar\dbar$}
\Line(35,30)(160,30)
\Line(35,100)(160,100)
\Text(165,30)[l]{$\u$}
\Text(165,100)[l]{$\ubar$}
\put(178,37.5){\oval(7,16)[r]}
\put(182,65){\oval(15,100)[r]}
\put(178,92.5){\oval(7,16)[r]}
\end{picture}
\hspace{5mm}
\begin{picture}(210,130)(0,-15)
\Text(110,-5)[]{(b)}
\LongArrow(0,60)(50,60)
\Text(0,70)[l]{$\pbar$}
\LongArrow(210,60)(160,60)
\Text(210,70)[r]{$\p$}
\LongArrow(100,60)(170,65)
\Text(175,70)[l]{$\ubar$}
\LongArrow(100,60)(160,90)
\Text(165,94)[l]{$\b$}
\LongArrow(100,60)(30,50)
\Text(26,47)[t]{$\u$}
\LongArrow(100,60)(80,10)
\Text(90,10)[r]{$\bbar$}
\LongArrow(100,60)(35,75)
\Text(35,87)[]{$\u\d$}
\LongArrow(100,60)(170,55)
\Text(175,49)[l]{$\ubar\dbar$}
\DashLine(170,65)(160,90){4}
\DashLine(30,50)(80,10){4}
\DashLine(35,75)(170,55){4}
\end{picture}\\[2mm]
\captive{Example of a string configuration in a $\p\pbar$ collision. 
(a) Graph of the process, with brackets denoting the final colour singlet
subsystems. (b) Corresponding momentum space picture, with dashed lines
denoting the strings. 
\label{fragstrings}}
\end{figure}

The string topology can be derived from the colour flow of the
hard process. For instance, consider the LO process $\u\ubar \to \b\bbar$
in a $\p\pbar$ collision. The colour of the incoming $\u$ is inherited
by the outgoing $\b$, so the $\b$ will form a colour-singlet together with the
proton remnant, here represented by a colour antitriplet $\u\d$ diquark. 
In total, the event will thus contain two strings, one $\b$--$\u\d$ and
one $\bbar$--$\ubar\dbar$. In $\g\g \to \b\bbar$ a similar inspection shows 
that two distinct colour topologies are possible. Representing the 
proton remnant by a $\u$ quark and a $\u\d$ diquark (alternatively $\d$ 
plus $\u\u$), one possibility is to have three strings $\b$--$\ubar$, 
$\bbar$--$\u$ and $\u\d$--$\ubar\dbar$, Fig.~\ref{fragstrings}, and the other 
is the three strings $\b$--$\u\d$, $\bbar$--$\ubar\dbar$ and $\u$--$\ubar$.

In processes with several possible colour topologies, the relative composition
may become nontrivial. For $\g\g \to \b\bbar$, the symmetry of the process 
gives an equal integrated --- but not differential --- rate for the two 
topologies. For a more illustrative example, consider e.g. $\u \g \to \u \g$, 
again in $\p\pbar$, which contains both $s$-, $t$- and $u$-channel graphs, 
including interference terms. There are again two possible colour topologies, 
$\u$--$\ubar$ plus $\u\d$--$\g$--$\ubar\dbar$ and $\u$--$\g$--$\ubar$ plus 
$\u\d$--$\ubar\dbar$.
The $u$-channel only contributes to the former and the $s$-channel to the
latter, but the $t$-channel contributes to both, meaning there is a nontrivial
kinematics dependence on the relative probability for the two topologies.
Furthermore, the cross section contains an interference contribution that
corresponds to an undetermined colour flow, where it is not possible to
subdivide the event into two separate colour singlets. Since the hadronic 
final state consists of singlets, clearly a collapse of this ambiguity must 
occur at some stage. We therefore subdivide the interference term, in a 
sensible but not unique way, between the two configurations above 
\cite{hansuno}. As should be expected, it is suppressed by a factor 
$1/N_C^2$, where $N_C=3$ is the number of colours. 

The above example carries over to flavour excitation $\Q \g \to \Q \g$, but 
additionally the colour flow in the initial- and final-state cascades has to 
be considered, at the very least the branching $\g \to \Q \Qbar$. Since we 
only work to leading order, where no interference contributions are explicit
(implicitly they have been used e.g. to introduce angular ordering in the 
shower evolution), this is straightforward: a new colour-anticolour pair is 
created and spanned between the daughters
in $\g \to \g\g$ and $\q \to \q \g$, while the existing colours are split in
$\g \to \q \qbar$. No special colour rules are needed for heavy flavours. The 
last vertex, although the most rare of the three, has a special r\^ole in 
subdividing one colour singlet into two. With increasing energy and 
parton-shower activity, 
it gives an increasing average number of separate singlets in an event.

\subsubsection{Hadronization}

Once the string topology has been determined, the Lund string 
fragmentation model \cite{AGIS} can be applied to describe the
nonperturbative hadronization. To first approximation, we assume that the 
hadronization of each colour singlet subsystem, i.e. string, can be considered 
separately from that of all the other subsystems. Presupposing that the 
fragmentation mechanism is universal, i.e. process-independent,
the good description of $\e^+\e^-$ annihilation data should carry over.
The main difference between $\e^+\e^-$ and hadron--hadron events is that 
the latter contain beam remnants which are colour-connected with the
hard-scattering partons. More about these remnants below, in 
Sect.~\ref{beamremnants}.

Depending on the invariant mass of a string, practical considerations
lead to the need to distinguish three hadronization prescriptions:
\begin{Enumerate}
\item {\em Normal string fragmentation}.
In the ideal situation, each string has a large invariant mass. Then the 
standard iterative fragmentation scheme, for which the assumption of a 
continuum of phase-space states is essential, works well. The average 
multiplicity increases linearly with the string `length', which means 
logarithmically with the string mass. In practice, this approach can be 
used for all strings above some cut-off mass of a few GeV. 
\item {\em Cluster decay}.
If a string is produced with a small invariant mass, maybe only two-body
final states are kinematically accessible. The continuum assumption
above then is not valid, and the traditional iterative Lund scheme is 
not applicable. We call such a low-mass string a cluster, and consider
it separately from above. When kinematically possible, a $\Q$--$\qbar$ 
cluster will decay into one heavy and one light hadron by the production 
of a light quark--antiquark pair in the colour force field between the 
two cluster endpoints, with the new quark flavour selected according to 
the same rules as in normal string fragmentation. The $\qbar$ cluster end 
or the new $\q\qbar$ pair may also  denote diquarks; for ease of notation 
we will not always enumerate all the possible combinations covered in the 
full description.
\item {\em Cluster collapse}.
This is the extreme case of the above situation, where the string 
mass is so small that the cluster cannot decay into two hadrons.
It is then assumed to collapse directly into a single hadron, which
inherits the flavour content of the string endpoints. The original 
continuum of string/cluster masses is replaced by a discrete set
of hadron masses, mainly $\D/\B$ and $\D^*/\B^*$ (or corresponding 
baryon states). This mechanism plays a special r\^ole, in that it allows 
large flavour asymmetries in favour of hadron species that can inherit 
some of the beam-remnant flavour content. 
\end{Enumerate}

We assume that the nonperturbative hadronization process does not change the 
perturbatively calculated total rate of charm production. By local duality 
arguments \cite{duality}, we further presume that the rate of cluster collapse
can be obtained from the calculated rate of low-mass strings. This is related 
to the argument used in the $\e^+\e^- \to \c\cbar$ channel, that the cross 
section in the $\mathrm{J}/\psi$ and $\psi'$ peaks is approximately equal to 
a purely perturbatively calculated $\c\cbar$ production cross section
restricted to the below-$\D\Dbar$-threshold region. Similar
relations have also been studied e.g. for $\tau$ decay to hadrons
\cite{taudecay}, and there shown to be valid to good accuracy.
In the current case, the presence of other strings in the 
event additionally allows soft-gluon exchanges to modify 
parton momenta as required to obtain correct hadron masses.
Traditional factorization of short- and long-distance physics would 
then also protect the charm cross section. Local duality and factorization,
however, do not specify \textit{how} to conserve the overall energy and 
momentum of an event, when a continuum of $\cbar\d$ masses is to be replaced 
by a discrete $\D^-$ one. This will therefore be one 
of the key points to be studied below.

A first step towards constructing a model is to decide which mass range
a string belongs to. We have above settled for $m_{\c} = 1.5$~GeV and
$m_{\b} = 4.8$~GeV. Light quarks are given constituent masses,
$m_{\d} = m_{\u} = 0.33$~GeV and $m_{\s} = 0.5$~GeV. Diquark masses are
essentially the sum of the constituent masses above, with a spin-splitting
term added. If the string invariant mass exceed the sum of the two string
endpoint masses by some margin, $\sim$1~GeV, the normal string fragmentation 
routine can be used. This routine can produce two, three or more hadrons from 
the string, with the actual multiplicity determined dynamically during the 
hadronization process. Close to the lower limit, the two-body states dominate, 
so there should be a smooth transition to the cluster decay description.

For smaller string masses, a special cluster fragmentation procedure is 
used. Whether this results in the production of one or two hadrons depends on
the assumed two-body threshold behaviour. Consider a $\c\ubar$ cluster, for
instance. In one extreme point of view, a $\D\pi$ pair should always be formed 
when above this threshold, and a single $\D$ never. In another extreme, the
two-body fraction would gradually increase at 
a succession of thresholds: $\D\pi$, $\D^*\pi$, $\D\rho$, $\D^*\rho$, etc.,
where the relative probability for each channel is given by the standard 
flavour and spin mixture in string fragmentation. (For instance, $\D^*$
and $\D$ are assumed to be produced in the 3:1 ratio implied by spin counting,
while the big $\rho$--$\pi$ mass splitting there gives a mixture more like 
1:1.) In our current default model, we have chosen to steer a middle course, 
by allowing two attempts to find a possible pair of hadrons. Thus a fraction 
of events may collapse to a single resonance also above the $\D\pi$ threshold,
but $\D\pi$ is effectively weighted up. For instance, a 2.2~GeV string mass 
might, in a first round, be chosen to 
decay to $\D^*\rho$, and thus fail. If a second attempt instead gives $\D\pi$,
this two-body state would be accepted, but if $\D^*\rho$ is selected again, 
the cluster would collapse to a single hadron. If a large number of
attempts had been allowed (this can be varied as a free parameter), 
collapse would only become possible for cluster masses below the $\D\pi$
threshold.

One might have chosen also to include a phase-space factor close to each 
two-body threshold, instead of the step function used here. However, 
measurements of $R$ in $\e^+\e^-$ above the charm and bottom thresholds
\cite{RPP} indicate that Coulomb final-state interaction effects cancel any
such suppression. (Actually, the same data could be used to argue for having
only two-body states above the $\D\pi$ threshold. However, there is a 
difference: in a hadronic environment there will be a competition between
the production of one or of two hadrons, while collapse to a single particle 
is not an option in $\e^+\e^-$ away from the $\c\cbar$ resonance masses.
Within the large error bars of the data, one might also read in some trend
towards a larger $R$ a bit further above the $\D\Dbar$ threshold.)

In a cluster decay to two particles, a simplified version of normal string
fragmentation is used, in a spirit similar to the machinery for joining
the fragmentation chain by the production of two final hadrons somewhere in 
the middle of a normal string. In the cluster rest frame, a string 
direction is defined by the momentum vector of the heavy quark $\Q$. 
As a starting distribution, the cluster is allowed to decay isotropically to 
the two hadrons. The hadronic transverse momentum with respect to this 
direction is then used to introduce a Gaussian suppression factor 
$\exp(-\pt^2/2\sigma^2)$, with $\sigma = 0.36$~GeV denoting the standard 
fragmentation $\pt$ width parameter.
At threshold the decay thereby remains isotropic,
but at (an imagined) large cluster mass one would reproduce the same $\pt$ 
spectrum as when a string breaks by the production of a new $\q\qbar$ pair.
The heavy hadron H could still equally likely be produced in the $\Q$ 
hemisphere as in the opposite one, however. In string fragmentation, these
two configurations enter with different relative weights, that can be derived
from the space-time history of the process. Applied to the current case,
this gives
\begin{equation}
\mathcal{P}_{\mathrm{opposite}} = \frac{1}{1 + e^{b\Delta}} \hspace{1cm}
\mathrm{with} \hspace{1cm} \Delta = 
\sqrt{(m^2 - m_{\perp\H}^2 - m_{\perp\mathrm{h}}^2)^2 - 
4 m_{\perp\H}^2 m_{\perp\mathrm{h}}^2} ~,
\end{equation}
where $m$ is the cluster mass, $m_{\perp\H}$ and 
$m_{\perp\mathrm{h}}$ are the heavy and light hadron transverse masses, and 
$b$ the string fragmentation parameter, $b \approx 0.9$~GeV$^{-2}$. 
$\Delta = \Gamma_2 - \Gamma_1 = \kappa^2 (\tau_2^2 - \tau_1^2)$ is the
difference in string area $\Gamma_i$ spanned for the two solutions, which 
can be related to a difference in decay proper times $\tau_i$ by the
string tension $\kappa$. After
this final correction, giving the `natural' ordering with the heavy hadron
usually close to the direction of the heavy quark when well above the 
threshold, the transition between normal string fragmentation and cluster 
decay is reasonably smooth. 

What is not so smooth is the cluster collapse mechanism. Here confinement 
effects have to project the continuum of string masses onto the observed
discrete hadron mass spectrum. Because of the aforementioned local duality 
and factorization arguments, the total area of the spectrum should 
be conserved in the process. How the projection should be done is not 
known from first principles, however. 

One conceivable strategy could be to introduce a weight function 
consisting of $\delta$ function peaks at the single-hadron masses, 
with suitably adjusted normalizations, and then step functions at the 
two-particle thresholds. This weight function, when multiplied with
the partonic mass spectrum, should then give the hadron-level mass 
spectrum. Such an approach is not well suited for
Monte Carlo simulation, since the string mass is a complicated 
function of a number of variables and therefore the $\delta$ function
cannot easily be integrated out. Conceptually, it would also suffer from 
the problem of having to have a non-universal weight function: the
coefficients would have to be adjusted somewhat as a function of energy
to ensure exact conservation of the total cross section, since the
cluster mass spectrum itself is somewhat energy dependent.
However, on general grounds,
we do not expect the overall distribution of event 
characteristics to differ significantly between events with a
$\cbar$--$\d$ string mass exactly equal to the $\D^-$ one, and 
events where the string mass is maybe 100~MeV off. An appealing
shortcut therefore is to accept all partonic configurations  
and thereafter introduce some `minimal' adjustments to the 
kinematics to allow hadrons to be produced on the mass shell.
Such a strategy would be consistent not only with local duality arguments,
but also with the presence of soft final-state interactions,
i.e. the exchange of nonperturbative gluons that can carry some
amount of momentum between the low-mass string and the surrounding
hadronic system. In the following we will therefore adopt the 
language of `gluons' transferring energy and momentum between the
strings in a collision, while leaving unanswered the question on the
exact nature of those `gluons'. Specifically, we will not address the 
possibility of changes in the colour structure of events by such 
`gluons'.

The basic strategy will be to exchange some minimal amount of momentum 
between the collapsing cluster and other string pieces in its neighbourhood.
Consider an event in its CM frame, with all partons emerging from an 
assumed common origin. Partons move out with close to the speed of light,
so if they move in the same direction they also stay close to each other
for a long time, and therefore have an enhanced chance to exchange momenta.
An exchange can also occur to the string pieces spanned between the
partons, quarks or gluons. The piece between two partons 1 and 2 
spans the set of velocity vectors
\begin{equation}
\mathbf{v}_{\mathrm{string}} = \alpha \mathbf{v}_1 + 
(1-\alpha) \mathbf{v}_2 ~,~ 0 \leq \alpha \leq 1 ~.
\end{equation} 
A closest `distance' between this string piece and the cluster can then 
be defined as
\begin{equation}
D^2 = \min_{0 \leq \alpha \leq 1} (\mathbf{v}_{\cl} - 
\mathbf{v}_{\mathrm{string}})^2 ~. 
\end{equation} 
Based on this measure, the string piece closest to the cluster is found.

The momentum transfer can be in either direction, depending on whether the 
hadron is heavier or lighter than the cluster it comes from,
$m_{\H} \gtrless m_{\cl}$. The hadron species, and 
thereby hadron mass, is selected according to the standard flavour selection 
rules. That is, there is no mass dependence, e.g. so that a lighter cluster 
could have been more likely to form a $\D/\B$ and a heavier a  
$\D^*/\B^*$; after all, the mass splitting is not so large that 
kinematics should come out particularly different for the two.

The simpler situation is when $m_{\H} < m_{\cl}$. Then one may split the 
cluster four-momentum into two parallel vectors, 
$p_{\H} = (m_{\H}/m_{\cl}) p_{\cl}$ and $p_{\g} = (1 - m_{\H}/m_{\cl}) 
p_{\cl}$. The latter momentum, for an imagined gluon, can be absorbed by the 
closest string piece found above, i.e. be inserted between the endpoint
partons 1 and 2. This gluon has $m_{\g}^2 > 0$, but not too large or a 
collapse would not have occurred. Such somewhat massive gluons are well
modeled by the standard string fragmentation framework \cite{stringwithg}.
One could have chosen a `decay' of the cluster into a massless gluon,
e.g. with an isotropic angular distribution in its rest frame, but such an
ansatz gives the same average behaviour as the one above, and only slight
differences in fluctuations.

A worse situation is when $m_{\H} > m_{\cl}$.
A negative-energy gluon could be defined and handled as above. Usually this
works fine, but it can lead to complete strings (not just string pieces)
with negative $m^2$, or even to hadrons with $|\xF| > 1$, so such an approach 
is not quite trustworthy. Instead we assume an exchange in the opposite 
direction, where the nearest string piece emits a gluon that can be absorbed 
by the cluster to give it the desired hadron mass. To be more precise, form a 
weighted sum of the endpoint momenta
\begin{equation}
  p_s = \alpha p_1 + \beta p_2 
  = \frac{p_2 p_{\cl}}{p_1 p_2} p_1     
  + \frac{p_1 p_{\cl}}{p_1 p_2} p_2  ~,     
\end{equation}
so that the end of the string that is closest to the cluster
is weighted up relative to the one further away. Thereafter
define
\begin{equation}
 p_{\H} = p_{\cl} + \delta p_s ~,
\end{equation}
with $\delta$ determined by the constraint $p_{\H}^2 = m_{\H}^2$.
The hadron will then have the correct mass, and the string endpoint
momenta are scaled down by factors $1 - \delta \alpha$ and 
$1- \delta \beta$, respectively. (Also the endpoint masses are scaled
down in the process. This is no problem, since the string fragmentation is not
dependent on having partons of a fixed mass.)

In the rare case that, e.g., $1 - \delta \alpha < 0$, the procedure has
to be extended. If the parton 1 is a gluon, the string does not end there 
but extends further to a parton 3. Then the gluon 1 four-momentum can be fully 
absorbed by the cluster, and the procedure above repeated for the partons
2 and 3. That way, one gluon after the next could be absorbed, at least
in principle. If instead a string endpoint is involved, this trick does not
work and we there revert to the old scheme, where four-momentum is shuffled
between the cluster and the parton furthest away from it, i.e. with the 
largest cluster+parton invariant mass. This scheme is more robust, and
normally requires only small four-momentum transfers, but physically it is not
so appealing, since it runs counter to the principle of locality in the
hadronization description. In practice, though, there is a good general 
agreement between results for the new and the older description \cite{our}.   

\subsubsection{Beam remnants}
\label{beamremnants}

A characteristic feature of hadronic collisions is the presence of a beam 
remnant. This remnant is defined by what is left behind of the hadron by 
the initial-state parton shower initiator. In the simplest case, when a 
valence quark is picked out 
of the incoming hadron, the remnant is a single antiquark, for a meson, or a   
diquark, for a baryon. In either case it is in a colour antitriplet state that
can be considered as a unit. For a baryon, simple flavour+spin SU(6) rules
can be used, e.g. to select between a $\u\d_0$ and a $\u\d_1$ diquark.

\begin{figure}[t]
\begin{center}
\epsfig{file=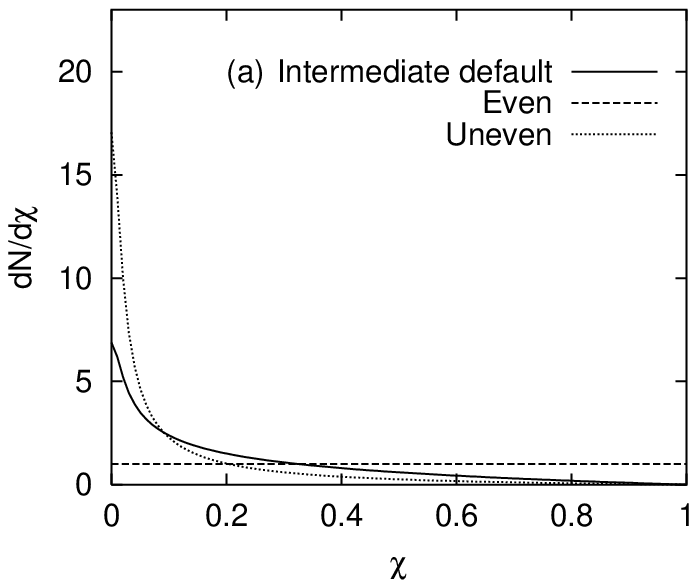}
\epsfig{file=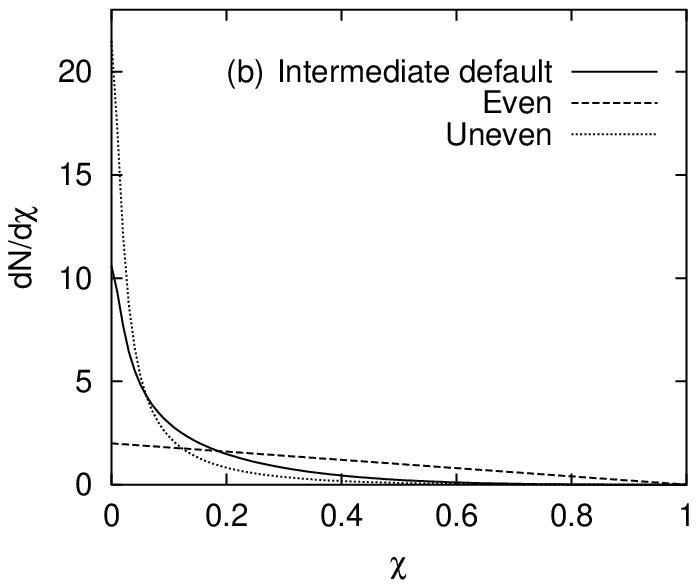}
\end{center}
\captive{Distribution of $\chi$ variable at 40~GeV for (a) a meson and (b) a
baryon. Full curves show the default intermediate option, dashed the even one and
dotted the uneven one, corresponding to $\langle \chi \rangle=$ 0.22, 0.50 and .12 for
mesons and .13, .33 and .076 for baryons.
\label{chidist}}
\end{figure} 

A more complex situation is when a gluon is picked out of the hadron, so that
the remnant is a colour octet, i.e. attached to two strings. A convenient
approach is to imagine this system split into two separate string endpoints,
one colour triplet and the other antitriplet. For a meson this would 
correspond to valence quark+antiquark, for a baryon to quark+diquark. 
The beam remnant distribution function (BRDF) is introduced to describe how 
the (light-cone) momentum of the remnant is shared between the two, in 
fractions $\chi$ and $1 - \chi$, respectively. For an octet meson remnant the 
$\chi$ distribution is always implicitly symmetrized between the $\q$ and 
$\qbar$, while for an octet baryon remnant one quark (picked at random among 
the three) takes the fraction $\chi$ and the remaining colour antitriplet 
diquark $1 - \chi$. There is no first-principles theory for BRDF's, so one 
has to rely on sensible ans\"atze. It turns out that asymmetries, e.g. 
between $\D$ and 
$\Dbar$ mesons, are very sensitive to the choices made here, especially in the 
baryon fragmentation regions \cite{our}. Therefore we will have reason to
compare different forms. As an intermediate, default, option we use
\begin{eqnarray}
\label{intermediate}
f(\chi) & \propto & \frac{1 - \chi}{\sqrt[4]{\chi^2 + c_{\mathrm{min}}^2}} 
\hspace{1cm} \mathrm{for~mesons,} \nonumber  \\
f(\chi) & \propto & \frac{(1 - \chi)^3}{\sqrt[4]{\chi^2 + c_{\mathrm{min}}^2}} 
\hspace{1cm} \mathrm{for~baryons,}  
\end{eqnarray} 
see Fig.~\ref{chidist}.
Here $c_{\mathrm{min}} = 0.6~\mathrm{GeV}/E_{\mathrm{CM}} \approx 
2m_{\q}/E_{\mathrm{CM}}$ provides an effective damping for $\chi$ values
so small that a parton ends up in the opposite hemisphere from its mother
hadron. Some arguments for the forms above, especially the $1/\sqrt{\chi}$ 
behaviour, can be found in reggeon phenomenology \cite{reggeon}, but basically
this is just a compromise between extremes. One such is that of an even 
sharing between all the valence partons,
\begin{eqnarray}
\label{even}
f(\chi) & = & 1 \hspace{1cm} \mathrm{for~mesons,} \nonumber  \\
f(\chi) & = & 2(1 - \chi) \hspace{1cm} \mathrm{for~baryons.}  
\end{eqnarray} 
Another is that of an uneven distribution,
\begin{eqnarray}
\label{uneven}
f(\chi) & \propto & \frac{1 - \chi}{\sqrt{\chi^2 + c_{\mathrm{min}}^2}} 
\hspace{1cm} \mathrm{for~mesons,} \nonumber  \\
f(\chi) & \propto & \frac{(1 - \chi)^3}{\sqrt{\chi^2 + c_{\mathrm{min}}^2}} 
\hspace{1cm} \mathrm{for~baryons,}  
\end{eqnarray}
reminiscent of the parton distributions encountered in the process of
perturbatively extracting a parton from a hadron. 

Also sea quarks/antiquarks may be emitted from a hadron. We here refer to the
lighter $\u$, $\d$ and $\s$ quarks that, unlike the heavier quarks, exist in 
the parton distributions at the lower shower cut-off scale $Q_0 \approx 1$~GeV.
In this case, the remnant is in an (anti)triplet state, which conveniently 
is subdivided into a colourless hadron plus a simple coloured remnant. 
For instance, a $\u\u\d\s$ remnant could become $\Lambda + \u$. The current 
default is to assign the hadron a $\chi$ lightcone fraction according to the 
normal string fragmentation function. Differences relative to using simpler 
expressions, in the spirit e.g. of the even sharing above, are minor for the 
quantities of interest to us. One reason is that the sea contribution is 
much smaller than the gluon one above. A questionable
but convenient approximation is to assume that any emitted quark that could
be a valence quark also is one; a better choice would be to split e.g. the
$\u$ distribution of the proton into one valence and one sea part that are to
be handled differently. This simplification is not so critical, for the same 
reason as above.

The partons entering the hard interaction are traditionally taken to
have a non-vanishing  primordial $\kt$. In a shower description, 
such a $\kt$ is instead assigned to the initial-state shower initiators.
These could be seen as having a purely nonperturbative Fermi motion inside 
the incoming hadrons. Typical values should thus be 300--400 MeV,
consistent with constituent quark masses and fragmentation transverse 
momenta ($\sigma$ above). The initial-state shower will add further activity, 
so the parton that enters the hard $2 \to 2$ subprocess could well have more. 
However, in many connections \cite{primkt}, also for charm production at
fixed-target energies \cite{pertcharm2}, it has been 
noted that much higher values are required, at or even above 1 GeV.
This remains somewhat of a mystery, which we do not attempt to solve 
here. We will use a Gaussian width of 1~GeV as default. The choice of 
primordial-$\kt$ distribution is of non-negligible importance, both by 
providing a $\pt$ kick to the produced heavy-flavour quarks and, by momentum 
conservation, an opposite kick to the beam remnants.   

When a remnant is split up in two, not only longitudinal but also transverse
momentum sharing has to be specified. If the large primordial $\kt$ comes
from a complicated multigluon emission process, there is no reason why 
all of it should be taken by one of the remnants. Instead it is assumed shared 
evenly between the two. Furthermore, a relative kick is added between them, 
picked according to the standard fragmentation $\pt$ width $\sigma$, 
for simplicity and in the lack of any experimental indication.

A further aspect of the beam-remnant physics is the possibility of having
multiple parton--parton interactions in an event. This could have
an impact in a number of ways. Some of these, like an increase of the
underlying event activity and more complicated string drawings, are included 
in the standard {\Py} framework \cite{maria}, but obviously with several
degrees of freedom in the description. Others, like the production of 
multiple heavy-flavour pairs in separate hard processes and the possibility
of even more complicated beam remnants than the ones above, have not (yet) 
been studied. 

Another area not addressed is that of QCD interconnection, wherein a given 
colour configuration may be rearranged by soft-gluon exchanges 
\cite{intercon}. Mechanisms in this spirit have been proposed e.g. to produce 
closed-heavy-flavour states (J/$\psi$ etc.) from colour-octet heavy-flavour 
pairs \cite{uppsala}.

These examples serve as useful reminders that the modelling, however 
sophisticated, cannot be considered as complete in the nonperturbative sector.
Therefore one cannot hope for perfect agreement between the  model and the 
data. In the following we will show, however, that the current experimental 
data can be 
understood qualitatively, and often also quantitatively. This gives some 
confidence that the modeling described above is a good first approximation,
that could also be used for predictions in processes or at energies not yet 
studied. 

\section{Simple model properties}

In this section we examine some properties of the model as presented
in the previous section. In the first part we study purely
perturbative properties of the model such as the total cross section,
$\pthat$ of the hard interaction and quark distributions.
In the second part we study the properties of the
nonperturbative fragmentation. Experimental observables will be presented
and confronted with data in the next section.

\subsection{Properties of the perturbative production}
\label{pert_prop}

Above, three different production channels have been distinguished 
in the parton-shower description: pair creation, flavour excitation 
and gluon splitting. In the following we will present their separate 
contributions, even though this subdivision of course is unobservable
and model-dependent. It will still provide helpful insights.

\begin{figure}[tbp]
\begin{center}
\epsfig{file=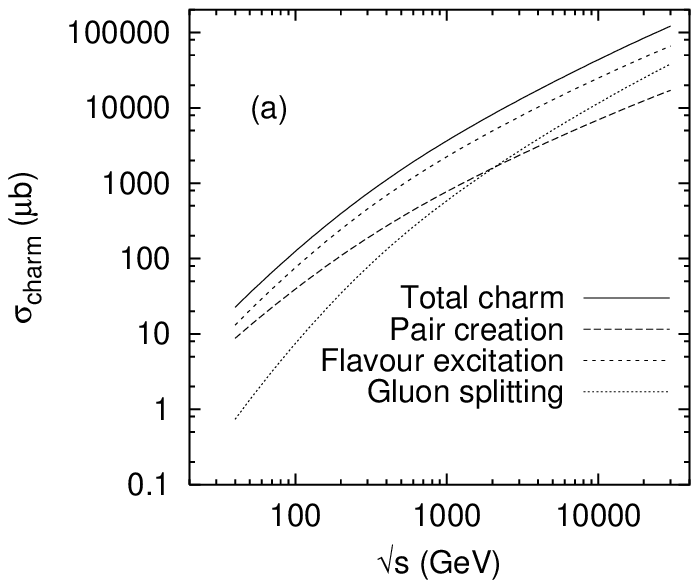}
\epsfig{file=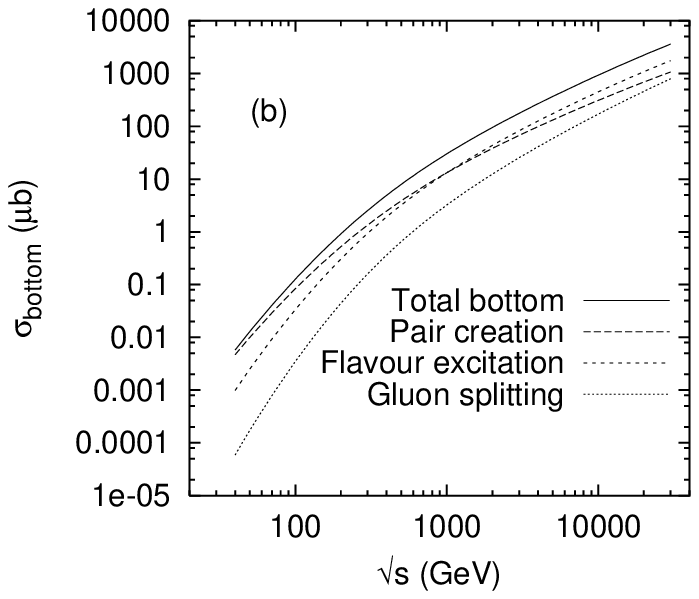}
\end{center}
\captive{The total (a) charm and (b) bottom cross sections for pp collisions 
as a function of $E_{\mathrm{CM}}=\sqrt{s}$. The contributions from pair 
creation, flavour excitation and gluon splitting are shown separately.
\label{fig:total_xsect}}
\end{figure}

The most basic and inclusive observable is the total heavy-flavour cross 
section. In Fig.~\ref{fig:total_xsect} we present it as a function of the 
pp center-of-mass energy, from the fixed-target r\'egime to LHC and beyond, 
both for charm and bottom. The cross section is divided into the 
contributions from the three perturbative production channels. As noted 
before, we assume that no nonperturbative effects contribute to the total 
cross section. The level of the total cross section is in sensible
agreement with the present data (not shown), indicating that there is no
need for any further significant production mechanism.

For small (fixed-target) energies the pair creation cross section 
is dominating the production, followed by a non-negligible fraction of
flavour excitation, whereas gluon splitting is very small. As the energy
is increased, flavour excitation overtakes pair production and gluon 
splitting is catching up. At very large energies gluon splitting
becomes the dominant production mechanism, so that the low-energy 
pattern is completely reversed.

The reason is not so difficult to understand. If we think of any partonic
process, it will only contain one hardest $2 \to 2$ scattering whatever
the energy, whereas the number of branchings in the associated initial- 
and final-state showers will increase with energy. This increase comes
in part from the the growing phase space, e.g. the larger rapidity
evolution range of the initial-state cascades, in part from the increase
in accessible and typical virtuality scales $Q^2$ for the hard subprocess.
The multiplication effect is at its full for gluon splitting, whereas
flavour-excitation topologies are more restrictive. At small energies,
however, the less demanding kinematical requirements for flavour 
excitation in a shower gives it an edge over gluon splitting.

\begin{figure}[tbp]
\begin{center}
\epsfig{file=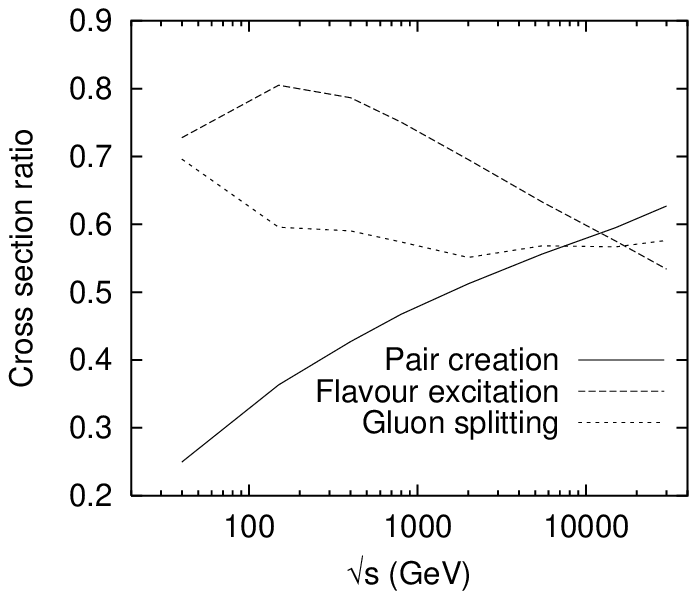}
\end{center}
\captive{Dependence of the charm cross section on model aspects, for pp 
collisions as a function of $E_{\mathrm{CM}}=\sqrt{s}$. Shown is the ratio of 
cross sections: pair creation for $m_{\c}=1.7$~GeV/$m_{\c}=1.3$~GeV,
flavour excitation for GRV 94L/CTEQ 5L parton distributions, and
gluon splitting for $Q^2_{\mathrm{max}} = M^2_{\mathrm{max}} = Q^2$/%
$Q^2_{\mathrm{max}} = M^2_{\mathrm{max}} = 4 Q^2$.
\label{fig:xsect_vary}}
\end{figure}

The total cross section is strongly dependent on QCD parameters such as 
the heavy-quark mass, parton distributions, and factorization and 
renormalization scales. It is not our aim here to present theoretical 
limits and errors --- this has been done elsewhere \cite{pertcharm2}. 
However, Fig.~\ref{fig:xsect_vary} gives some examples of how much
results may vary. Clearly, the quark-mass choice is very important,
especially for charm. Maybe surprisingly, the charm parton distributions 
in the proton do not differ by that much, probably reflecting a convergence 
among the common parton distributions and in the scheme adapted for 
$\g \to \Q\Qbar$ branchings in the evolution equations. Among the 
examples given, the largest uncertainty comes from the choice of the
heavy quark mass. However, it should be remembered that the 
variations above have no formal meaning of a `$1\sigma$' range of
uncertainty, but merely reflects some more or less random variations.

To gain further insight into the properties of the perturbative production
processes, one may study `non-observables' that characterize the 
hard-scattering process associated with the production, such as the 
$\pthat$ of the hard interaction. We also show kinematical 
distributions, like the rapidity and transverse momentum of the heavy 
quarks, and correlations between them, in order to quantify in which 
regions the different production processes contribute. As an example,
$\b\bbar$ production is studied at a 2~TeV $\p\pbar$ collider, where
the three production mechanisms are of comparable magnitude. Since the
valence-quark-dependent contribution to hard subprocesses is small
at this energy, there is no significant difference between $\p\p$ and 
$\p\pbar$.

\begin{figure}[tbp]
\begin{center}
\epsfig{file=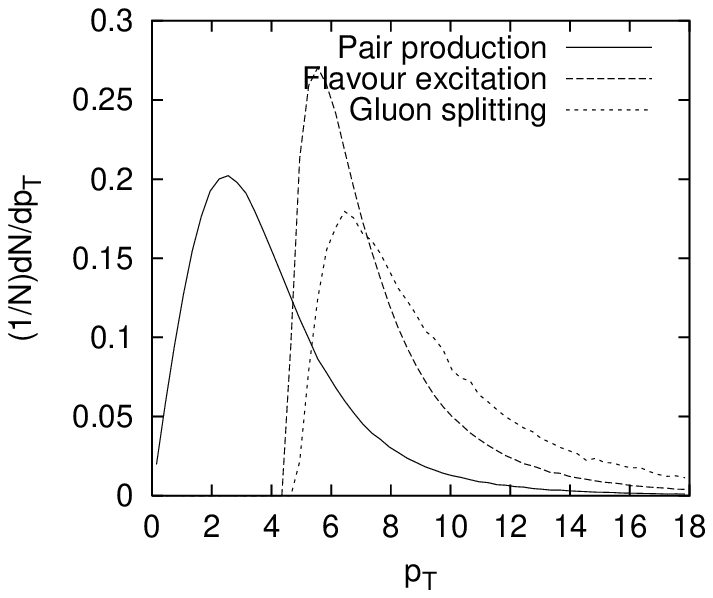}
\end{center}
\captive{$\pthat$ distributions for the hard interaction
associated with a $\b\bbar$ event at a 2 TeV $\p\pbar$ collider. The
pair creation, flavour excitation and gluon splitting curves are each
normalized to unit area to simplify comparisons of the shape.
\label{fig:pthat}}
\end{figure}

Fig.~\ref{fig:pthat} shows the $\pthat$ distribution of the hard 
interaction, where $\pthat$ is the transverse momentum of the outgoing 
partons evaluated in the hard-interaction rest frame.  
The main difference is in the behaviour at small  $\pthat$. For the 
pair creation process, massive LO matrix elements are used, so that
$\pthat$ goes all the way down to zero. The differential cross section 
is not divergent and no explicit $\ptmin$ cut is needed. For the other two
processes, massless matrix elements are used as a starting point ---
implying a divergent cross section in the limit $\pthat \to 0$ ---
and mass constraints are introduced the back door. To be able to resolve 
a heavy quark inside a hadron, i.e. flavour excitation and initial-state 
gluon splitting, a virtuality $Q^2 = \pthat^2 > m_{\mathrm{Q}}^2$ is 
needed, so the quark mass sets $\ptmin$ in this case. 
(With a massive quark in the final state, the actually reconstructed $\pt$
of the hard scattering is always smaller than the nominal $\pthat$ one.)
Also the final-state shower contribution to gluon splitting begins at
$\ptmin = m_{\mathrm{Q}}$. Here the shower evolution scale is set by 
$M^2_{\mathrm{max}} = 4 \pthat^2$, but the threshold for gluon splitting is
at $M^2 = 4 m_{\mathrm{Q}}^2$, so the two factors of 4 cancel. 
We remind that the cuts fill a well-defined function: the 
heavy-flavour-producing part of the graph cannot be the most virtual 
one in flavour excitation or gluon splitting, or one would double-count 
with pair creation. Nevertheless, the very sharp thresholds may be
somewhat of an artefact, and are certainly smeared when the effects of
further QCD emissions are included.

\begin{figure}[tbp]
\begin{center}
\epsfig{file=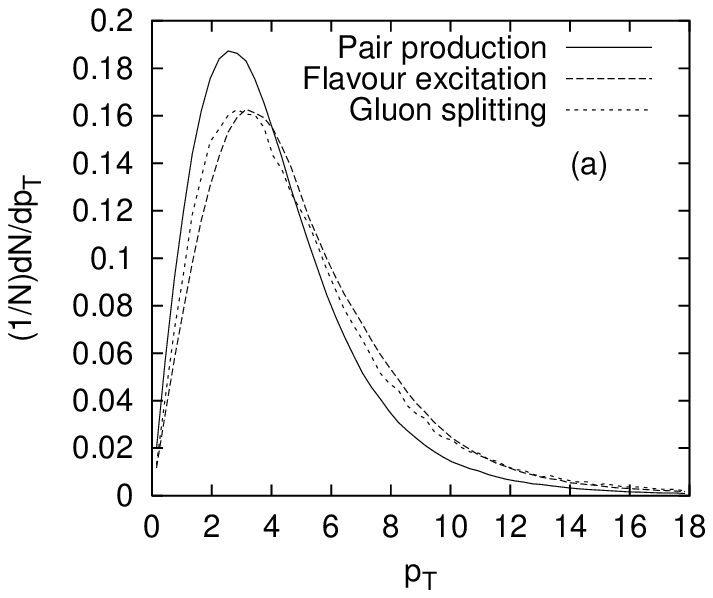}
\epsfig{file=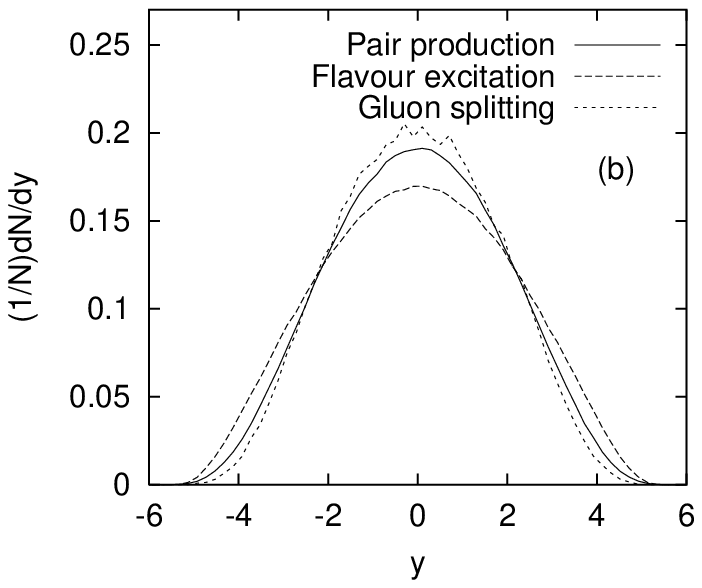}
\end{center}
\captive{Single $\b/\bbar$ quark (a) $\pt$ and (b) $y$ distributions 
at a 2 TeV $\p\pbar$ collider. The curves are each normalized to unit 
area to simplify comparisons of the shape.
\label{fig:single}}
\end{figure}

As an illustration, consider Fig.~\ref{fig:single}, where we show the 
single $\b/\bbar$ transverse momentum ($\pt$) and rapidity ($y$)
distributions of the produced quarks for the three production channels.
Here the full parton-shower and intrinsic-$\kt$ smearing effects are
included. Now all the $\pt$ spectra extend down to $\pt = 0$, and the
shapes are surprisingly similar, although pair creation remains somewhat
softer than the other two mechanisms. The rapidity
spectra agree even better between the three mechanisms, although flavour
excitation gives somewhat more production at larger rapidities and gluon
splitting more at central ones, as could be expected. This seems to 
indicate that the heavy-quark production part of the process is more or 
less independent of what goes on in the rest of the event. There are 
indeed also similarities in the descriptions, e.g. gluon splitting
$\g \to \Q\Qbar$ is equivalent to the $s$-channel graph of pair
creation $\g\g \to \Q\Qbar$, while flavour excitation is closely
related to the $t$-channel graph of pair creation. Furthermore, 
compensation mechanisms are at play: the $\pt$ spectrum of gluon
splitting is softened by the $\Q$ and $\Qbar$ having to share the
$\pt$ of the gluon between them, but this is compensated by the relative
$\pt$ in the $\g \to \Q\Qbar$ branching itself.

\begin{figure}[tbp]
\begin{center}
\epsfig{file=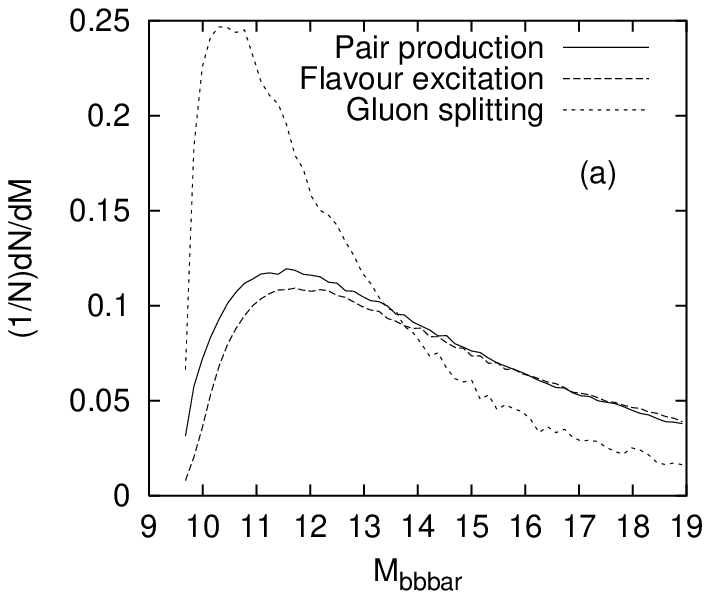}
\epsfig{file=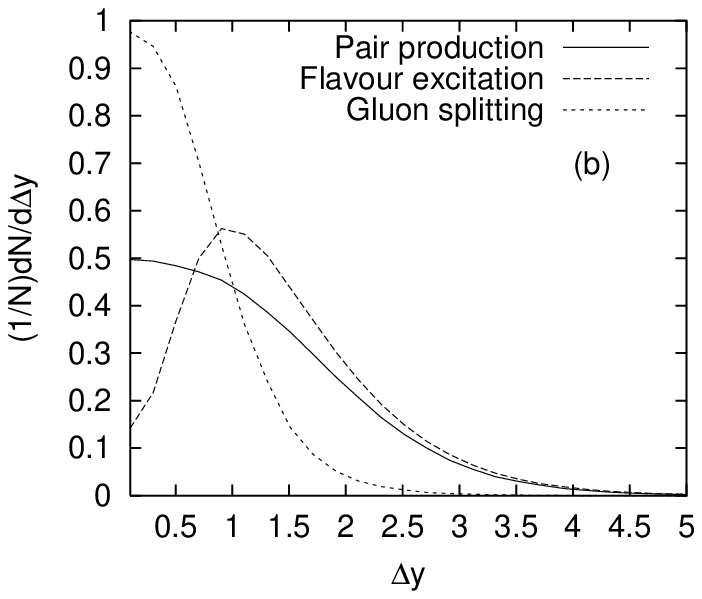}
\epsfig{file=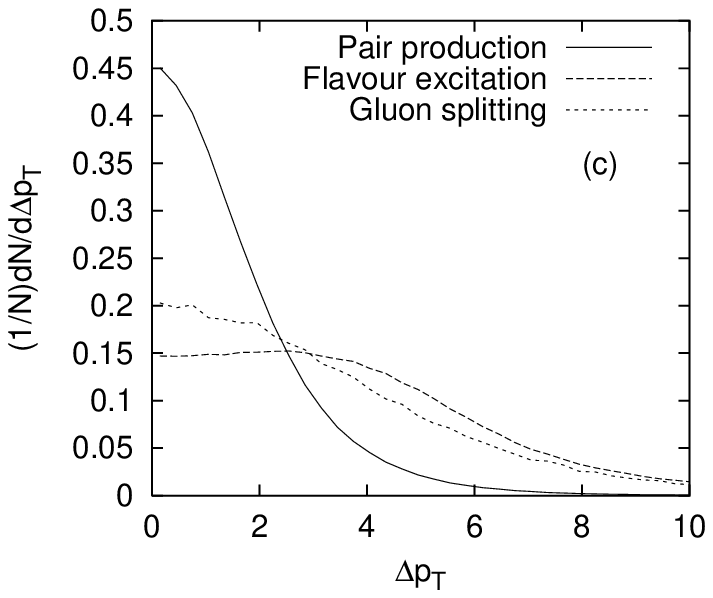}
\epsfig{file=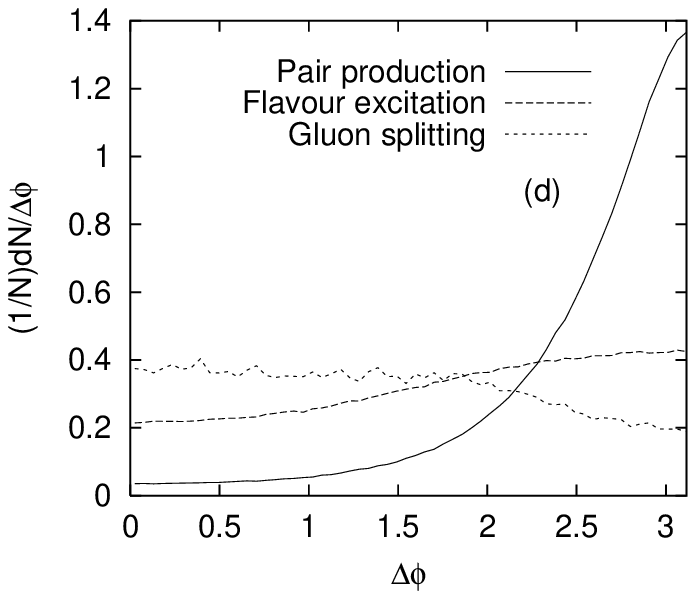}
\end{center}
\captive{Correlations between $\b$ and $\bbar$ at a 2 TeV $\p\pbar$ 
collider: (a) $m_{\b\bbar}$, (b) $\Delta y = |y_{\b}-y_{\bbar}|$, 
(c) $\Delta\pt = |p_{\perp\b}-p_{\perp\bbar}|$ and (d)
$\Delta\phi=|\phi_{\b}-\phi_{\bbar}|$. The curves are each normalized to 
unit area to simplify comparisons of the shape.
\label{fig:correlations}}
\end{figure}

Correlations between the produced heavy quarks turn out to be more 
interesting, since here the difference between the three production
channels are better visible. In Fig.~\ref{fig:correlations} we present 
the distributions of $m_{\b\bbar}$, $\Delta y = |y_{\b}-y_{\bbar}|$, 
$\Delta\pt = |p_{\perp\b}-p_{\perp\bbar}|$ and 
$\Delta\phi=|\phi_{\b}-\phi_{\bbar}|$.
The invariant mass spectrum is appreciably more peaked for gluon
splitting than for the other two mechanisms. Given that gluon splitting
is equivalent to the $s$-channel exchange of a gluon, while the other
two are dominated by $t$-channel contributions, it is clear why the
gluon splitting is more suppressed at large $m_{\b\bbar}^2 =$`$\hat{s}$'. 
As a logical consequence, also the $y$ correlation is more narrow for
gluon splitting.

In the $\Delta y$ distribution the differences are even more marked.
Here flavour excitation is depleted at small rapidity differences and
approaches the pair production spectrum only at large $\Delta y$.
The explanation of this involves several mechanisms. When a gluon
in the parton-distribution evolution
splits into a $\b\bbar$ pair this gives them a small initial rapidity
separation, with a distribution which is centered around zero much like
the gluon splitting distribution in Fig.~\ref{fig:correlations}b. One of
the heavy quarks then enter the hard interaction and is back-scattered
by a parton from the other beam. Since the minimum $\pthat$ of the hard
interaction here is $m_{\b}$ and this is the largest scale of the
process, the rapidity shift can be fairly large. An additional smearing
is introduced by further gluon emissions in the parton shower, but
not enough to hide the underlying behaviour.

Differences also appear in transverse momentum correlations. In the
$\Delta\pt$ and $\Delta\phi$ distributions, pair creation is the one 
most peaked in the region of a heavy-quark pair with opposite and 
compensating $\pt$. Thus the basic LO-process behaviour largely 
survives showers and primordial $\kt$. In the other two processes 
the correlations are more smeared. Especially discerning is the 
$\Delta\phi$ distribution, where gluon splitting gives an almost
flat curve, pair creation a clear peak near $180^{\circ}$, and
flavour excitation is somewhere in between. In gluon splitting,
the $\pt$ of the hard scattering favours small angles and the
$\pt$ of the splitting itself large angles, so the near-flat curve 
is the result of a non-trivial balance. Needless to say, a cut on
the $p_{\perp\b}$ and $p_{\perp\bbar}$ values would distort the
$\Delta\phi$ distribution significantly: at large $\pt$'s, pair
creation becomes more peaked at large angles and gluon splitting
peaked at small angles.

To summarize, we note that flavour excitation and gluon splitting
give significant contributions to the total heavy quark cross
section at large energies and thus must be considered.
NLO calculations probably do a better job on the total b cross section
than the shower approach, whereas for the lighter c quark, production in
parton showers is so large that the NLO cross sections are more questionable.
The shapes of single heavy quark spectra are not altered as much as the correlations
between $\Q$ and $\Qbar$ when flavour excitation and gluon splitting is added
to the leading order result. Similar observations have been made when
comparing NLO to LO calculations \cite{pertcharm1,Nason_and_CO}.

\subsection{Properties of the fragmentation}

We now proceed to describe properties of the fragmentation process.
In the Lund string fragmentation model no new heavy flavours are
produced during the fragmentation, so the model can `only' map the
momentum of a heavy quark onto the momentum and species of a
heavy hadron in the final state. This can, however, be dramatic 
enough, e.g. with hadrons formed at larger momenta than the 
perturbatively produced quark, or with flavour asymmetries 
favouring the production of heavy hadrons sharing light valence 
quark flavours with the incoming beam particles. Such topics will be 
covered in detail in the following section, so here we only mention 
a few of the basic aspects. 

\begin{figure}[tbp]
\begin{center}
\epsfig{file=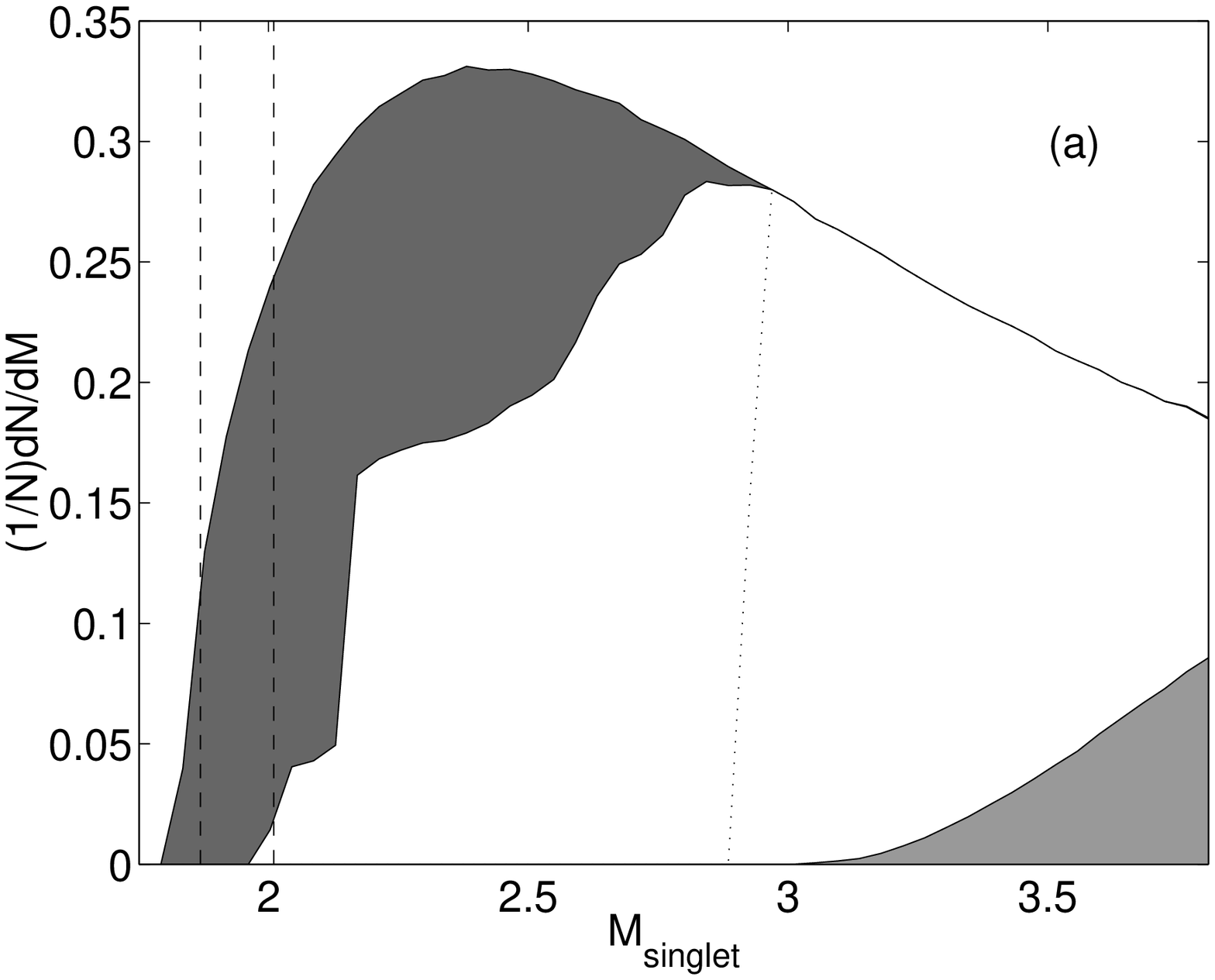,width=7.5cm}
\epsfig{file=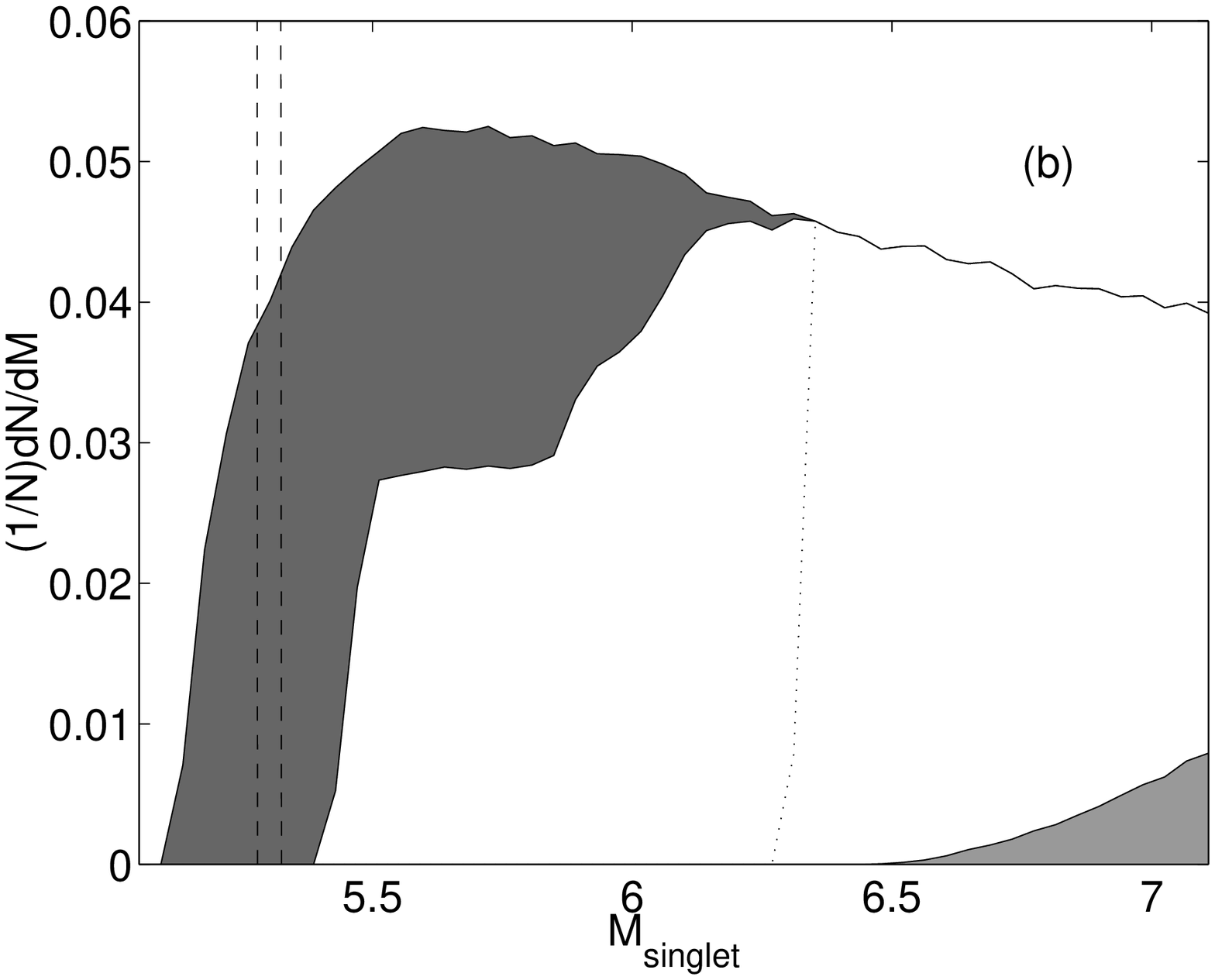,width=7.5cm}
\end{center}
\captive{The fate of a cluster/string as a function of its mass,
(a) for charm in $\pi^-\p$ collisions with a $\pi^-$ beam momentum 
of 500 GeV, and (b) for bottom in $\p\pbar$ collisions at 2 TeV.
The full curve represents the original mass spectrum, for simplicity 
only from pair creation. Clusters within the gray area to the left 
collapse to a single particle, predominantly the 
$\D/\D^*$ or $\B/\B^*$ states indicated by 
dashed vertical lines. The white middle area gives two primary
hadrons, with cluster decay to the left of the dotted curve and
with string fragmentation to the right. The rightmost gray area
corresponds to the production of three or more primary hadrons 
from the string.  
\label{fig:clustermass}}
\end{figure}

The fate of a colour singlet system in the string model depends
on its mass and on its flavour content.
The mass spectrum of strings/clusters containing one heavy flavour and
a u or d quark at the other end is shown in Fig.~\ref{fig:clustermass} for two 
typical processes. Here we are only interested in the low-mass
behaviour where only few primary hadrons are produced. (Secondary
decays, of everything from $\rho$ to $\B^*$, is not considered here.)
In the high mass region traditional string fragmentation should work well.
Technically, the description is split into clusters, giving one
or two hadrons, and strings, giving two or more. In total, the 
area under the curve in Fig.~\ref{fig:clustermass} splits into these 
four contributions. The transition from one to two hadrons comes in 
a set of steps, somewhat smeared e.g. by the $\rho$ Breit-Wigner shape. 
That from two to three particles is continuous, but still slightly 
tailor made. Only beyond that is the multiplicity determined fully 
dynamically, by fragmenting off hadrons of random energies until 
all has been used up.

\begin{figure}[tbp]
\begin{center}
\epsfig{file=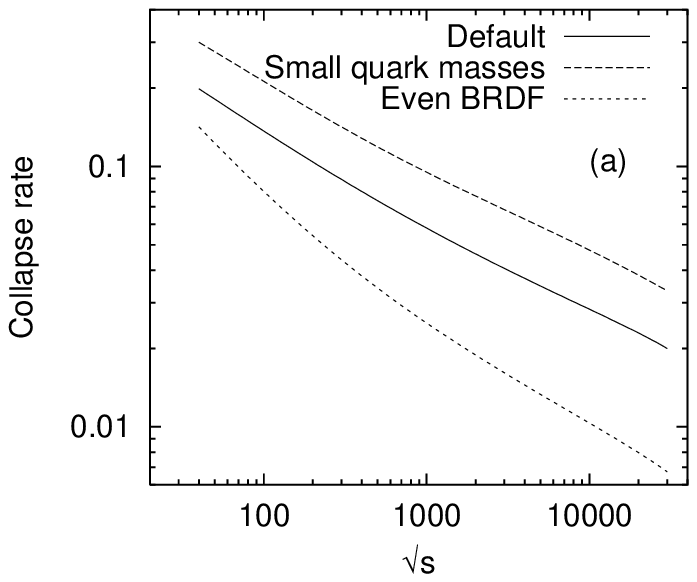}
\epsfig{file=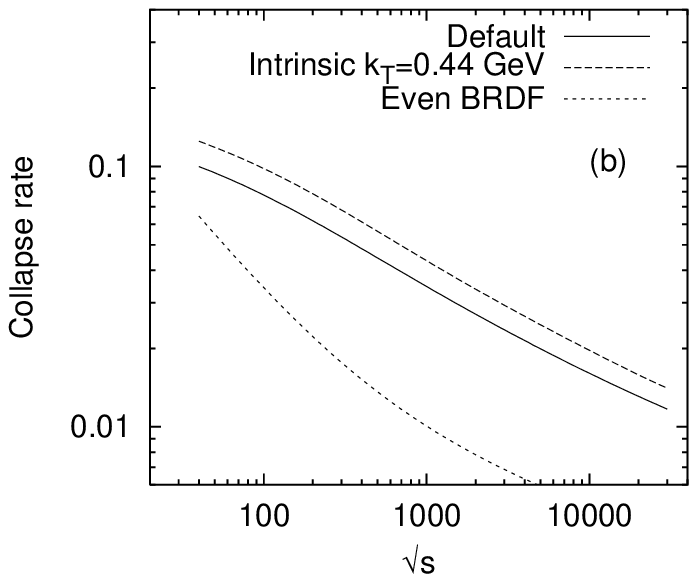}
\end{center}
\captive{The average number of cluster collapses to a single heavy
hadron per heavy-quark event, for (a) charm and (b) bottom in a pp
collision as a function of $E_{\mathrm{CM}}=\sqrt{s}$.
For simplicity, only pair creation is included.
\label{fig:collapserate}}
\end{figure}

The mass spectrum near threshold, and thus the amount of collapses 
to a single hadron, is sensitive to a large number of parameters, 
such as the heavy and light quark/diquark masses, the average 
primordial $\kt$, and the beam-remnant description \cite{our}.  
Some of these are constrained by information from other
processes, but a significant uncertainty remains. By introducing
some piece of experimental data, such as the flavour asymmetries in
$\pi^-\p$ collisions, a reasonable overall set of parameters has
been found. The energy dependence of the collapse rate is then 
predicted, Fig.~\ref{fig:collapserate}. The drop with larger energies
is a natural consequence of the string mass spectrum then extending to
larger values. The collapse rate can be shifted up or down e.g. by varying 
the charm mass, and shifted in shape by the beam-remnant description, 
but always follows the same qualitative behaviour. No input has been 
used from $\B$ physics, so here measurements would directly test the 
universality of the model. Note that the collapse rate is expected to
be lower for bottom than for charm, since the mass spectrum near
threshold scales roughly with $m/m_{\Q}$, i.e. the bulk of the mass
spectrum is higher above threshold for heavier quarks, while the upper
limit for collapse goes like $m_{\Q} +$constant.

\begin{figure}[tbp]
\begin{center}
\epsfig{file=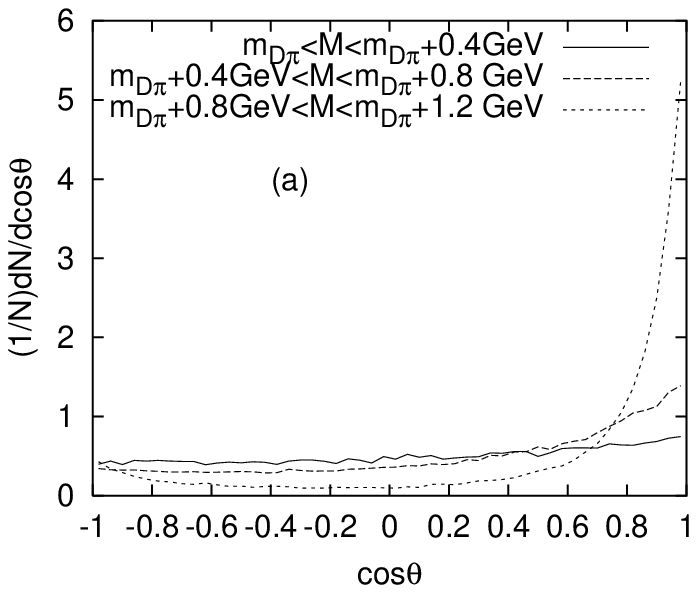}
\epsfig{file=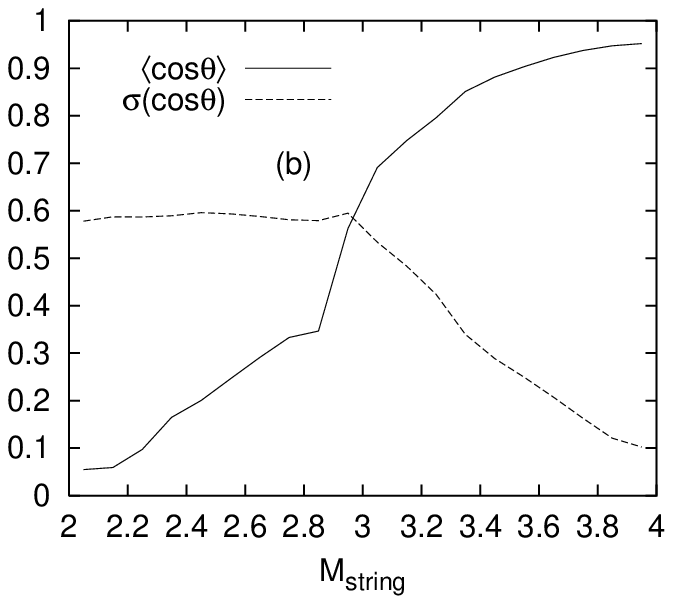}
\end{center}
\captive{Direction of a $\D/\D^*$ hadron in the decay of a $\c\dbar$ 
string/cluster at rest.
(a) The distribution $\d n/\d(\cos\theta)$, 
with $\theta$ the angle between the $\c$ and $\D/\D^*$, for a few
masses.
(b) $\langle \cos\theta \rangle$ and $\sigma(\cos\theta)$ as a
function of the string/cluster mass.
\label{fig:clusterdecay}}
\end{figure}

The transition between the cluster and string two-hadron scenarios
is purely artificial, and in the best of worlds the treatments should
smoothly match at the crossover. In the new {\Py} version, an 
attempt has indeed been made to ensure that. Specifically, the cluster 
decays anisotropically in a fashion that should mimic the string 
scheme for the larger cluster masses. We have studied the angular 
distributions in the transition region and found this to be reasonably well
fulfilled, Fig.~\ref{fig:clusterdecay}.

\begin{figure}[tbp]
\begin{center}
\epsfig{file=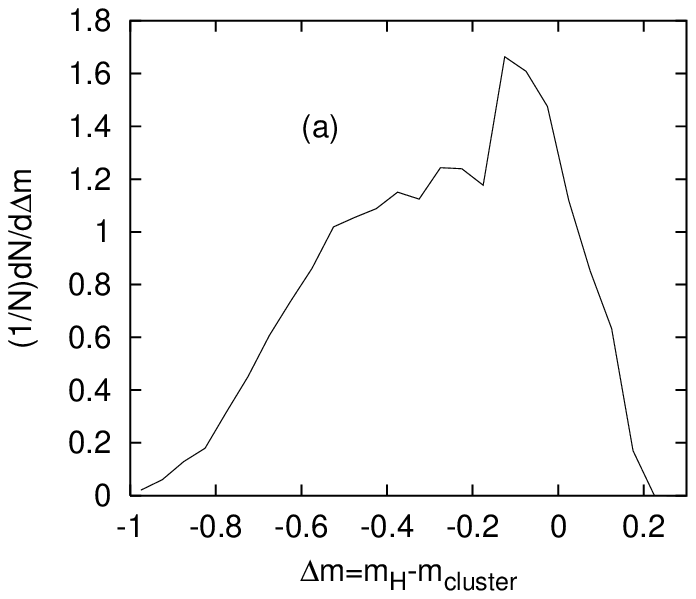}
\epsfig{file=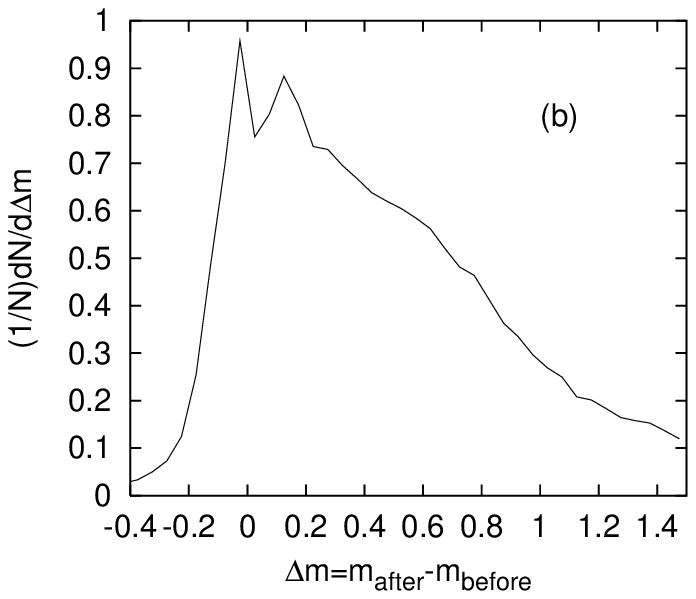}
\end{center}
\captive{Distribution of mass shifts induced by the collapse of clusters
to a single charm hadron, in $\pi^-\p$ collisions with a $\pi^-$ beam 
momentum of 500 GeV, and for simplicity only including pair creation.
(a) For the collapsing system, i.e. 
$\Delta m = m_{\H} - m_{\mathrm{cluster}}$.
(b) For the string taking the momentum recoil,
$\Delta m = m_{\mathrm{after}} - m_{\mathrm{before}}$.
\label{fig:collapseshift}}
\end{figure}

When a cluster collapses to a single hadron, energy and momentum is
redistributed between this system and the rest of the event. The mass
shift in the collapsing cluster is implicit in the shape of the 
leftmost gray area in Fig.~\ref{fig:clustermass}, and an explicit
illustration is given in Fig.~\ref{fig:collapseshift}a. With the
current default set of quark masses and form of the two-particle threshold,
it is more likely that the 
produced hadron has a smaller mass than the original cluster.
For lower quark masses,
it is possible to reverse this asymmetry, but then at the price of 
a cluster collapse rate in excess of what is indicated by data.
The string system that takes the energy/momentum recoil of the
collapse clearly will see its mass shifted in the opposite direction,
Fig.~\ref{fig:collapseshift}b. If the cluster and string are at 
relative rest, the two mass shifts are exactly compensating, 
but a relative motion tends to distort this, in the direction of a
larger (opposite) string mass shift than cluster mass shift.
The reasonably narrow distribution in Fig.~\ref{fig:collapseshift}b
then indicates that the compensation algorithm is working well.

\begin{figure}
\begin{center}
\epsfig{file=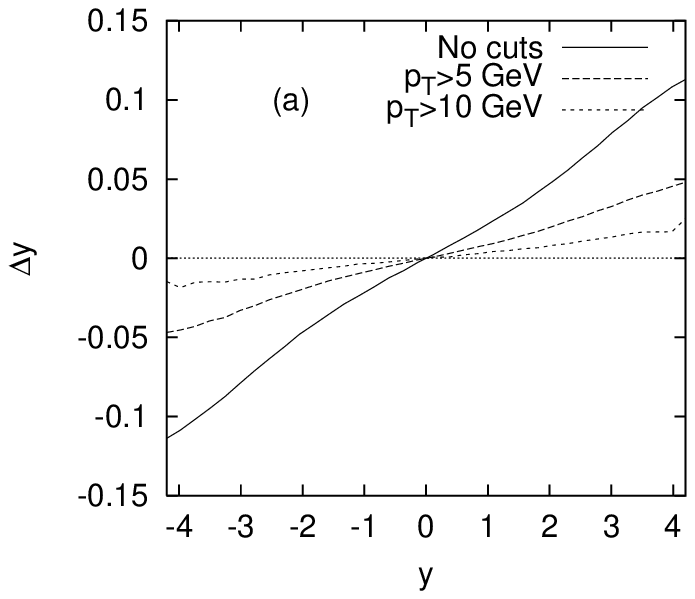}
\epsfig{file=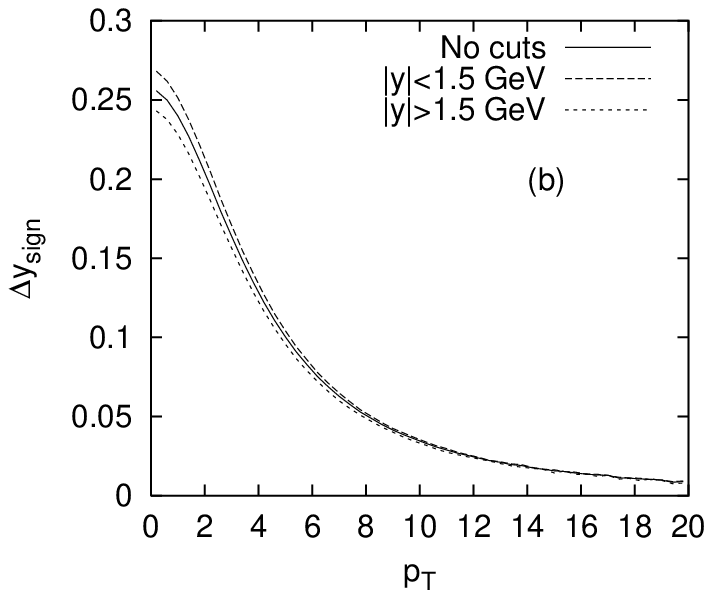}
\end{center}
\captive{
(a) Average rapidity shift $\Delta y$ as a function of $y$ for some different
$\pt$ cuts for a $\p\pbar$ collider at 2 TeV.
(b) Average rapidity shift $\Delta y_\mathrm{sign}$ as a function of $\pt$
for some different rapidity cuts.
\label{drag}}
\end{figure}

The colour connection between the produced heavy quarks and the beam
remnants in the string model gives rise to an effect called
beam remnant drag. In an independent fragmentation scenario, a quark jet
fragments symmetrically around the quark direction. The light cone
(along the quark axis) energy-momentum of the quark
is then simply scaled by some factor, picked from a fragmentation function,
in order to give the momentum of the hadron.
Thus, on average, the rapidity would be conserved in the fragmentation process.
This is not necessarily so in string fragmentation where
both string ends contribute to the four-momentum of the produced
heavy hadron. If the other end of the string is a beam remnant,
the hadron will be shifted in rapidity in the direction of the beam remnant,
often resulting in an increase in $|y|$. This beam-drag is shown
qualitatively in Fig.~\ref{drag}, where the rapidity shift for bottom
hadrons in a 2 TeV $\p\pbar$ collision is shown as a
function of rapidity and transverse momentum. We use two different measures of
the rapidity shift. The first is the average rapidity shift
$\Delta y = \langle y_\mathrm{B} - y_\mathrm{b} \rangle$. Here the heavy quark
can be connected to a beam remnant on either side of the event, giving rise
to shifts in both directions which tend to cancel in inclusive
measures. A better definition is therefore
\begin{equation}
\Delta y_\mathrm{sign} = \langle (y_\mathrm{\B}-y_\mathrm{\b})
\cdot \mathrm{sign}(y_\mathrm{other~end}) \rangle,
\end{equation}
which measures the rapidity shift in the direction of the other end of the string.
This shift should almost always be positive. The rapidity shift is not directly accessible
experimentally, only indirectly as a discrepancy between the shape of perturbatively
calculated quark distributions and the data.

\subsubsection{High-$\pt$ asymmetries}

There is another possible asymmetry which occurs at large transverse momentum,
involving the collapse and drag of scattered valence quarks and heavy quarks
produced by gluon splitting in the parton shower. As an illustration, consider
Fig.~\ref{highpt} where a valence u quark is scattered to high transverse momentum
in a high energy $\p\p$ collision. In such high-$\pt$ jets, parton showering will
be profuse. If a gluon close to the scattered u quark splits into a heavy quark
pair, the heavy antiquark could be in a colour singlet system together with the scattered
u quark. If this singlet has a small mass it could collapse into a single heavy
hadron. Heavy hadrons with its light quark constituent in common with the beam
will thus be favoured. The effect then is due to the asymmetry in the composition
of jet flavours. At reasonably low  $\pt$, where gluon (and second to that seaquark)
jets dominate, effects therefore are vanishingly small.

\begin{figure}
\begin{center}
\begin{picture}(200,145)(0,0)
\SetOffset(0,-30)

\Text(45,70)[r]{p}
\Line(50,73)(60,73)
\Line(50,67)(60,67)
\GOval(65,70)(15,5)(0){.7}

\Text(155,70)[l]{p}
\Line(140,73)(150,73)
\Line(140,67)(150,67)
\GOval(135,70)(15,5)(0){.7}

\Gluon(100,70)(130,70){4}{3}
\ArrowLine(70,70)(100,70)
\Text(91,75)[b]{u}

\Gluon(100,70)(90,40){4}{3}
\ArrowLine(100,70)(110,100)

\Gluon(110,100)(125,125){4}{3}
\ArrowLine(110,102)(97,142)

\ArrowLine(125,125)(110,150)
\ArrowLine(125,125)(140,150)
\Text(96,145)[b]{u}
\Text(105,150)[b]{$\bbar$}
\Text(141,153)[bl]{b}

\SetWidth{1.}
\Oval(99,151)(8,16)(35)
\Text(80,166)[l]{$\B^+$}
\end{picture}

\end{center}
\captive{
Illustration of the high-$\pt$ asymmetry.
\label{highpt}}
\end{figure}

This mechanism was studied in \cite{Basym} and the size of the effect
was found to be at the $10^{-3}$ level. Here we would like to study
if the modifications to the model has changed this result.
To increase the effectiveness we study high-$\pt$ quark jets 
in a 14 TeV pp collision and look at B mesons produced within the jets.
To be specific, and hint at a possible experimental procedure to study
the effect, we generate events
containing the subprocess $\q\q'\to\q\q'$ (where q is a scattered valence quark)
with $\hat{p}_{\perp,min} > 500$ GeV and look for events containing two high-$\pt$
jets. We use a cone algorithm to find the jets and
then look for B mesons within these jets which carry at least 20\% of the
jet $E_\perp$. The asymmetry between $\B^0$ and $\Bbar^0$ fulfilling these
criteria was found to be
$0.019 \pm 0.005$ and the asymmetry between $\B^+$ and $\B^-$
$0.011 \pm 0.005$. The size of possible collapse asymmetries is limited by
the probability for a b hadron within a jet to be produced in a collapse
between a scattered valence quark and a b quark.
This probability was found to be at the $10^{-3}$ level. This indicates that
another mechanism is at play giving rise to a larger asymmetry.
Furthermore, the two asymmetries above might have been expected to be of opposite
sign, by equality between the number of $\b$ and $\bbar$ quarks,
at least if strangeness and baryon production can be neglected.

A possibility is that gluon splittings on the perimeter of the jet-cone
give rise to $\b\bbar$ pairs where the $\bbar$ is colour connected to the
scattered valence quark and the $\b$ is connected to the beam remnant diquark.
In the string fragmentation process the $\bbar$ could be dragged towards
the scattered high-$\pt$ quark at the center of the jet and the $\b$ towards
the low-$\pt$ beam remnant, i.e. away from the jet, thus lowering the rate
of $\Bbar^0$ and $\B^-$ within the jet and at the same time giving rise to a
slightly harder $\pt$ spectrum for leading B mesons at high $\pt$. This is
simply a variation of the drag effect already discussed, only this time
the drag is in the transverse direction instead of the longitudinal one.

The total asymmetry then is a
convolution of the asymmetry in the composition of jet flavours with
the asymmetry in the b hadronization mechanisms. To get an estimate of
the total asymmetry the result above must be diluted with all other
non-asymmetry-generating QCD processes contributing to high-$\pt$ jets
containing B mesons, e.g. $\g\g\to\g\g$ and flavour excitation.
The ratio between the cross sections for producing a B meson within
a valence quark jet and within any jet is approximately 0.035
in this specific case, so the
diluted $\B^0/\Bbar^0$ asymmetry is $(6.5 \pm 1.7) \cdot 10^{-4}$.
The lesson to be learned is that asymmetries can turn up also when not
expected and will depend on the procedure used in studying the effects,
like jet $\pt$, jet clustering algorithms, and B hadron selection criteria.

\section{Applications}

In this section we apply the model presented in the preceding sections
to some typical current and future experiments at both small and large
energies. No attempt will be made to be exhaustive, instead different
examples will be picked as illustrations of the basic ideas.
At low energies the most striking effect is the flavour 
asymmetries already observed in several experiments
\cite{oldobs,WA82, WA92,E769, E791}. At large energies the most important 
aspect may be the beam drag effect, suggested in HERA data
both for photoproduction \cite{HERA_photo} and deep inelastic scattering (DIS)
\cite{HERA_DIS}.

Since the model is available in Monte Carlo form, further studies are 
left to the interested reader. However, there are some caveats. In 
particular, nuclear-target effects are not simulated. Instead such a 
target has to be represented by a single proton or neutron, with total 
cross sections suitably rescaled. One does not expect large nuclear 
effects in the heavy-flavour production characteristics, but effects 
may be non-negligible.

\subsection{Fixed-target \boldmath{$\pi^-$}p}\label{fixed}

Charm production in fixed-target $\pi^-\p$ collisions was already
studied in \cite{our,HQ98proc}, but only using the pair creation
mechanism of charm production. Here we will extend the results to the other
production mechanisms and to correlations between the two charm hadrons in the event.

\begin{figure}
\begin{center}
\epsfig{file=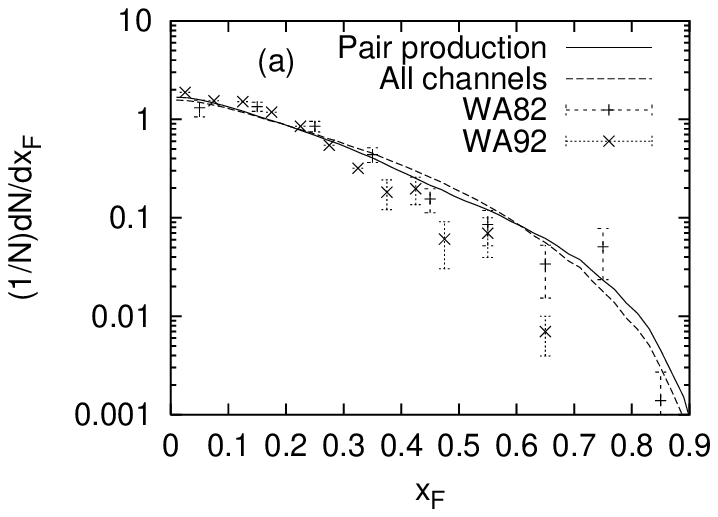}
\epsfig{file=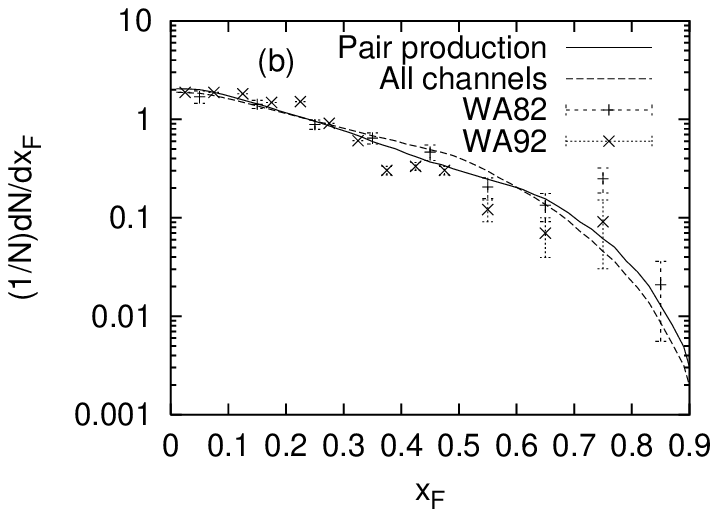}
\epsfig{file=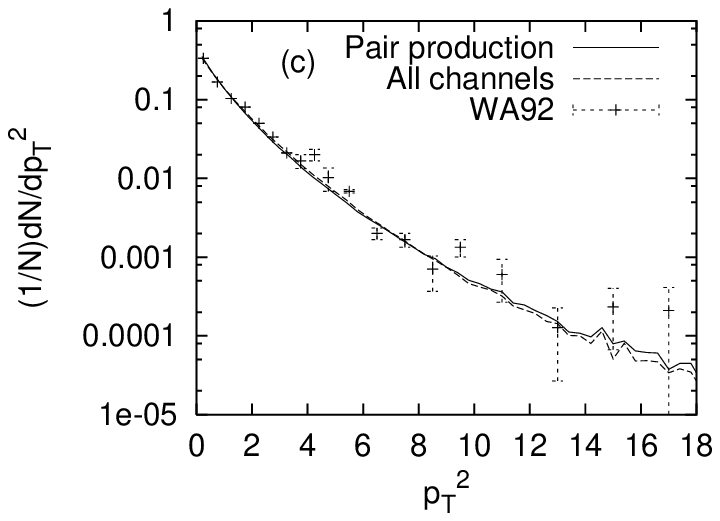}
\epsfig{file=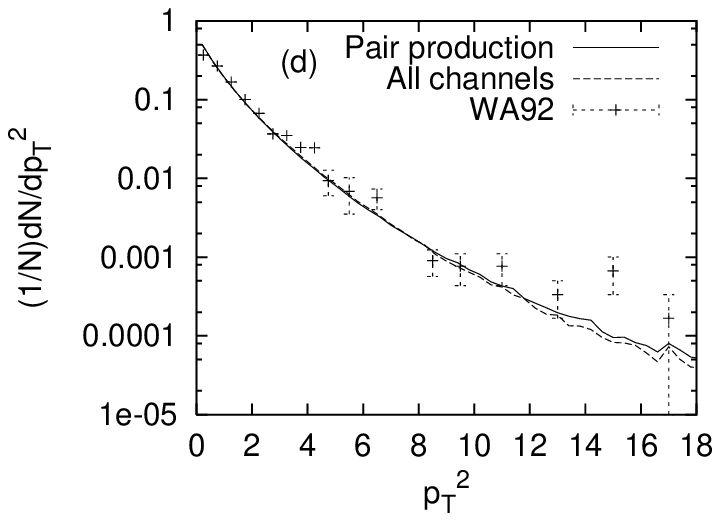}
\end{center}
\captive{$\D^+/\D^-$ meson spectra in a $\pi^- \p$ collision
with $\sqrt{s} = 26$ GeV.
(a) Single $\D^+$ $\xF$ distributions.
(b) Single $\D^-$ $\xF$ distributions.
(c) Single $\D^+$ $\pt^2$ distributions.
(d) Single $\D^-$ $\pt^2$ distributions.
The distributions are normalized to the sum of the experimental $\D^+$ and $\D^-$
cross sections in each case.
\label{fixed:single}}
\end{figure}

\begin{figure}
\begin{center}
\epsfig{file=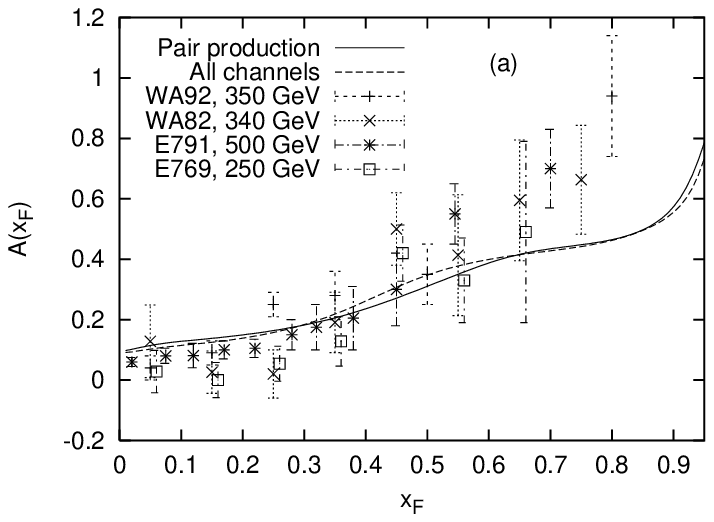}
\epsfig{file=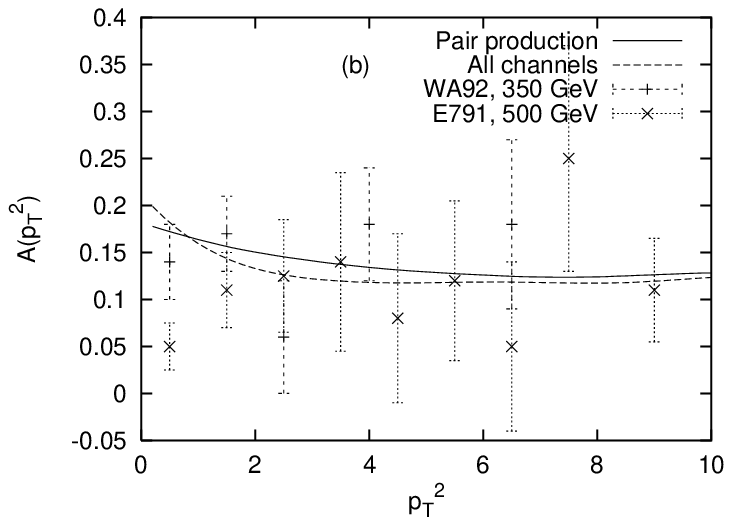}
\end{center}
\captive{The resulting asymmetry as a function of $\xF$ and $\pt^2$. The theoretical curves
are model results for a 340 GeV $\pi^-$ beam on a proton target ($\sqrt{s}=$ 26 GeV).
\label{fixed:asym}}
\end{figure}

We see from Fig.~\ref{fig:total_xsect} that gluon splitting does not give a significant
contribution to the total charm cross section for fixed-target energies around
30 GeV. Flavour excitation, on the other hand, gives a contribution to charm production
which is as large as the pair production one even at these low energies.
Fig.~\ref{fixed:single} shows the single heavy hadron $\xF$ and $\pt^2$ distributions
as predicted by the model, using the default parameters, compared to data from the WA82
\cite{WA82} and WA92 \cite{WA92} experiments. The leading order pair creation result
and the result with all production channels added together are shown separately.
The agreement with
data is reasonable, though the $\xF$ spectra are slightly harder than the data,
especially for $D^+$
which is non-leading. The resulting asymmetry as a function of $\xF$ and $\pt^2$ is
shown in Fig.~\ref{fixed:asym}. The data are taken at slightly different energies
but the energy dependence of the model is small within the experimental energy range.
This seems to be true also for the data.
As expected from the study in Sec.~\ref{pert_prop}, single
charm spectra and asymmetries are not significantly altered by the addition of
flavour excitation and gluon splitting.

\begin{figure}
\begin{center}
\epsfig{file=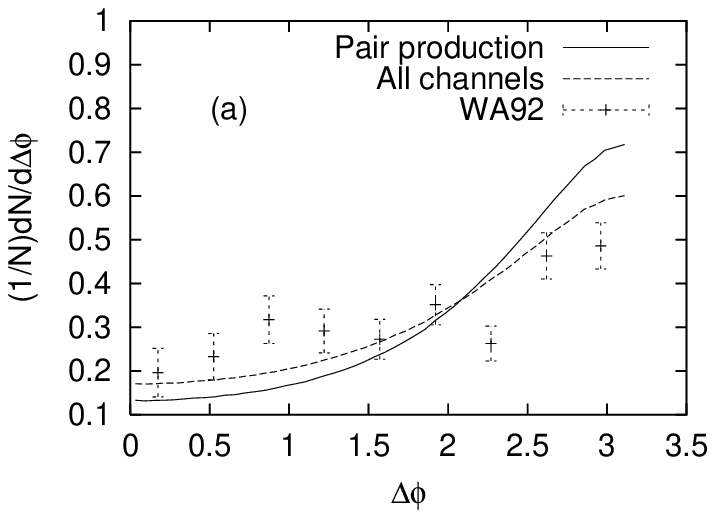}
\epsfig{file=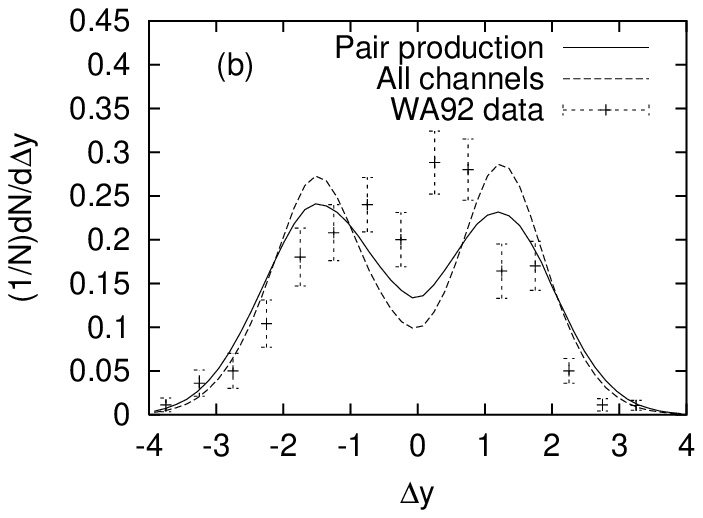}
\end{center}
\captive{Correlations between $\D^+$ and $\D^-$ when at least one D meson is
produced with $\xF>0$ for a $\pi^-\p$ collision with $\sqrt{s}=26$ GeV.
The figure shows the distribution of
(a) $\Delta\phi = |\phi_{\D^+} - \phi_{\D^-}|$ and
(b) $\Delta y = y_{\D^+} - y_{\D^-}$.
The distributions are normalized to the total $\D^+\D^-$ cross section in each case.
The data is taken from the WA92 experiment \cite{WA92corr}.
\label{fixed:corr}}
\end{figure}

\begin{figure}
\begin{center}
\epsfig{file=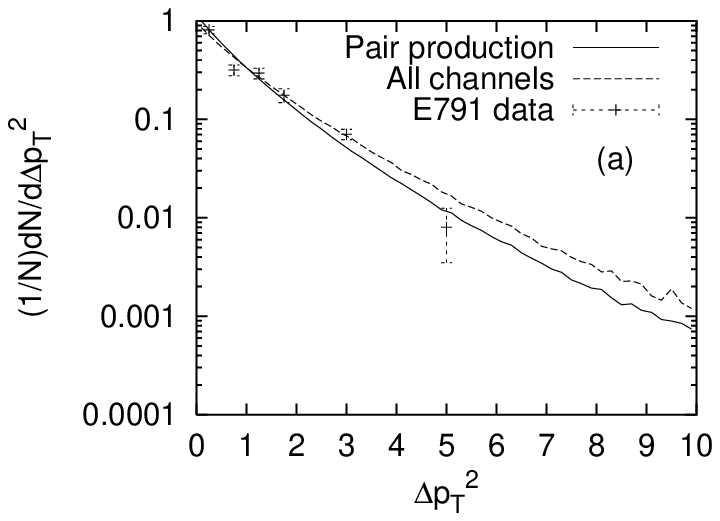}
\epsfig{file=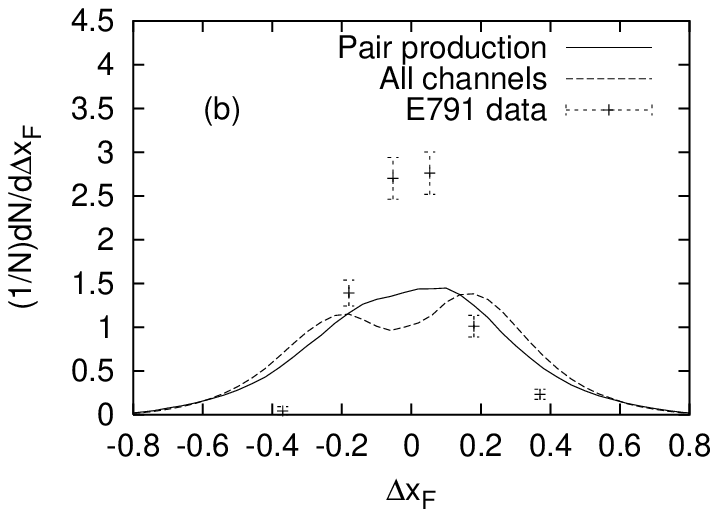}
\end{center}
\captive{Correlations between $\D$ and $\Dbar$ mesons when both are produced with
rapidity $-0.5 < y < 2.5$ for a $\pi^-\p$ collision with $\sqrt{s}=30$ GeV.
The figure shows the distribution of
(a) $\Delta p_\perp^2 = |p_{\perp,\D}^2 - p_{\perp,\Dbar}^2|$ and
(b) $\Delta \xF = x_{\mathrm{F},\D} - x_{\mathrm{F},\Dbar}$.
The distributions are normalized to the total $\D\Dbar$ cross section in each case.
The data is taken from the E791 experiment \cite{E791corr}.
\label{E791_corr}}
\end{figure}

Fig.~\ref{fixed:corr} shows the correlation in $\phi$, the angle between $\D^+$ and $\D^-$
in the transverse direction, and rapidity, $y$ compared to data from the WA92
experiment \cite{WA92corr}. The model prediction for correlations is
more sensitive to the addition of flavour excitation and gluon splitting than single
charm spectra are, again as expected from Sec. \ref{pert_prop}. In this case the description
of data is improved for the transverse correlation ($\Delta\phi$) but not for the longitudinal
($\Delta y$) one. The large separation in the rapidity distribution is a consequence of
the colour connection between the produced charm quarks and the beam remnants, which tend
to shift the charm momenta in the direction of the respective beam remnant, thus increasing
the rapidity separation. The same pattern is repeated when comparing correlation
distributions to the E791 collaboration \cite{E791corr} in Fig.~\ref{E791_corr}.

There are two mechanisms which could decrease the rate of connection.
The first is gluon splitting into light quarks ($\g \rightarrow \q\qbar$)
in the parton shower and the other is colour reconnections \cite{intercon,uppsala}.
Gluon splitting would
split the string in two and each would then hadronize independently. The memory of
the colour connection is decreasing for each splitting of the string, thus decreasing
the drag in the direction of the beam remnants. However, at these low energies the phase space
for gluon splitting is limited.
Colour reconnection by soft gluon exchange could change the colour structure of the
event, thus making the charm quark lose its colour connection to the beam remnant.
In the simplest scenario the charm quark pair could be reconnected to form a
$\c\cbar$ colour singlet, causing them to pull
each other closer, resulting in a drastic decrease in $\langle \Delta y \rangle$.
However, this mechanism would increase the production of $\mathrm{J}/\psi$
from collapses of low-mass $\c\cbar$ colour singlet
systems. Given the experimental ratio of $\mathrm{J}/\psi$ to $\D\Dbar$
production, $\sigma(\mathrm{J}/\Psi)/\sigma(\D\Dbar) \sim 0.02$ \cite{JPsi_ratio},
and a collapse probability of $\sim$50\% in reconnected events (not all to $\mathrm{J}/\Psi$),
the colour reconnection probability would thus be limited to 5 -- 10 \%. More complicated
colour reconnections could be imagined, involving also gluons from the parton shower,
where the charm quarks are not always in a colour singlet but still not connected directly to
the beam remnants. This kind of more sophisticated colour reconnection models have been
used with some success to describe $\mathrm{J}/\psi$ production and rapidity gaps in
hadronic collisions and DIS \cite{uppsala}. The colour reconnection rates
needed to describe the longitudinal correlations would also significantly soften
the single charm distributions as well as lower the asymmetry. This is not favoured
by the data so in the following we only consider single heavy quark distributions
which are well described by the model.

We also compare results on some cross section ratios
shown in Table \ref{table:ratios} and again the description is
reasonable but not perfect. The cross sections are interrelated by
several model aspects. Consider e.g. the ratio of $\D_s$ to $\D_{\q}$. The
main parameter that determines the rate of strangeness production in the
fragmentation is the ratio between u, d and s production, which by default is set
to 1:1:0.3. This number has been fitted to $\ee$ data and can not be changed appreciably.
The $\D_s$ production ratio is also sensitive to the collapse rate of colour
singlets containing charm and a light quark from a beam remnant.
A large collapse rate decreases the ratio, because $\D_s$ is depleted in favour
of $\D_\q$ when the beam consists of non-strange particles (like in this case $\pi^-$).

\begin{table}
\begin{center}
\begin{tabular}{|l|c|c|c|c|} \hline
 		& $\frac{\sigma({\D^+,\D^-})}{\sigma({\D^0,\Dbar^0})}$
		& $\frac{\sigma({\D_s^+,\D_s^-})}{\sigma({\D^0,\Dbar^0,\D^+,\D^-})}$
		& $\frac{\sigma(\D^-)}{\sigma(\D^+)}$
		& $\frac{\sigma(\D^0)}{\sigma(\Dbar^0)}$\\ \hline
Model			& 0.33	& 0.11	& 1.39	& 0.94\\ \hline
Experimental average
\cite{WA92}		& 0.415	& 0.129	& 1.35	& 0.93\\ \hline
\end{tabular}
\end{center}
\caption{Cross section ratios for $\pi^-\p$ collisions around 26 GeV.}
\label{table:ratios}
\end{table}

To summarize, we find good agreement with several fixed target experiments
when it comes to single charm spectra and asymmetries. The only case where
the model does not perform well is for longitudinal correlations. Similar
results were obtained \cite{HQ98proc} in comparisons with the E791 experiment
\cite{E791corr}. In that study the contradiction between single charm and correlation
data from different experiments was observed, a problem that as of yet has not been
resolved, as we have seen.

\subsection{HERA-B}

\begin{figure}
\begin{center}
\epsfig{file=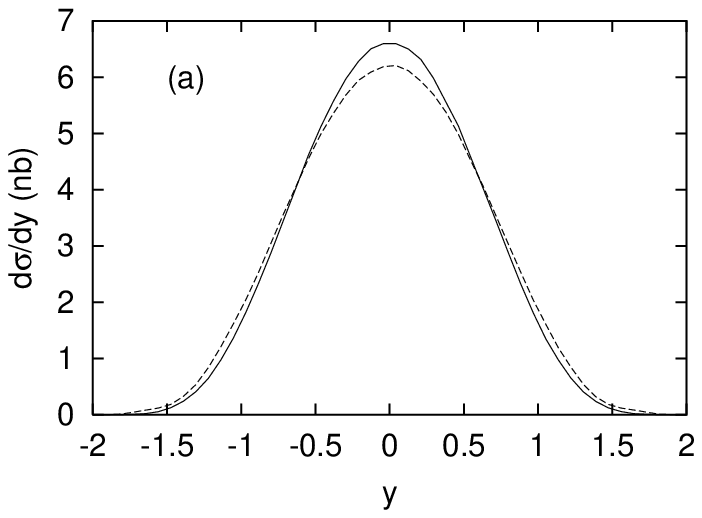}
\epsfig{file=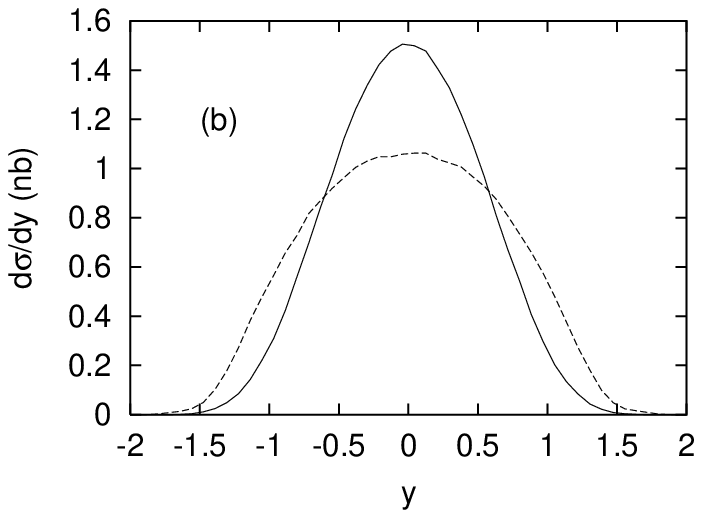}
\epsfig{file=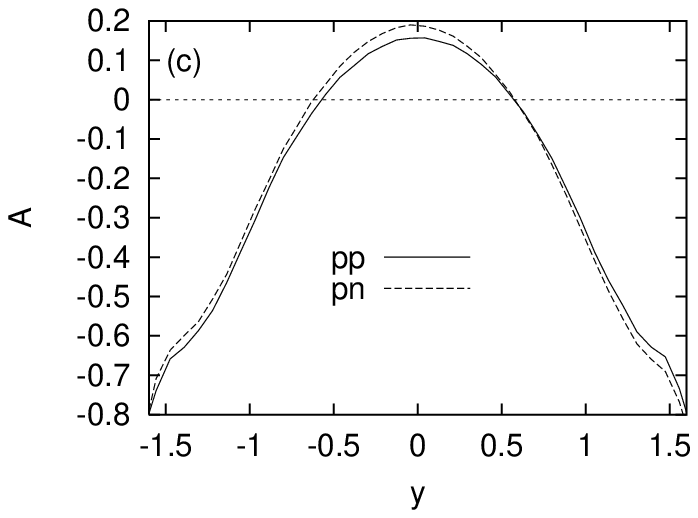}
\epsfig{file=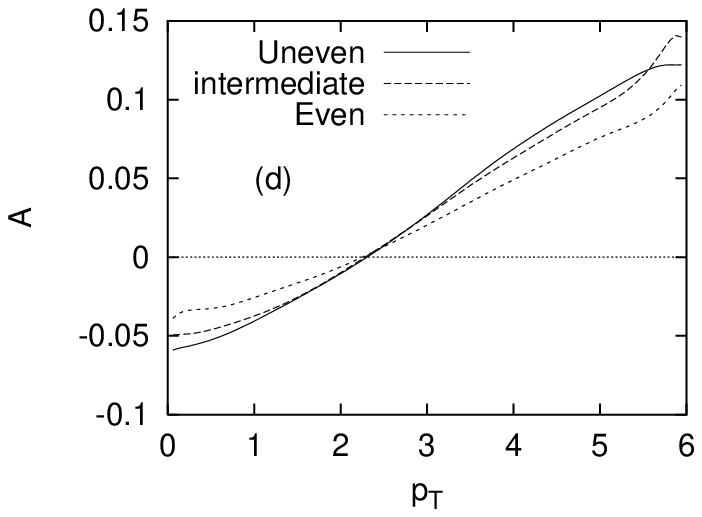}
\end{center}
\captive{Bottom production in a pA collision at HERA-B energies,
neglecting nuclear effects.
(a) Rapidity distribution of bottom quarks (full)
and the B hadrons produced from them (dashed).
(b) Rapidity distribution of $\B^0$ (full) and $\Bbar^0$ (dashed).
(c) The asymmetry
$A=\frac{\sigma(\B^0)-\sigma(\Bbar^0)}{\sigma(\B^0)-\sigma(\Bbar^0)}$
as a function of rapidity comparing pp and pn collisions.
(d) The asymmetry as a function of $\pt$ for three different parameterizations
of the BRDF of the proton. For simplicity, only pair production is included.
\label{herab:Bhadrons}}
\end{figure}

The HERA-B experiment at DESY is a fixed-target experiment built
especially for bottom studies. The experiment will study $\p\mathrm{A}$
collisions at a center of mass energy of about 40 GeV. It is therefore
an ideal experiment to test the results of Section~\ref{fixed} for
bottom quarks. Predictions for $\B\Bbar$ asymmetries and spectra follow directly
from the model using the new updated set of parameters and hadronization mechanisms.
The HERA-B experiment collides protons with nuclei but we do not include any simulation
of nuclear effects. We take into account the neutrons in the nuclei by simulating
pp and pn events separately and use the mean to produce the plots.
The only missing pieces are then the bottom quark mass and the proton beam remnant
distribution. By the simple ansatz of eq.~(\ref{bmass}) we obtain $m_\b=4.8$ GeV.
The BRDF of the proton is more problematic because there is no fundamental
understanding of its structure. There are some indications that an even sharing
of energy-momentum is favoured by experiments~\cite{E791,photoprod}.
We have tried different parameterizations but found no significant
qualitative differences (see e.g. Fig.~\ref{herab:Bhadrons}).

Fig.~\ref{herab:Bhadrons} shows the distribution of bottom mesons at HERA-B, showing
both the size of the drag effect and asymmetries. The asymmetry is significant
at all rapidities, not only large ones, and can reach as high as 20\% even in the
central rapidity region. When the kinematical limit at large rapidities is
approached, the asymmetry changes sign for small
$\pt$ because of the drag-effect; $\b$ quarks connected to diquarks
from the proton beam remnant which carry most of the remnant energy often produce
$\Bbar^0$ hadrons which are shifted more in rapidity than the $\mathrm{B}^0$'s are.
Cluster collapse, on the other hand, tend to
enhance the production of `leading' particles (in this case $\mathrm{B}^0$) so
the two mechanisms give rise to asymmetries with different signs. Collapse is
the main effect at central rapidities while eventually at very large $y$,
the drag effect dominates. This is also reflected in the $\pt$ dependence
of the asymmetry which exhibits a sign shift at small $\pt$. The $\pt$ dependence
is, however, partly a consequence of the fact that large $\pt$'s imply small $|y|$.
Compared to pp, the asymmetry is slightly larger for pn collisions in the negative
rapidity region. This is a natural consequence of the larger amount of d quarks in
the neutron beam remnant.

\subsection{The Tevatron Collider}
The Tevatron $\p\pbar$ collider operating at CM energies up to 2 TeV represent a
significant step up the energy ladder and offers a good opportunity to check the
energy dependence of our results. We first show some generic distributions
for the Tevatron and then consider a scenario where
very forward, low-$\pt$ bottom hadrons can be detected. We use $2.5<|y|<4$ and
$\pt<5$ GeV. This would be ideal for studying the drag effect which is inherently a
low-$\pt$/high-$y$ phenomenon, see Fig.~\ref{drag}.

\begin{figure}
\begin{center}
\epsfig{file=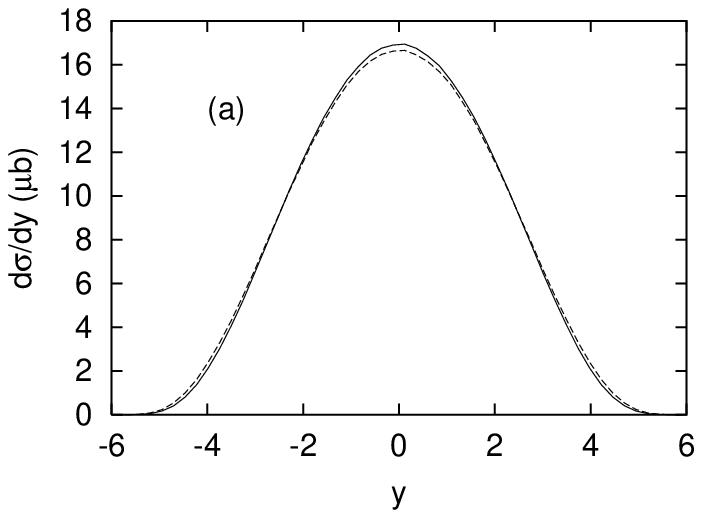}
\epsfig{file=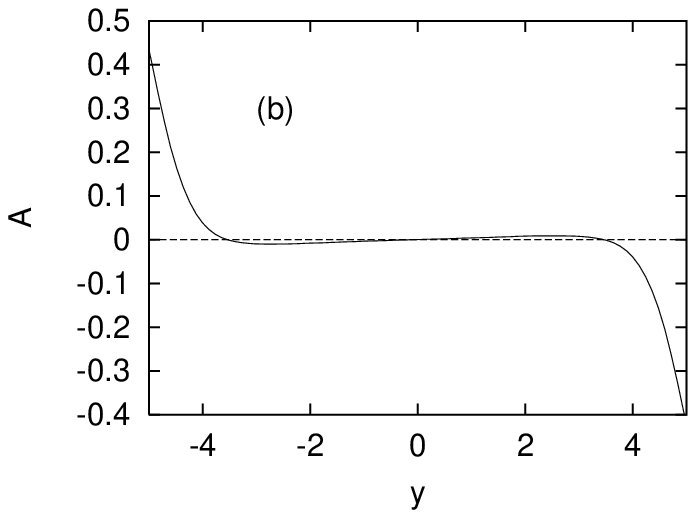}
\end{center}
\captive{Bottom production at the Tevatron.
(a) Rapidity distribution of bottom quarks (full)
and the B hadrons produced from them (dashed).
(b) The asymmetry $A=\frac{\sigma(\B^0)-\sigma(\Bbar^0)}{\sigma(\B^0)-\sigma(\Bbar^0)}$
as a function of rapidity. For simplicity, only pair production is included.
\label{tev_generic}}
\end{figure}

Fig.~\ref{tev_generic} shows the distribution of bottom quarks and the hadrons
produced from them, as well as the asymmetry between $\B^0$ and $\Bbar^0$ without
any kinematical cuts. The trend is similar to that of HERA-B, but the asymmetry is
antisymmetric because of the asymmetry of the initial state.
Therefore the asymmetry
is zero at y=0 and increasing in different directions for increasing/decreasing
rapidities. As expected from Fig.~\ref{fig:collapserate}, the integrated
asymmetry has decreased significantly.

\begin{figure}
\begin{center}
\epsfig{file=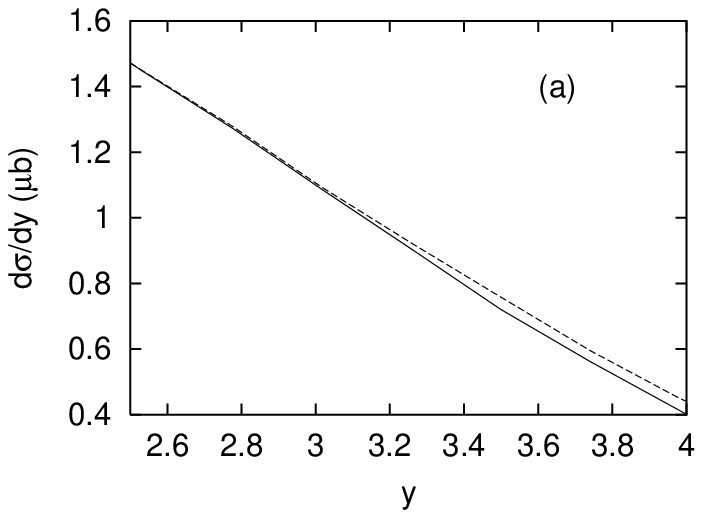}
\epsfig{file=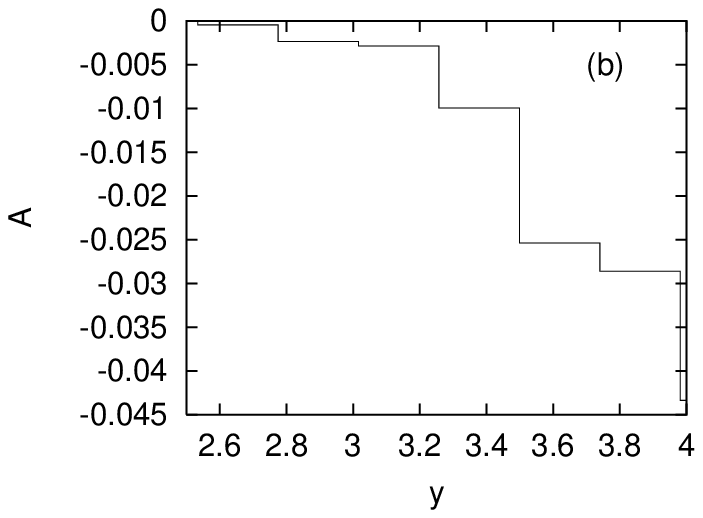}
\end{center}
\captive{Bottom production at the Tevatron for $2.5<|y|<4$ and $\pt<5$ GeV.
(a) $\B^0$ (full) and $\Bbar^0$ (dashed) rapidity spectra.
(b) The asymmetry $A=\frac{\sigma(\B^0)-\sigma(\Bbar^0)}{\sigma(\B^0)-\sigma(\Bbar^0)}$
as a function of rapidity. For simplicity, only pair production is included.
\label{tev_btev}}
\end{figure}

In Fig.~\ref{tev_btev} we introduce cuts in order to study the region of large
rapidities and small $\pt$. $\Bbar^0$ is shifted slightly more in the direction
of the beam remnant than $\B^0$, but the size of the effect is quite small, approaching
4\% at very large rapidities. Still, if large precision is desired in CP violation
studies, this effect could be non-negligible.

\subsection{LHC}
The difference between the Tevatron and the LHC collider is mainly that the energy
at the LHC is one order of magnitude larger and both colliding particles are protons.
Due to the similarities we only give some generic results on asymmetries and try to
assess the theoretical uncertainty of the model by looking at some parameter variations.

\begin{figure}
\begin{center}
\mbox{\epsfig{file=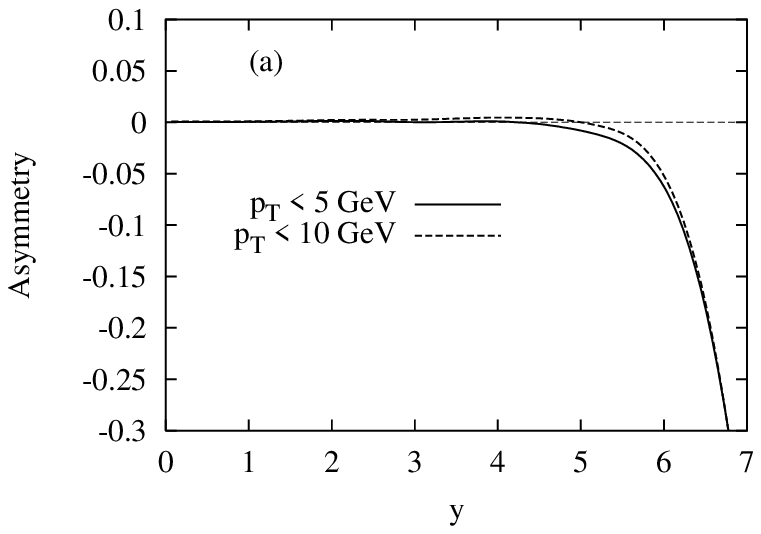}}
\mbox{\epsfig{file=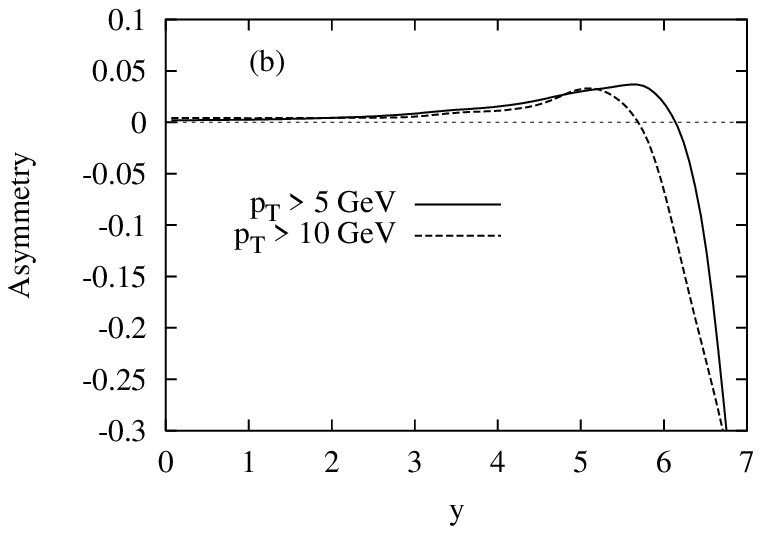}}
\end{center}
\caption{The LHC asymmetry, $A=\frac{\sigma(\mathrm{B}^0) - \sigma(\Bbar^0)}
{\sigma(\mathrm{B}^0) + \sigma(\Bbar^0)}$, as a function of rapidity for
different $\pt$ cuts: (a) $\pt < 5,10$ GeV and (b) $\pt > 5, 10$ GeV using
parameter set 1 as described in the text. For simplicity, only pair production is
included.}
\label{asym:string_asym}
\end{figure}

Fig.~\ref{asym:string_asym}
shows the asymmetry between $\mathrm{B}^0$ and $\Bbar^0$ as a function of $y$ for several
$\pt$ cuts in the string model. The asymmetry is essentially zero for central rapidities,
where the beam remnant flavour content is not felt so much. At intermediate
rapidities it is then positive (except at small $\pt$) only to turn negative at
larger rapidities. The reason is the same as for HERA-B,
but here the switch over is closer to the kinematical limit at large rapidities.

In Table~\ref{table:parameterdep} we study the parameter dependence of the asymmetry by
looking at the integrated asymmetry for different kinematical regions using three
different parameter sets:
\begin{Itemize}
\item {\bf Set 1} is the new default as described in Sec. \ref{sec:model}.
\item {\bf Set 2} The same as Set 1 except it uses simple counting rules in the beam remnant
splitting, i.e. each quark get on average one third of the beam remnant energy-momentum.
\item {\bf Set 3} The old parameter set, before fitting to fixed-target data, is included as a
reference. This set is characterized by current algebra masses, lower intrinsic $\kt$,
and an uneven sharing of beam remnant energy-momentum.
\end{Itemize}

\begin{table}
\begin{center}
\begin{tabular}{|l|c|c|c|} \hline
Parameters	& $|y|<2.5$, $\pt>5$ GeV& $3<|y|<5$, $\pt>5$ GeV& $|y|>3$, $\pt<5$ GeV \\ \hline
Set 1		& 0.003(1)		& 0.015(2)		& $-$0.008(1) \\ \hline
Set 2		& $-$0.000(2)		& 0.009(3)		& $-$0.005(2) \\ \hline
Set 3		& 0.013(2)		& 0.020(3)		& $-$0.018(2) \\ \hline
\end{tabular}
\end{center}
\caption{Parameter dependence of the asymmetry in the string model. The statistical
error in the last digit is shown in parenthesis (95\% confidence).
For simplicity, only pair creation is included.}
\label{table:parameterdep}
\end{table}

We see that in the central region the asymmetry is generally very small whereas for
forward (but not extremely forward) rapidities and moderate $\pt$ the asymmetry
is around 1--2\%. In the very forward region at small $\pt$, drag effects dominates,
which can be seen from the change in sign of the asymmetry. The asymmetry is fairly
stable under moderate variations in the parameters even though the difference between
the old and new parameter sets (Set 1 and 3) are large in the central region.
Set 1 typically gives rise to smaller asymmetries. Note also that asymmetries
on the perturbative level, calculated to NLO \cite{pertcharm1}, could become relatively
more important at
LHC energies, where the collapse asymmetries are small. Other non-perturbative
effects, such as intrinsic bottom, are also expected to be small at LHC energies
\cite{LHCbottom}.

To summarize, we find small asymmetries
at the LHC except at large rapidities. Unless the other b in the event is
unambiguously tagged, the asymmetry is still not completely negligible
for high-precision CP asymmetry studies, especially at LHC-B.

\subsection{Photoproduction}
The model can also be used in the photoproduction of heavy quarks in $\gamma\p$
collisions. Here we wish to apply the model to $\gamma\p$ physics
at HERA. The asymmetries are small in this case because of the high energy
and the flavour neutral photon beam. Instead we study beam-drag
effects, consequences of the photon structure and higher-order effects.

The photon is a more complicated object than a hadron because it has two components,
one {\em direct} where the photon interacts as a whole and one {\em resolved} where it has
fluctuated into a $\q\qbar$ pair before the interaction.
This will result in very different event structures
in the two cases. This study is constrained to real photons (photoproduction) as modeled by
Schuler and Sj\"ostrand \cite{SaS} and implemented in the \Py~\cite{Pythia} event generator.
We include the photon flux and use cuts close to the experimental ones. We first
examine the leading-order charm spectra for direct and resolved photons,
estimate the cross section
in the two cases, and study how the fragmentation process alters the charm
spectra in the string model. Then we add flavour excitation and gluon splitting
and find that also in this case they give a significant contribution to the charm
cross section, especially for resolved photons.

We consider charm photoproduction in an $\mathrm{e}^{\pm}\p$ collision
(820 GeV protons and 27.5 GeV electrons) with real photons ($Q^2<1$ GeV) and rather
large energy in the $\gamma\p$ CMS system ($130<W_{\gamma \mathrm{p}}<280$) using
some different $\pt$-cuts. The analysis is done in
the $\gamma\p$ center of mass system using true rapidity
($y = \frac{1}{2}\ln(\frac{E+p_z}{E-p_z})$) as the main kinematical variable.
The photon (electron) beam is incident along the negative z-axis.

\begin{figure}
\vspace*{2mm}
\epsfig{file=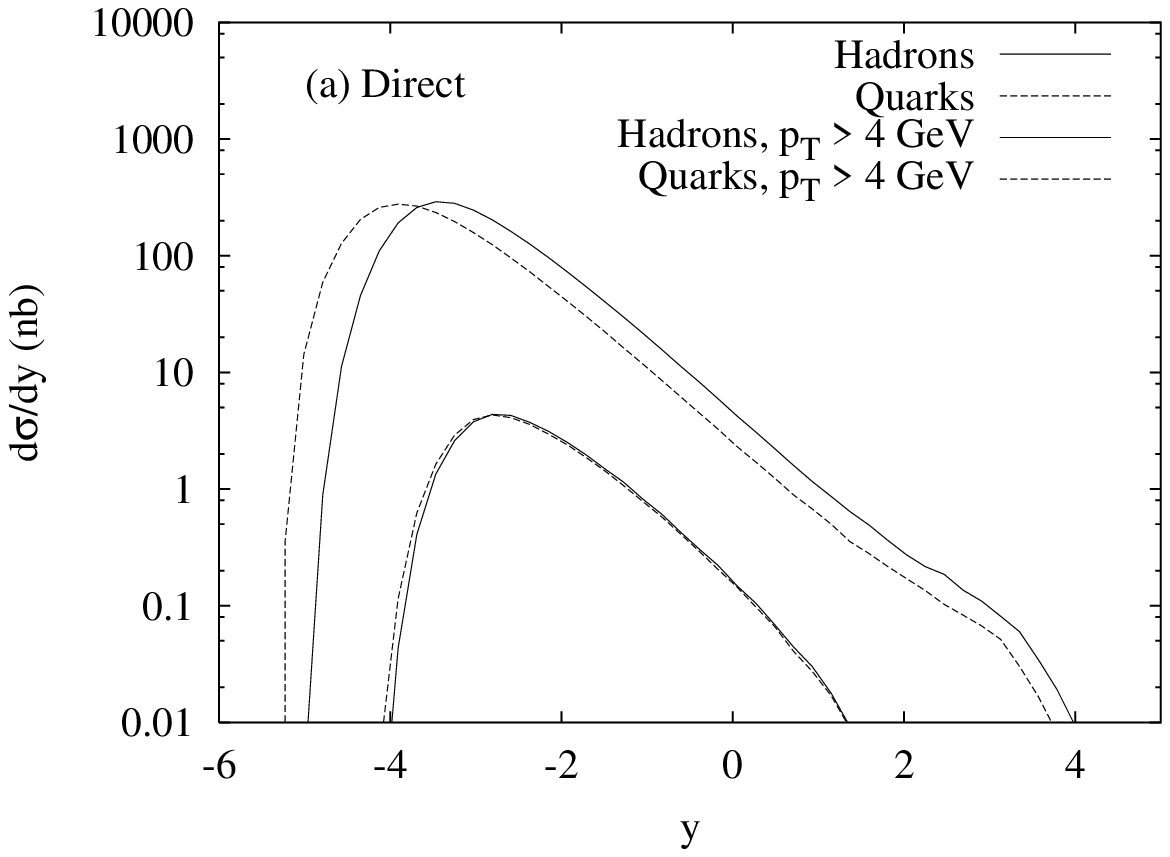, width=80mm}
\epsfig{file=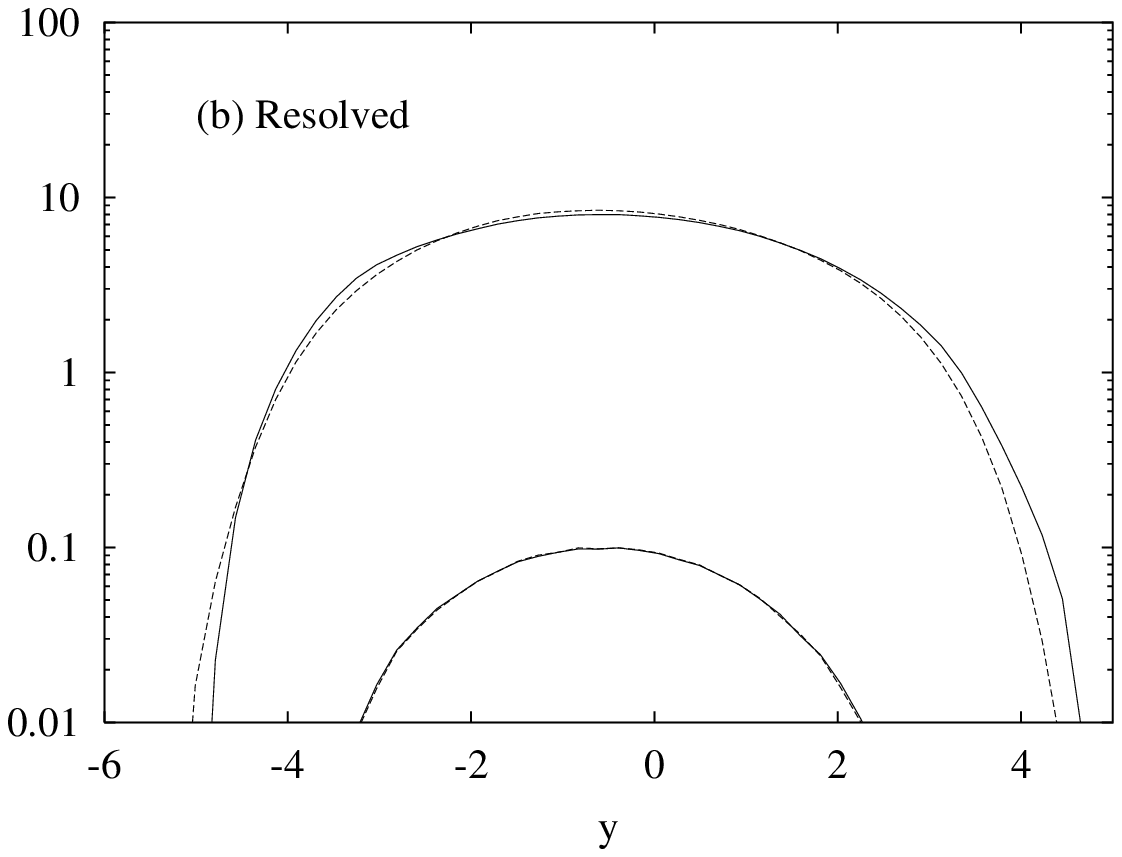, width=80mm}
\epsfig{file=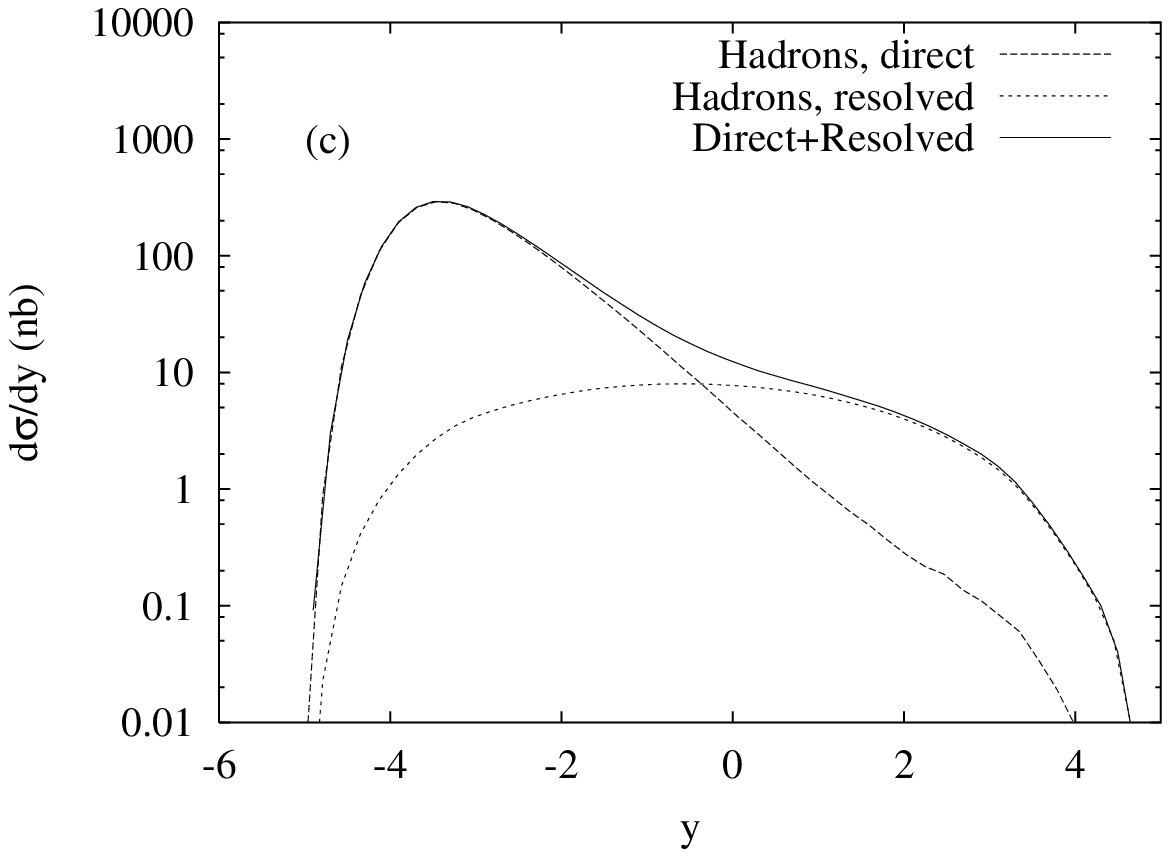, width=80mm}
\epsfig{file=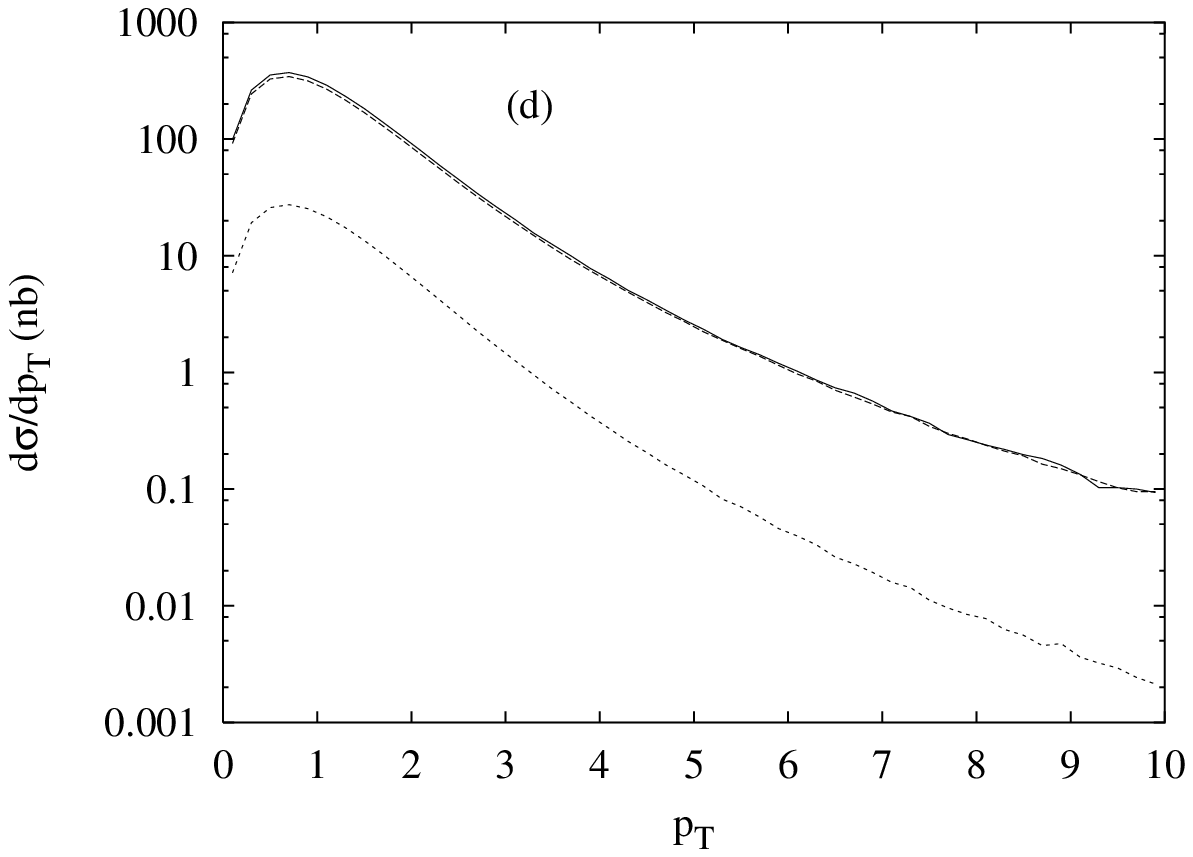, width=80mm}
\caption[junk]{
Leading order distribution of charmed hadrons and quarks in rapidity:
(a) direct and (b) resolved photons.
Comparison of resolved and direct processes in (c) rapidity and (d) transverse momentum.
}
\label{LO}
\end{figure}

To leading order, the massive matrix elements producing charm are the fusion
processes $\g \gamma \rightarrow \c \cbar$ (direct), $\g \g \rightarrow \c \cbar$ and
$\q \qbar \rightarrow \c \cbar$ (resolved). Fig.~\ref{LO} shows the distribution of
charmed quarks and charmed hadrons separated into these two classes. For direct photons
the hadrons are shifted in the direction of the proton beam, since both charm quarks
are colour-connected to the proton beam remnant. In a resolved event the photon also has
a beam remnant, so the charmed hadron is shifted towards the beam remnant it is
connected to. Also in this case the drag effect is a small-$\pt$ phenomenon.

\begin{figure}
\vspace*{2mm}
\mbox{
\epsfig{file=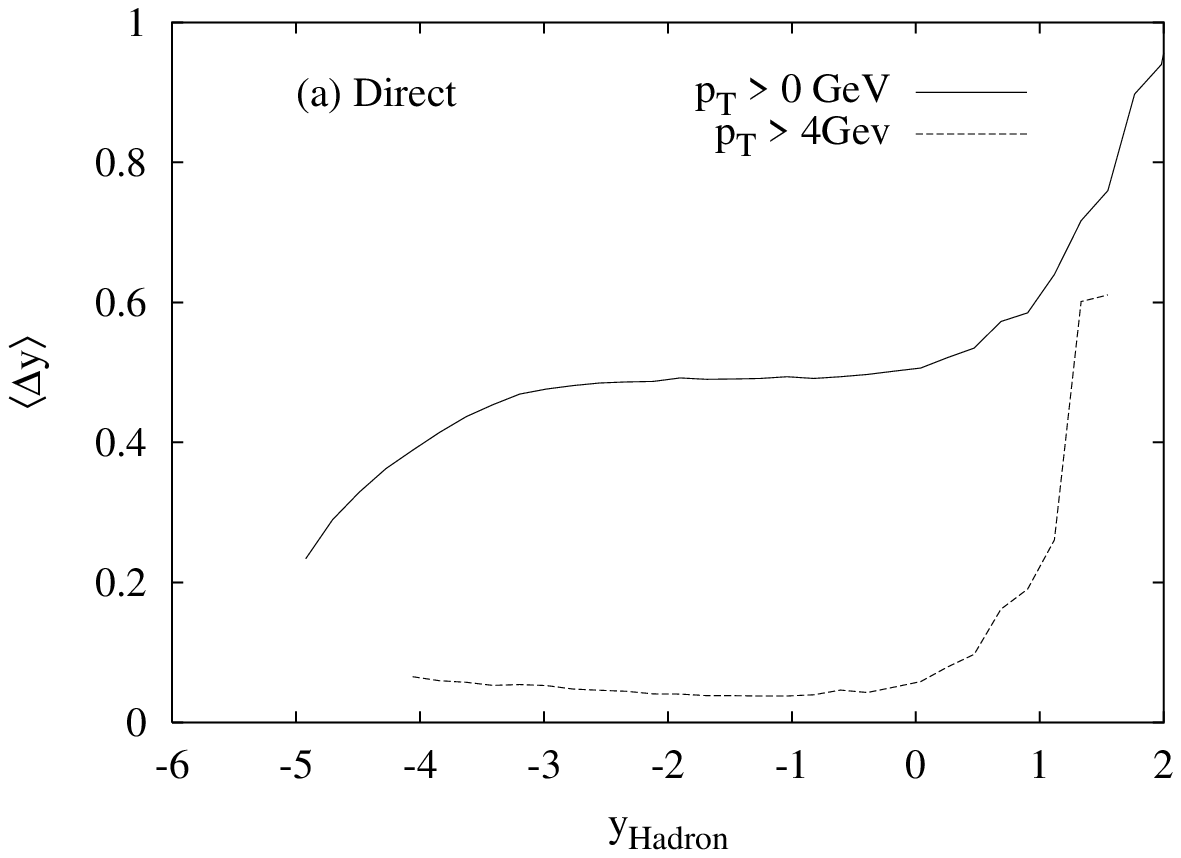, width=80mm}
\epsfig{file=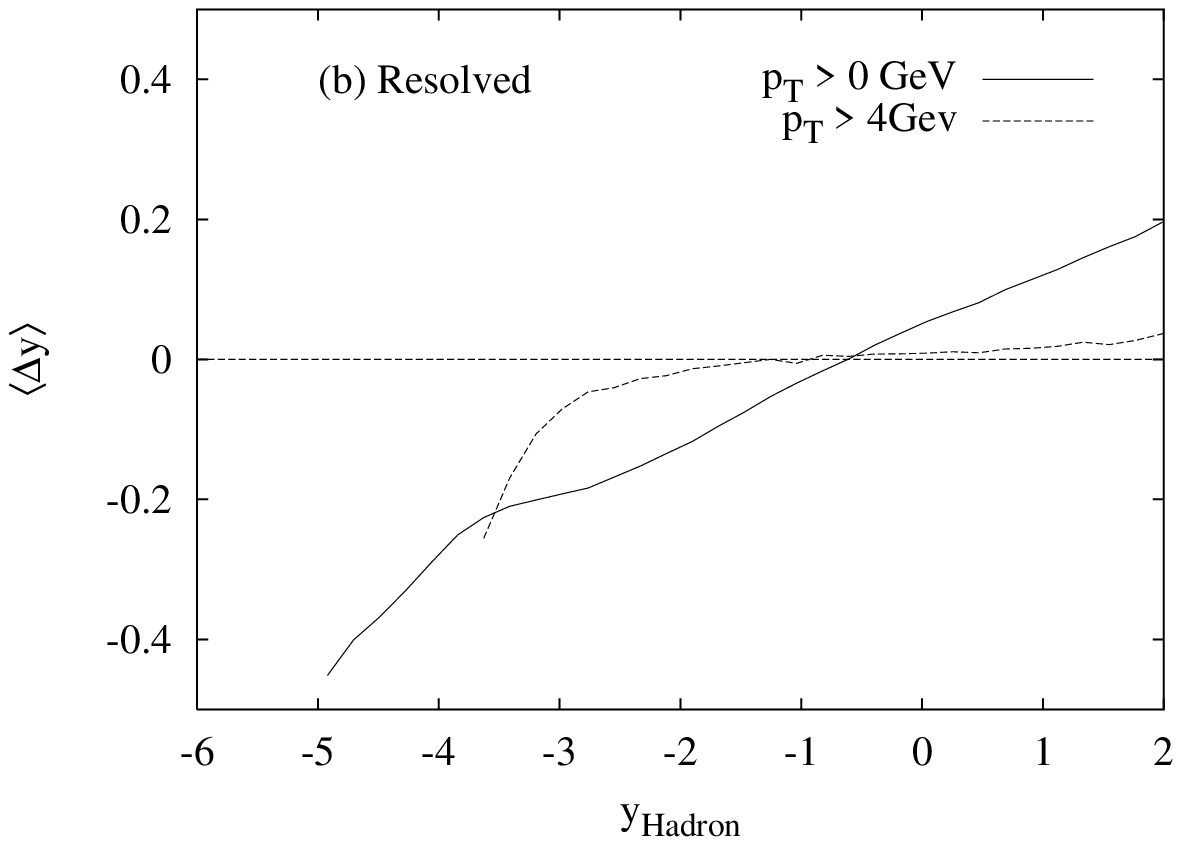, width=80mm}
}
\caption[junk]{
Rapidity shift
$\langle \Dy \rangle = \langle y_\mathrm{Hadron} - y_\mathrm{Quark} \rangle$
for (a) direct and (b) resolved photons as a function of rapidity.
}
\label{drag1}
\end{figure}

The drag effect is illustrated in Fig.~\ref{drag1} where the average rapidity shift in
the hadronization,
$\langle \Dy \rangle = \langle y_\mathrm{Hadron} - y_\mathrm{Quark} \rangle$,
is shown as a function of $y_\mathrm{Hadron}$.
For direct photons and central rapidities the shift is approximately constant. The
increasing shift for large rapidities is due to an increased
interaction between the proton remnant and the charmed quark when
their combined invariant mass is small. At large negative rapidities there is no
corresponding effect because there is no beam remnant there. The drop of
$\langle\Dy\rangle$ in this region is a pure edge effect;
only those events with below-average
$\Dy$ can give a very negative $y_\mathrm{Hadron}$. For resolved photons
the shift is in the direction of the proton and photon beam remnants. Note that
what is plotted is only the mean. The width of $\Dy$ is generally larger than the
mean, so the shift can go both ways. For example the quarks at very small rapidities ($y\lsim -5$)
in Fig.~\ref{LO}b will almost all be shifted with $\Dy>0$ but hadrons produced there will,
on the average, come from quarks produced at larger rapidities (i.e. $\Dy<0$).
Hence the apparent contradiction with Fig.~\ref{drag1}b by these edge effects.
The differences between these figures and Fig.~\ref{drag} stem exclusively from
differences in the event topology.

\begin{figure}
\vspace*{2mm}
\mbox{
\epsfig{file=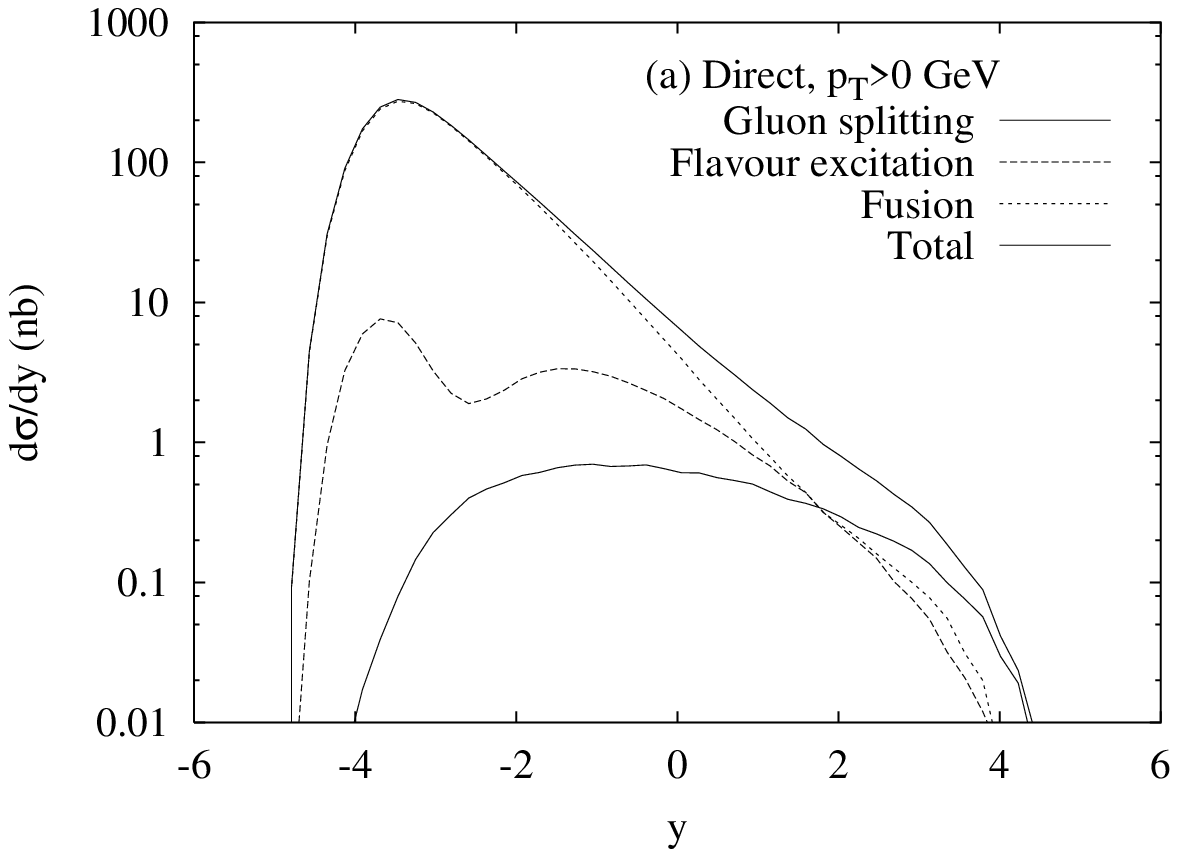, width=80mm, height=50mm}
\epsfig{file=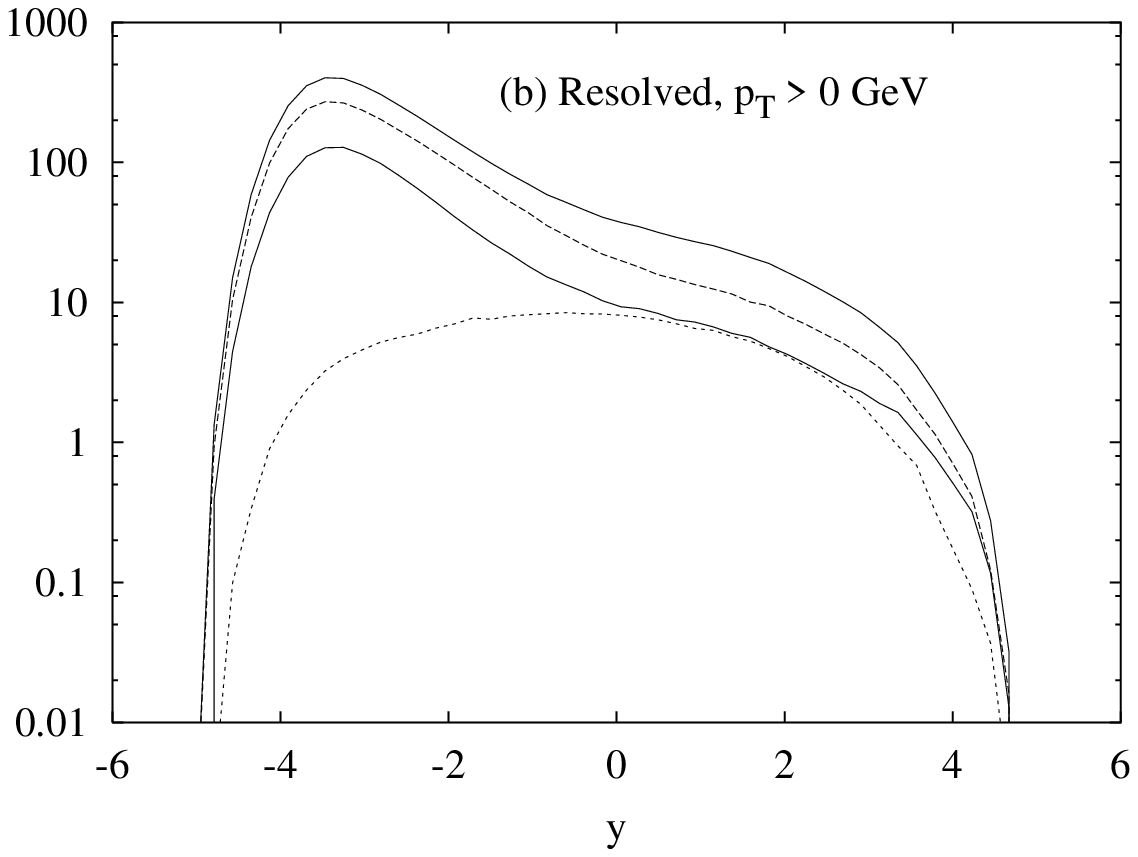, width=80mm, height=50mm}
}
\mbox{
\epsfig{file=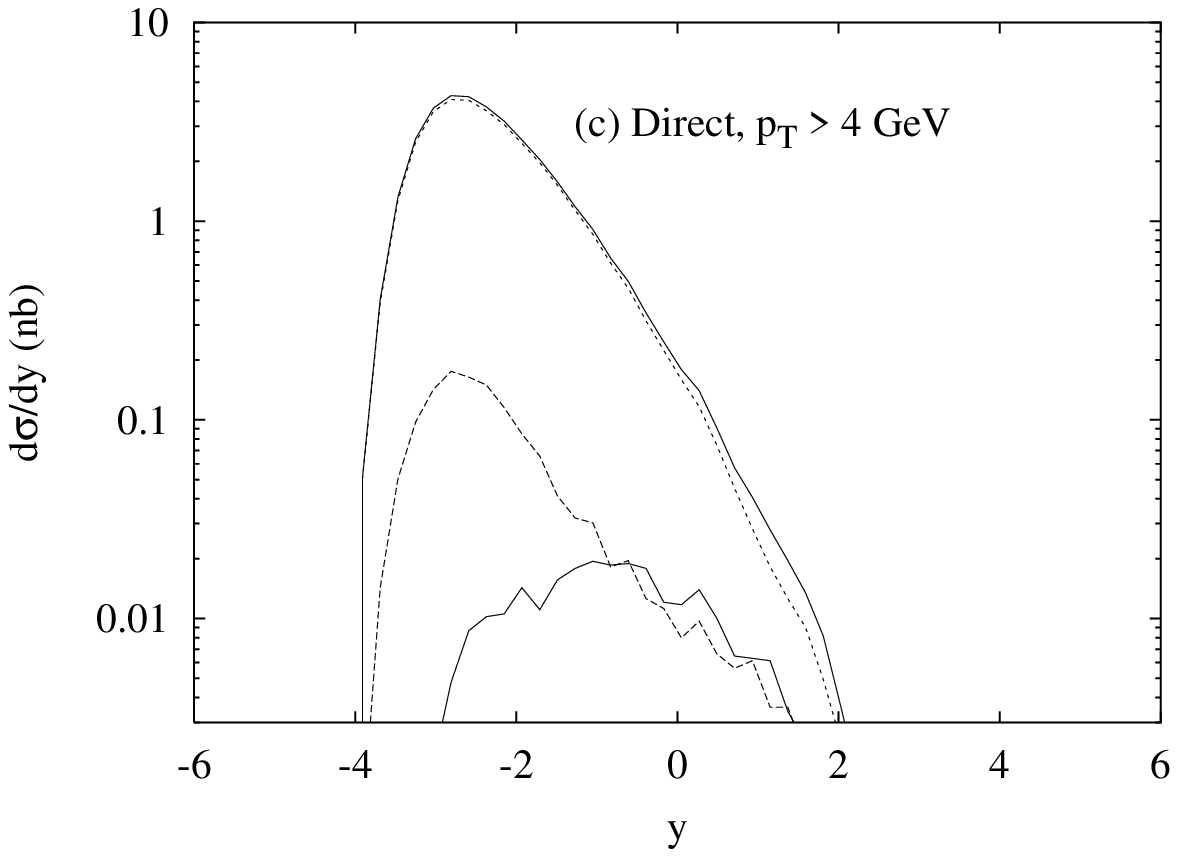, width=80mm, height=45mm}
\epsfig{file=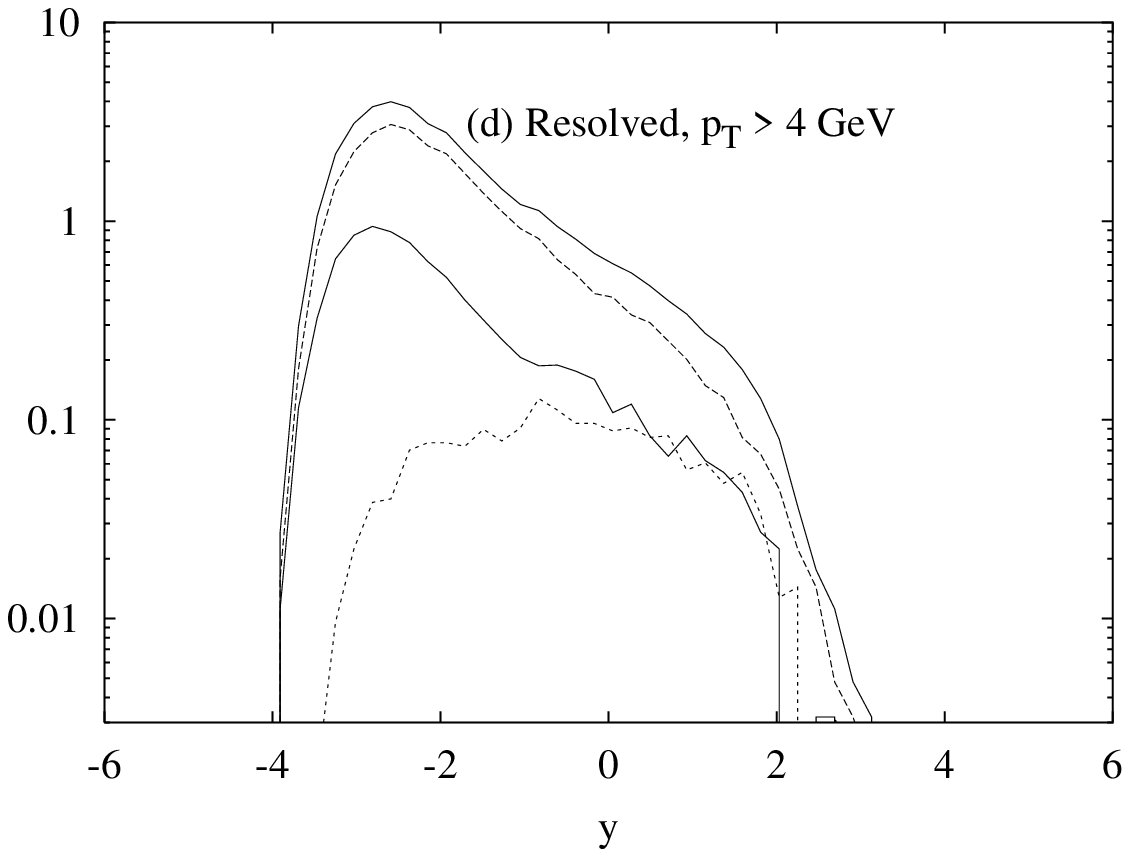, width=80mm, height=45mm}
}
\mbox{
\epsfig{file=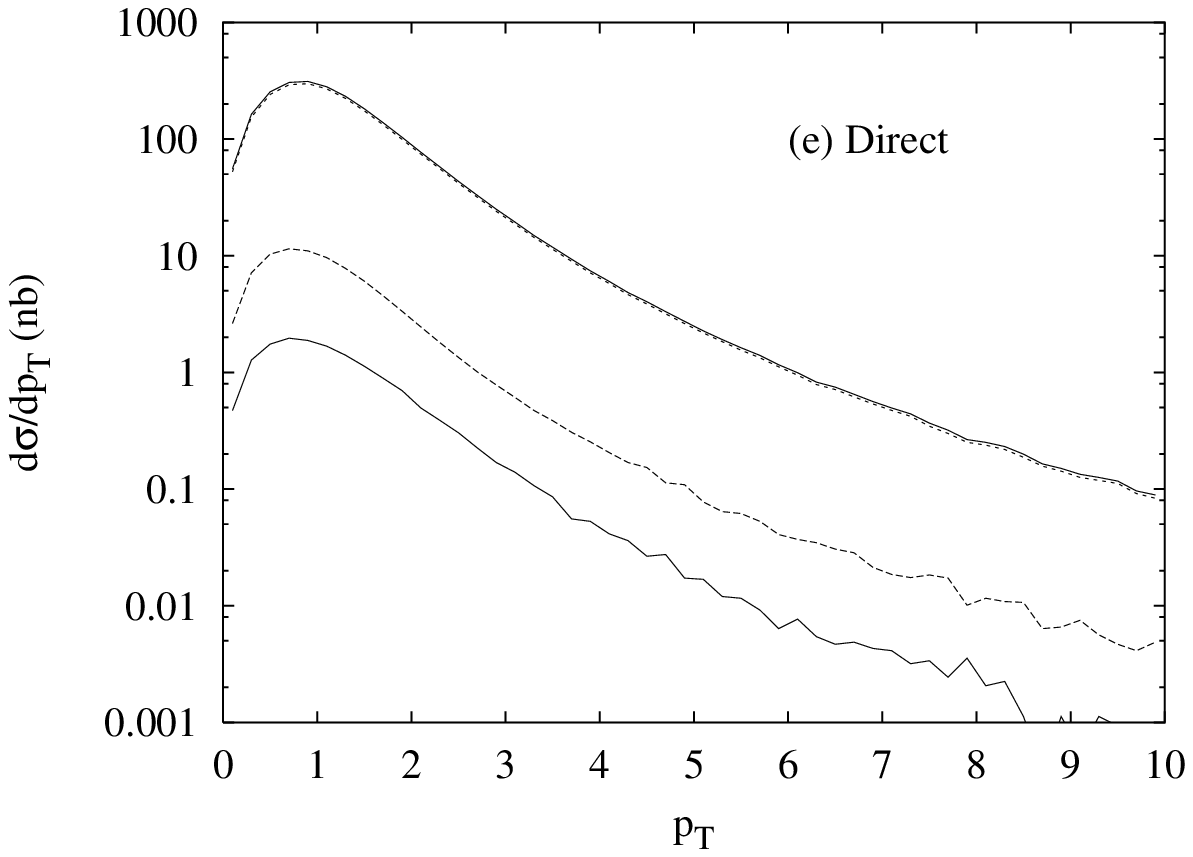, width=80mm, height=50mm}
\epsfig{file=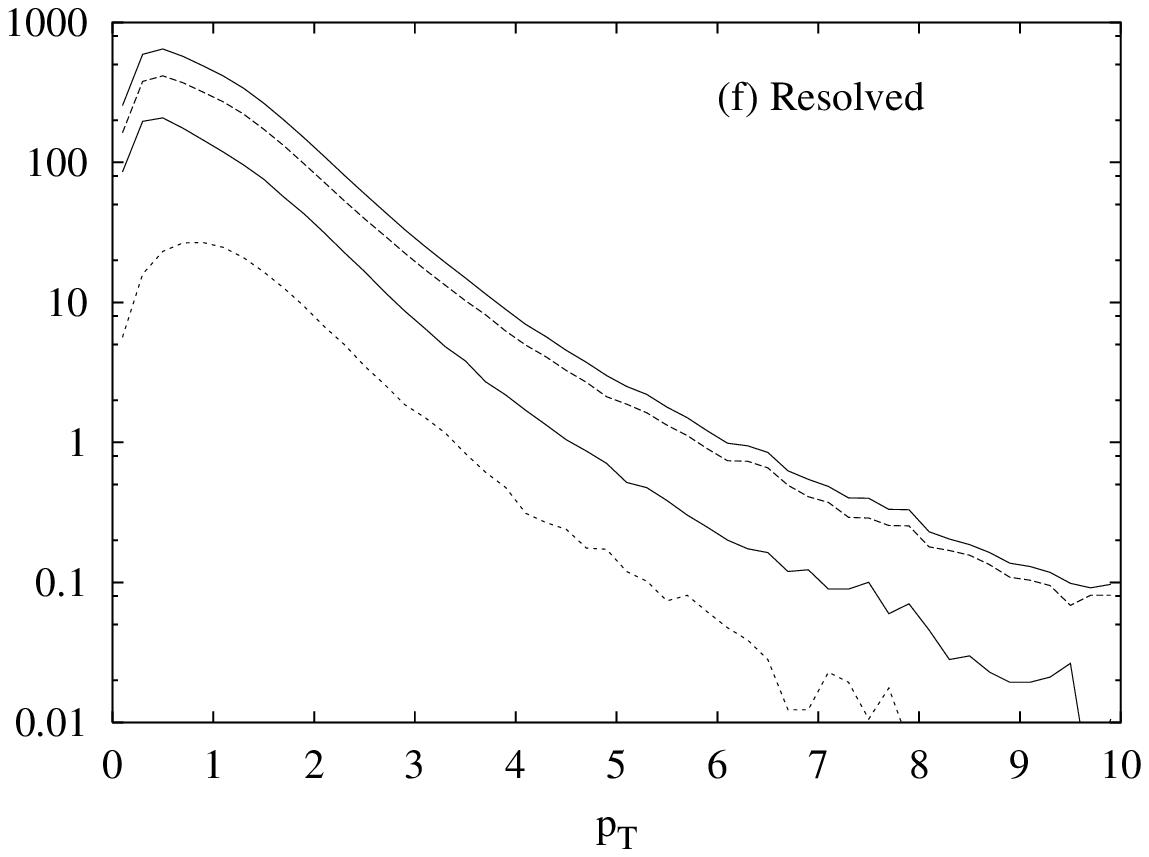, width=80mm, height=50mm}
}
\mbox{
\epsfig{file=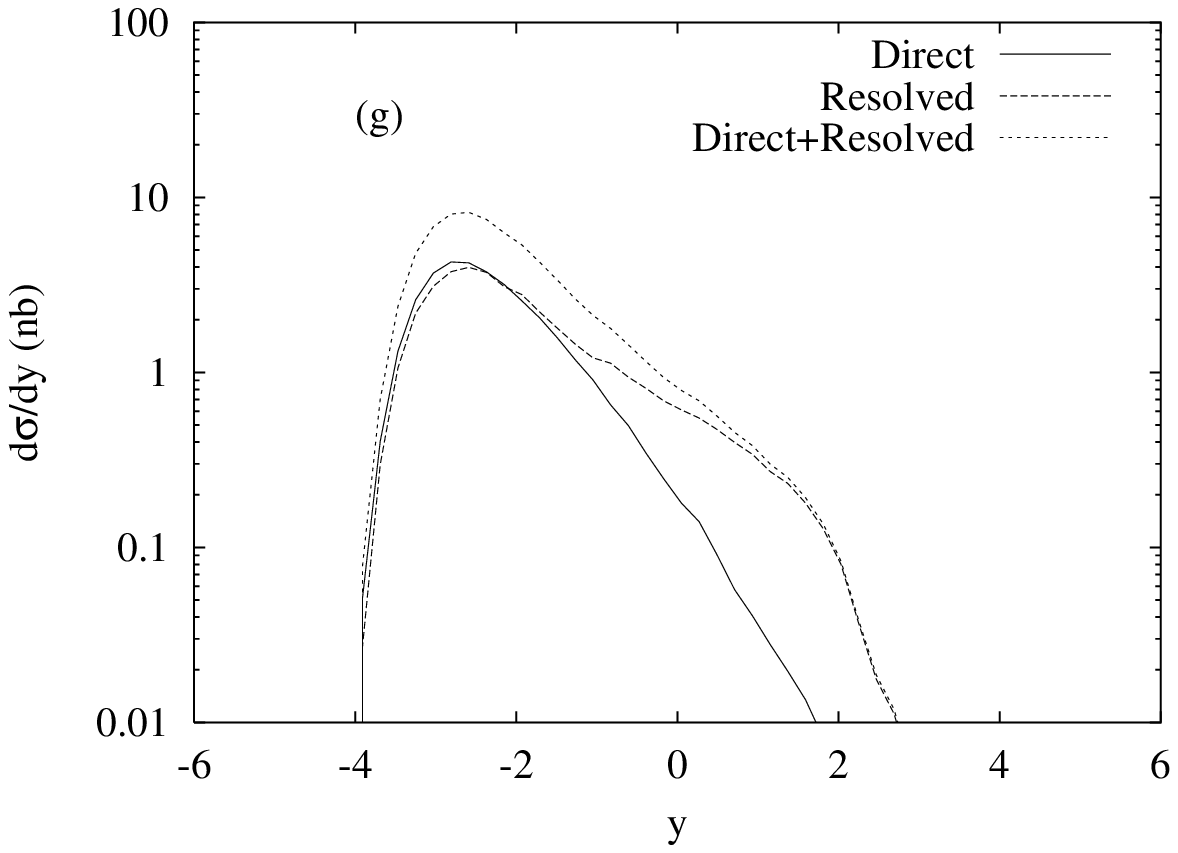, width=80mm, height=50mm}
\epsfig{file=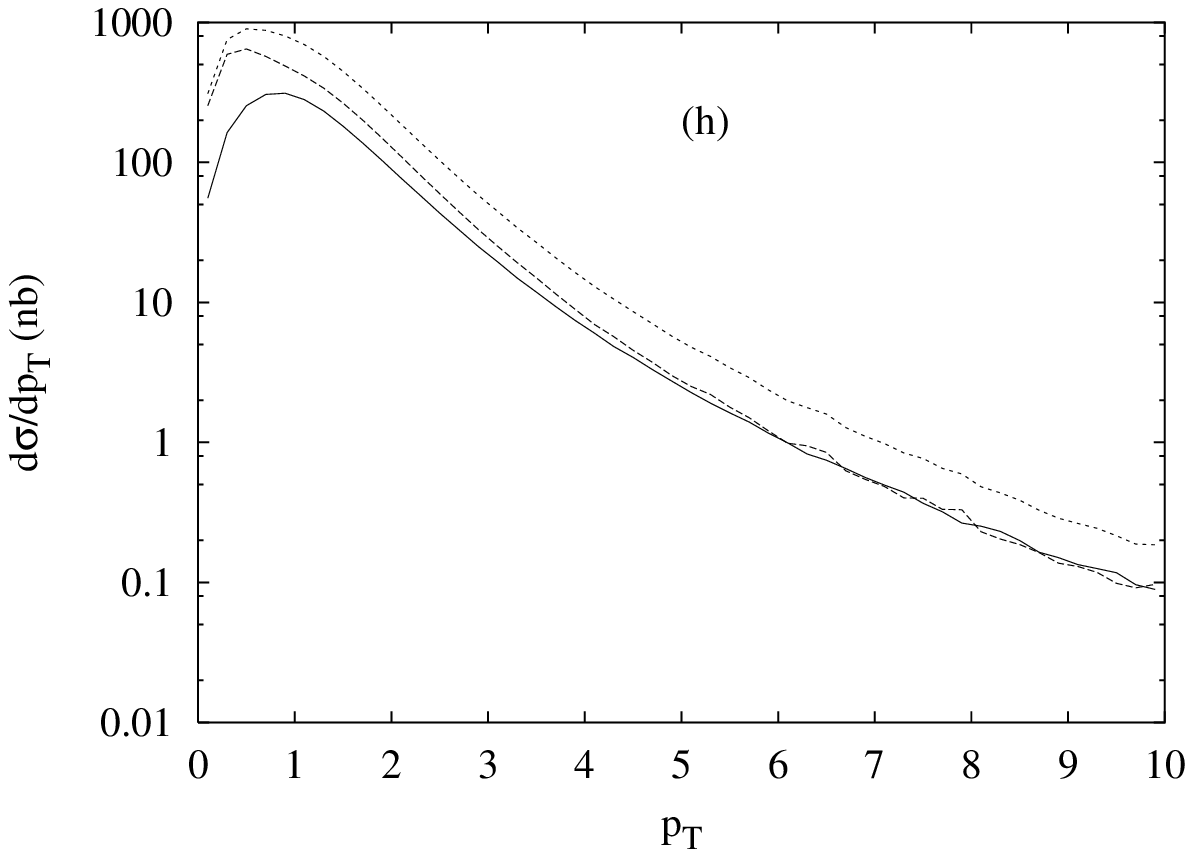, width=80mm, height=50mm}
}
\caption[junk]{The cross section for charm hadrons divided into different
production mechanisms and different photon structure.
(a) Direct and (b) resolved photons with $\pt > 0$ \textup{GeV}.
(c) Direct and (d) resolved photons with $\pt > 4$ \textup{GeV}.
(e) Direct and (f) resolved photons in $\pt$.
We add the components together for (g) rapidity ($\pt > 4$ \textup{GeV}) and (h) transverse momentum.
}
\label{HO}
\end{figure}

At HERA energies, flavour excitation and gluon splitting give large contributions
to the cross section.
In Fig.~\ref{HO} the cross section is divided into different production channels
for direct and resolved photons. We note that the cross sections are of the same
order of magnitude, unlike the results in lowest order, and the major contribution
in the resolved case is flavour excitation.
The details of course depend on the parameterization of the photon structure.

The double peak structure in the flavour excitation process for direct photons
is because the charm quark in the beam remnant at low $\pt$ is also included.
This peak disappears when a $\pt$ cut is introduced (Fig.~\ref{HO}c). 

\section{Summary and outlook}

In this study, we have further developed a model for the production and
hadronization of heavy quarks in hadronic collisions. While the emphasis 
lies on the modelling of the nonperturbative phenomena, the two cannot
be fully separated and therefore have to be considered in common.
Thus the road we take for the production stage --- using a three-component
picture of pair creation, flavour excitation and gluon splitting 
together giving the heavy-flavour cross section --- is not very
economical if viewed in isolation. A next-to-leading-order matrix
elements description could do the job much better, at least at current 
energies; only at higher energies could the possibility of extensive
showering histories make a fixed-order approach inferior to our 
leading-log showering one. The real difference is instead that our 
approach also defines the environment in which the heavy quarks are 
produced: partons from the hard interaction, from its associated showers
and from beam remnants, joined in a specific order by colour-confinement 
strings. And, as we have attempted to show, it is essential to 
have that background if one wants to understand the production of
the heavy hadrons, not only the heavy quarks.

This article represents the third main one on heavy flavours from the 
Lund group. The basic ideas were already established almost twenty years 
ago, so the path taken since has been evolutionary rather than 
revolutionary. At the time of the first study \cite{ABG}, neither the 
model nor the data were good enough more than to hint at the validity 
of the basic principles. In the subsequent years the modelling was 
gradually improved \cite{Pythia}, and fits to some model parameters
were performed by at least one experimental collaboration \cite{E791}.  
In our more recent study \cite{our} it was therefore possible to 
start at a higher level, and introduce a technically somewhat more
sophisticated re-implementation of the same basic ideas. This trend is
continued in the current article, where some further model details have 
been improved. The main difference, however, is that we here have 
considered a wider range of observables, for more different production 
channels, and for several experimental configurations.
  
The basic ideas are not particularly controversial today, but the outcome 
may often be unexpected and counterintuitive. For instance, it is well
known, from LEP and other $\e^+\e^-$ machines, that the heavy hadron 
only takes a fraction of
the heavy-quark energy, i.e. that the pull of the string in the
hadronization stage `slows down' the heavy quark. However, before one 
has studied the colour topology of heavy-flavour production in hadronic 
events, and done some trivial Lorentz boost brain gymnastics, it is not 
equally obvious that exactly the same phenomenon could `speed up'
the heavy quark here. Or: it is not unreasonable that a heavy quark could
pick up one of the beam valence flavours to form a hadron, but the extent 
to which this can happen over a wide range of longitudinal and transverse
momenta may come as a surprise.

The studies here have also put the finger on a few other interesting 
phenomena, such as:

\begin{Itemize}
\item The extrapolation from charm to bottom quarks.
\item The importance of heavy flavour production through
flavour excitation and gluon splitting.
\item The kinematics of backward evolution in the initial state shower.
\item Scale choices in the parton shower.
\item The importance of colour flow.
\item The matching of cluster decay and string fragmentation of low-mass colour singlets.
\item The details of the collapse mechanism.
\item The choice and importance of beam remnant distribution functions.
\item The choice and importance of intrinsic $\kt$ smearing.
\item High-$\pt$ asymmetries.
\item The influence of the photon structure.
\item The limitations of the model.
\end{Itemize}

While not as spectacular as the colour drag and flavour asymmetry ones 
above, they help to put flesh on the bones of our understanding of 
hadronization. Experimental results on the yet untested features clearly 
would be welcome.

A topic not discussed in this paper is cosmic ray physics where the momentum
spectra of charm and bottom hadrons has implications for the rate of prompt leptons
and neutrinos in atmospheric cascades \cite{thunman}. Also here our improved modelling may
affect the traditional flux calculations.

It is important to remember that the predictions can be wrong.
Hopefully not in a qualitative fashion, but quite possibly in a 
quantitative one. For instance, it is all well to assume we can 
control the colour topologies at fixed-target energies, where there
is a very small number of participating partons and thereby of separate 
string pieces. At high energies, the more extensive parton showers and
especially the increased rate of multiple parton--parton interactions 
could well mess up our tidy picture of colour flow, and thereby of 
single-particle spectra and correlations. If so, we would like to 
believe that heavy flavours here could be used as a probe for such 
effects.

Of course, our approach is not the only one that has been proposed
for the hadronization of heavy quarks \cite{variousmodels}. Especially
the intrinsic charm model \cite{intrinsic} is still very actively
pursued. It is not unlikely that several mechanisms may coexist,
but we have also encountered no evidence to indicate that the
ones outlined by us are not the dominant ones. However, as always,
more data may provide new insights also in this regard.
   
A final reflection is that B physics will remain a major 
topic of study for many years to come, because of the interest in CP 
violation studies. Furthermore, many of these studies will require
a tagging of both beauty hadrons in an event. One can therefore foresee
large data sets that will allow many detailed tests, beyond the ones 
shown here. It also appears plausible that charm hadron
samples can be extracted as a by-product, where, as we have seen, 
many of the predicted effects are larger. Hopefully we will therefore 
enjoy a gradually improved level of understanding.

\end{document}